\documentclass[11pt,a4paper,nofootinbib,showpacs,preprint,utf8]{article}
\usepackage{jheppub}
\pdfoutput=1
\usepackage{amsmath,amssymb}
\usepackage{amsfonts}
\usepackage{epsfig}
\usepackage{graphicx}
\usepackage{amssymb}
\usepackage{bbm}
\usepackage{pifont}
\usepackage[usenames,dvipsnames]{xcolor}
\definecolor{warmblack}{rgb}{0.0, 0.26, 0.26}
\definecolor{mediumtealblue}{rgb}{0.0, 0.33, 0.71}
\usepackage[compat=1.1.0]{tikz-feynman}
\usepackage{tikzsymbols}
\usepackage{comment}
\usepackage[normalem]{ulem}
\usepackage{orcidlink}

\usepackage{dcolumn}
\usepackage{bm}
\usepackage{graphicx}
\usepackage{amssymb,amsmath}
\usepackage{multirow}
\usepackage{afterpage}
\usepackage{bbm}
\usepackage{enumerate}   
\PassOptionsToPackage{colorlinks=true
,urlcolor=magenta
,anchorcolor=blue
,citecolor=blue
,filecolor=blue
,linkcolor=blue
,menucolor=blue
,linktocpage=true
,pdfproducer=medialab
,pdfa=true
}{hyperref}

\usepackage{tikz-feynman}
\tikzfeynmanset{compat=1.1.0}
\usetikzlibrary{trees}
\usetikzlibrary{decorations.pathmorphing}
\usetikzlibrary{decorations.markings}

\usepackage{array,booktabs}

\usepackage{slashed}
\usepackage{epsfig,psfrag,rotating,soul}
\usepackage{rotfloat}

\allowdisplaybreaks

\newcommand{\beq}{\begin{equation}}
\newcommand{\eeq}{\end{equation}}

\usepackage{xspace}

\newcommand{\ba}{\begin{array}}
\newcommand{\ea}{\end{array}}
\newcommand{\bd}{\begin{displaymath}}
\newcommand{\ed}{\end{displaymath}}
\newcommand{\besub}{\begin{subequations}}
\newcommand{\eesub}{\end{subequations}}
\newcommand{\bea}{\begin{eqnarray}}
\newcommand{\eea}{\end{eqnarray}}

\def\q2 {q^2}

\def\bt{\begin{table}}
\def\et{\end{table}}

\tikzfeynmanset{every dot/.style={minimum size=5pt}}

\definecolor{mygray}{gray}{0.85} 
\definecolor{myblue}{cmyk}{0.65, 0.37, 0.0, 0.19}

\title{Electroweak phase transition in SMEFT: Gravitational wave and collider complementarity}

\author[a,b]{{Sahabub Jahedi,}}
\author[c]{{Indrajit Saha}}
\author[c]{{and Abhik Sarkar}}

\affiliation[a]{State Key Laboratory of Nuclear Physics and Technology, Institute of Quantum Matter, South China Normal University, Guangzhou 510006, China}
\affiliation[b]{Guangdong Basic Research Center of Excellence for Structure and Fundamental Interactions of Matter, Guangdong Provincial Key Laboratory of Nuclear Science, Guangzhou 510006, China}
\affiliation[c]{Department of Physics, Indian Institute of Technology Guwahati, Assam 781039, India}

\emailAdd{sahabub@m.scnu.edu.cn}
\emailAdd{s.indrajit@iitg.ac.in}
\emailAdd{sarkar.abhik@iitg.ac.in}

\abstract{We study the first-order electroweak phase transition (FO-EWPT) within the Standard Model Effective Field Theory (SMEFT) framework induced by dimension-6 operators. Such phenomena can be probed independently via \textit{di}-Higgs production at the collider experiments as well as via the detection of gravitational waves (GW). There are three (one) dimension-6 SMEFT operators that simultaneously modify the Higgs potential at tree (1-loop) level and contribute to the \textit{di}-Higgs production at the hadron colliders. With \textit{di}-Higgs production being suppressed at current LHC runs, we aim to probe this production at high luminosity (HL) and high energy (HE) runs of the LHC to achieve better sensitivity of dimension-6 SMEFT operators. The correlations among these operators are analyzed in the context of probing FO-EWPT, emphasizing the complementarity between future GW observations and upgraded LHC searches.}

\makeatletter
\gdef\@fpheader{}
\makeatother

\begin{document}
\maketitle

\section{Introduction}
\label{sec:intro}
The discovery of the Higgs boson at the Large Hadron Collider (LHC) \cite{ATLAS:2012yve,CMS:2012qbp} and the precise measurement of its properties, which closely align with the predictions of the Standard Model (SM), have further reinforced confidence in the SM. However, the exact shape of the Higgs potential remains unknown, making the determination of the nature of the electroweak phase transition (EWPT) a crucial objective. In particular, a first-order EWPT (FO-EWPT) furnishes the necessary conditions to generate the observed baryon asymmetry of the universe (BAU) \cite{Kuzmin:1985mm,Gavela:1993ts} within the framework of electroweak baryogenesis. The SM, with its precisely measured parameters from particle collision experiments, can only accommodate an adiabatic crossover transition at the weak scale \cite{Gurtler:1997hr,Laine:1998jb,Csikor:1998eu,Aoki:1999fi}. Within the SM, the FO-EWPT occurs only if the Higgs boson mass is approximately below 70 GeV \cite{Kajantie:1996mn}.  Therefore, a FO-EWPT serves as a natural testing ground for physics beyond the SM (BSM). The FO-EWPT proceeds through the nucleation of true vacuum bubbles within the false vacuum background. Once nucleated, these bubbles expand and eventually collide with each other, leading to the percolation of the true vacuum phase and completion of the transition. This cosmological FO-EWPT generates gravitational wave (GW) spectrum, which can be detected by current and future interferometric experiments such as Laser Interferometer Space Antenna (LISA) \cite{Caprini:2015zlo,LISA:2017pwj}, DECi-hertz Interferometer Gravitational wave Observatory (DECIGO) \cite{Seto:2001qf,Kawamura:2006up}, and Big Bang Observer (BBO) \cite{Crowder:2005nr,Corbin:2005ny}. This makes FO-EWPTs a valuable complementary probe of BSM physics, accessible both at the colliders and through cosmic observations.

With no signs of new particles at the LHC, an effective field theory (EFT) approach provides a sensible framework to incorporate the influence of unknown new physics (NP) through higher-dimensional operators, which are otherwise excluded from the SM due to its renormalizability constraint. The Standard Model Effective Field Theory (SMEFT) \cite{Buchmuller:1985jz,Grzadkowski:2010es} provides a systematic approach to encoding potential NP effects through higher-dimensional operators constructed from SM fields. The SMEFT framework is one of the well-motivated scenarios to searching NP in a model-independent way. Dimension-6 SMEFT operators can modify the Higgs potential in a way that triggers FO-EWPT. Under this framework, a substantial amount of work has been devoted to studying the FO-EWPT \cite{Zhang:1992fs, Grojean:2004xa, Bodeker:2004ws, Damgaard:2015con, Harman:2015gif, deVries:2017ncy, DeVries:2018aul, Kanemura:2020yyr, Zhu:2025pht}. In addition to FO-EWPT, GW analyses have been performed in Refs.~\cite{Delaunay:2007wb, Cai:2017tmh, Chala:2018ari, Ellis:2019flb, Zhou:2019uzq, Banerjee:2024qiu, Gazi:2024boc}\footnote{A comprehensive review of the EWPT and GW phenomenology is provided in \cite{Athron:2023xlk}.}. Investigations of EWPT using a gauge-invariant approach have been carried out in Refs.~\cite{Patel:2011th, Camargo-Molina:2021zgz, Qin:2024idc}. Studies of the EWPT based on dimensionally reduced 3-dimensional high-temperature EFTs have been presented in Refs.~\cite{Qin:2024dfp, Camargo-Molina:2024sde, Chala:2025aiz, Chala:2025cya}, which are generally regarded as more accurate for making quantitative statements about the EWPT.  In this work, since our primary goal is to access the collider complementarity of the FO-EWPT and GW allowed parameter space, we first revisit how the Higgs potential modifications influence the EWPT within traditional finite-temperature calculations in 4 spacetime dimensions in the early Universe and lead to the generation of the GW. We then study the prospect of probing these scenarios at the high Luminosity (HL) and high energy (HE) run of the LHC via \textit{di}-Higgs production. The \textit{di}-Higgs production at the HL- and HE-LHC serves as a crucial probe of the trilinear Higgs coupling, which remains weakly constrained by the ATLAS \cite{ATLAS:2022jtk} and CMS \cite{CMS:2022dwd} experiments at the current LHC. Precise measurement of this coupling is essential for understanding the structure of the Higgs potential, the nature of electroweak symmetry breaking (EWSB), and potential BSM signatures. By employing an Artificial Neural Network (ANN)-based~\cite{Feickert:2021ajf} signal-background discriminator, we aim to enhance the sensitivity to the dimension-6 effective couplings along with their correlations and assess the viability of the EWPT within this improved sensitivity regime and complementarity with the GW signal.

Our paper is organised as follows: in Section \ref{sec:eft}, we list relevant dimension-6 SMEFT operators that modify the Higgs potential. The pedagogy of FO-EWPT is summarised in Section \ref{sec:FO-EWPT}. The GW spectrum originated from FO-EWPT is discussed in Section \ref{sec:gw}. Existing constraints on the effective couplings are presented in Section \ref{sec:collider}. In Section \ref{sec:di-higgs}, we illustrate the sensitivity of the SMEFT operators via \textit{di}-Higgs production at the HL-LHC and HE-LHC and the feasibility to probe FO-EWPT. We finally summarize our findings and conclude in Section \ref{sec:con}.

\section{Higgs potential and SMEFT}
\label{sec:eft}
The SMEFT is based on the assumption that any new particles introduced by extensions of the SM have masses larger than the electroweak scale. Under this framework, the effective theory at the electroweak scale consists of the SM supplemented by a series of gauge-invariant, higher-dimensional operators constructed from SM fields. These operators account for the effects of integrating out the heavy particles beyond the SM. The complete Lagrangian associated with all the higher-dimensional effective operators is written as
\beq
\mathcal{L}=\mathcal{L}_{\text{SM}}+\sum_{i,d}\frac{C_i}{\Lambda^{d-4}}\mathcal{O}_i,
\eeq
where $C_i$'s are the dimensionless Wilson coefficients (WCs), $\Lambda$ is scale of NP, and $\mathcal{O}$'s are higher dimensional effective operators of dimension $d$, constructed out of the SM fields. The relevant dimension-6 operators  responsible for the modification of the Higgs potential at tree level are given by
\begin{align} \label{eq:1}
\begin{split}
\mathcal{O}_{H} =(H^{\dagger}H)^3,\hspace{0.5cm}
\mathcal{O}_{H \Box}=(H^{\dagger}H)\Box (H^{\dagger}H),\hspace{0.5cm}
\mathcal{O}_{H D}=|H^{\dagger}D^{\mu} H|^2,
\end{split}
\end{align}
where $H$ is the SM Higgs doublet, $D_{\mu}$ is the covariant derivative, and $\Box =\partial^{\mu} \partial_{\mu}$ is the d'Alembert operator. The dimension-6 contribution relevant to the Higgs potential is written as
\begin{align}
\begin{split}
\mathcal{L}^{6}_{\text{EFT}}=& \frac{C_{H\Box}}{\Lambda^{2}}\; |H|^2 \Box |H|^2 +
  \frac{C_{HD}}{\Lambda^{2}}\; |HD_\mu H|^2 + \frac{C_H}{\Lambda^{2}}\; |H|^6,\\
=&C_{\text{kin}} \varphi^2 (\partial_{\mu} \varphi)^2 + \frac{1}{8} C_H \varphi^6,
\end{split}
\label{eq:l6}
\end{align}
where $\sqrt{2}\,H^{T}=(0 \;\ \varphi)=(0 \;\ v+h)$ and $C_{\text{kin}}=(C_{HD}/4\Lambda^{2}) - (C_{H\Box}/\Lambda^{2})$. With an appropriate redefinition of canonical Higgs field $\varphi =\varphi +\frac13 C_{\rm
  kin} \varphi^3 +{\cal O} (C_{\rm kin}^2  \varphi^5)$, eq.~\eqref{eq:l6} can be expressed as
\beq
\mathcal{L}^{6}_{\text{EFT}} \simeq \frac{1}{2} (\partial \varphi)^2- \left(
\frac{1}{2} a_2 \varphi^2 +\frac{1}{4} a_4 \varphi^4 + \frac{1}{6} a_6 \varphi^6 \right),
\eeq
with
\begin{align}
a_2 =\mu^2, \quad
      a_4 =\lambda- \frac43  C_{\rm kin} \mu^2,\quad
  a_6= -\frac{3C_H}{4\Lambda^2} - 2C_{\rm kin} \lambda.
  \label{eq:a.exp}
\end{align}
The parameters $a_2$ and $a_4$ are determined by the measured values of the Higgs vacuum expectation value $v = 246$ GeV and the Higgs mass $m_{h^0} = 125$ GeV through the relation:
\beq
  \partial_\varphi V|_{\varphi=v} =0, \quad\partial_\varphi^2 V|_{\varphi=v} =m_{h0}^2.
\label{eq:params}
\eeq
Therefore, the tree-level Higgs potential in terms of these physical quantities is given by
\beq
V_{\text{tree}} = -\frac{1}{4}(m_{h0}^2 -2 a_6 v^4) \varphi^2 +\frac{1}{4}\left(\frac{m_{h0}^2}{2v^2} -2 a_6 v^2\right)\varphi^4 +\frac{1}{6} a_6 \varphi^6.
\eeq

On the other hand, dimension-6 operators involving the Higgs field and top quarks induce modifications to the Higgs potential at the one-loop level. The relevant operators are given by
\begin{align}
\begin{split}
&(\mathcal{O}_{uH})_{ij}=(H^{\dagger}H)(\bar{q}_i u_j \tilde{H}),\\
&(\mathcal{O}_{Hq}^{(1)})_{ij}=(H^{\dagger}i\overleftrightarrow{D}_{\mu}H)(\bar{q}_i \gamma^{\mu}q_j),\\
&(\mathcal{O}_{Hq}^{(3)})_{ij}=(H^{\dagger}i\overleftrightarrow{D}_{\mu}^IH)(\bar{q}_i \tau^I \gamma^{\mu}q_j),\\
&(\mathcal{O}_{Hu})_{ij}=(H^{\dagger}i\overleftrightarrow{D}_{\mu}H)(\bar{u}_i  \gamma^{\mu}u_j), 
\end{split}
\label{eq:top_higgs_ops}
\end{align}
with the derivative
\begin{align}
    H^{\dagger}\overleftrightarrow{D}^I_{\mu} H=H^{\dagger} \tau^I D_{\mu} H-(D_{\mu}H)^{\dagger}\tau^I H.
\end{align}
Here, $q$ is the $SU(2)_L$ quark doublet, $u$ the right-handed up-type quark, quark-flavor indices $i,j$, an $SU(2)_L$ index $I$, and $\tau^I$ are the Pauli matrices. The correction due the operator $(\mathcal{O}_{uH})_{ij}$ to the Higgs potential is 
written as
\begin{align}
 V_{uH}= \frac{3(C_{uH})_{33}}{32\pi^2\Lambda^2}Y_t  \left\{\left(7-3\ln \left[\frac{m_t^2}{v^2}\right]\right)\cdot\varphi^2(\partial_{\mu}\varphi)^2 +Y_t^2\left(-1+\ln\left[\frac{Y_t^2 \varphi^2}{2v^2}\right]\right)\varphi^6\right\},
\label{eq:CuHloop}
\end{align}
where $Y_t$ is top-Higgs Yukawa coupling and $m_t$ is the top quark mass.
For other three operators $\mathcal{O}_{Hq}^{(1)}$, $\mathcal{O}_{Hq}^{(3)}$, and $\mathcal{O}_{Hu}$, although they contain both Higgs and top-quark fields, the neutral Higgs field in the derivative of the these operators cancel out. Consequently, they do not contribute to the Higgs potential \cite{Hashino:2022ghd}. Since, only the top quark flavored diagonal combination of $(\mathcal{O}_{uH})_{ij}$, i.e., $(\mathcal{O}_{uH})_{33}$ is relevant here, so to simplify the operator notation to $(\mathcal{O}_{uH})_{33} \equiv \mathcal{O}_{tH}$ and the corresponding WC to $(C_{uH})_{33}/\Lambda^{2} \equiv C_{tH}/\Lambda^{2}$ and corresponding modified Higgs potential is denoted as $V_{uH} \equiv V_{tH}$. Therefore, total SMEFT contribution to the Higgs potential is $V_{\text{tot}}=V_{\text{tree}}+V_{tH}$. It is worthwhile to note that, beyond these four SMEFT operators, additional operators can also contribute to the realization of the EWPT. However, their impact on di-Higgs production at hadron colliders is expected to be subleading and is therefore not relevant for the present analysis.

\section{First order electroweak phase transition}
\label{sec:FO-EWPT}
The dynamics of the EWPT are captured by the effective potential, accounting for Coleman-Weinberg quantum effects as well as finite-temperature contributions:
\begin{align}
V_{\rm eff}\left(\varphi, T\right) = V_{\text{tot}} + V_{\rm CW} + V_{\text{ct}}+V_{\rm th} + V_T^{\rm ring}.
\label{eq:veff}
\end{align}
The Coleman-Weinberg (CW) potential \cite{Coleman:1973jx} under $\overline{\mathrm{MS}}$ scheme and dimensional regularisation is expressed as
\beq 
V_{\rm CW} =
\sum_i (-1)^{n_f} \frac{n_i}{(8\pi)^2}
\left( M_i(\varphi)^4\left(\ln \left[\frac{M_i^2(\varphi)}{\mu^2}\right]-\mathcal{C}_i\right)\right),
\label{eq:vcw}
\eeq
where $i$ runs over particle species, $n_i$ and $m_i$ are the degrees-of-freedom (dof) and field-depended masses of $i$'th particle, respectively. $(-1)^{n_f}=+(-)1$ for bosons (fermions). $\mu=246$ GeV is the renormalization scale and $\mathcal{C}_i=5/6~(3/2)$ for gauge bosons (scalars and fermions). Inclusion of CW potential (eq.~\eqref{eq:vcw}) induces a shift in Higgs vev and mass, which can be compensated by introducing a suitable counterterm ($V_{\text{ct}}$) into the effective potential. The counter term is
\begin{equation}
    V_{\text{ct}}=-\frac{\delta m^2}{2}\varphi^2+\frac{\delta \lambda_{\phi}}{2}\varphi^4,
\end{equation}
which is derived by solving
\begin{equation}
    \frac{\partial(V_{\text{CW}}+V_{\text{ct}})}{\partial \varphi}\Bigg|_{\varphi=v}=0, \quad \frac{\partial^2(V_{\text{CW}}+V_{\text{ct}})}{\partial \varphi^2}\Bigg|_{\varphi=v}=0.
\end{equation}
Owing to the modification of the Higgs potential within the SMEFT, field-dependent masses of the relevant dofs take the form
\begin{align}
\begin{split}
&m^2_{W}(\varphi) = \cfrac{1}{4}\,g^2\varphi^2,\\ 
&m^2_{Z}(\varphi) = \cfrac{1}{4}\left(g^2+g'^2\right) \left(1+\frac{C_{HD}}{2 \Lambda^2}\varphi^2\right)\varphi^2,\\
&m^2_{G^\pm}(\phi) = m_H^2+\cfrac{\lambda}{2}\phi^2 - \cfrac{3}{4\Lambda^2}\,C_H\,\phi^4,\\
&m^2_{G^0}(\phi) = m_H^2+\cfrac{\lambda}{2}\phi^2 -\cfrac{m_H^2}{2\Lambda^2}\,C_{H\mathcal{D}}\,\phi^2-\cfrac{3}{4\Lambda^2}\,C_H\phi^4-\cfrac{\lambda}{4\Lambda^2}\,C_{H\mathcal{D}}\,\phi^4,\\
&m^2_h(\varphi) = m_H^2+\cfrac{3\lambda}{2}\varphi^2 -\cfrac{m_H^2}{2\Lambda^2}\,(C_{HD}-4C_{H\square})\varphi^2 
    -\cfrac{3}{4\Lambda^2}(5C_H+\lambda\,(C_{HD}-4C_{H\square}))\varphi^4,\\
 &m^2_t(\varphi) =\frac{1}{2} \left(Y_t +\frac{C_{tH}}{2\Lambda^2} \varphi^2\right)^2\varphi^2,
\end{split}
\end{align}
where $g'$ and $g$ are the $U(1)_Y$ and $SU(2)_L$ gauge couplings, respectively. We neglect all fermions except the top quark due to their smaller Yukawa couplings compared to $Y_t$. The CW potential is gauge-dependent. Under the Landau gauge adopted in this work, the Goldstone boson masses vanish at the physical minimum. Consequently, the loop-corrected effective potential suffers from infrared (IR) divergences in the limit where the field-dependent Goldstone mass approaches zero. These divergences can be consistently handled through the appropriate resummation techniques established in Refs.~\cite{Elias-Miro:2014pca,Espinosa:2016uaw}. As discussed in Ref.~\cite{Elias-Miro:2014pca}, the leading singularities are tamed by shifting the tree-level mass $G$ inside loop expressions to a radiatively corrected version, $\overline{G} = G + \kappa\Pi_G(0)$ with $\kappa=1/16\pi^2$, where the zero-momentum self-energy $\Pi_G(0)$ is evaluated by accounting only for the SM heavy state in the loop (Higgs, top quark, and vector bosons) to keep the potential's derivatives finite. A corresponding treatment in the generic Fermi gauge ($\xi \neq 0$) is outlined in Ref.~\cite{Espinosa:2016uaw}.

The expression of thermal effects \cite{Dolan:1973qd} in the effective potential under the Landau gauge is given by
\beq
V_{\rm th}=\frac{T^4}{2\pi^2}\left\{\sum_{i}n_{B_i} I_B\left(\frac{m_{B_i}^2}{T^2}\right)-n_{F_i} I_F\left(\frac{m_{{F_i}}^2}{T^2}\right)\right\},
\label{eq:vth}
\eeq
where $n_{B_i}$ ($n_{F_i}$) and $m_{B_i}$ ($m_{F_i}$) are the dof and mass of the bosons (fermions), respectively. For scalar (massive vector) bosons, the number of dof is $n_{B_i}=1~(3)$, while for massless vector bosons and fermions $n_{B_i}=n_{F_i}=2$. The integrals in the above expression are defined as
\begin{align}
\begin{split}
I_B(a_i^2)=&\int^{\infty}_{0} dx x^2 \ln\left[1-\exp\left(-\sqrt{x^2+a_i^2}\right)\right],\\
I_F(a_i^2)=&\int^{\infty}_{0} dx x^2 \ln\left[1+\exp\left(-\sqrt{x^2+a_i^2}\right)\right],
\end{split}
\end{align}
with $a_i=M_i/T$. Thermal effects arising from the SMEFT operators have also been included in our numerical computations.
\begin{figure}[t]
    \centering
    \includegraphics[width=0.6\textwidth]{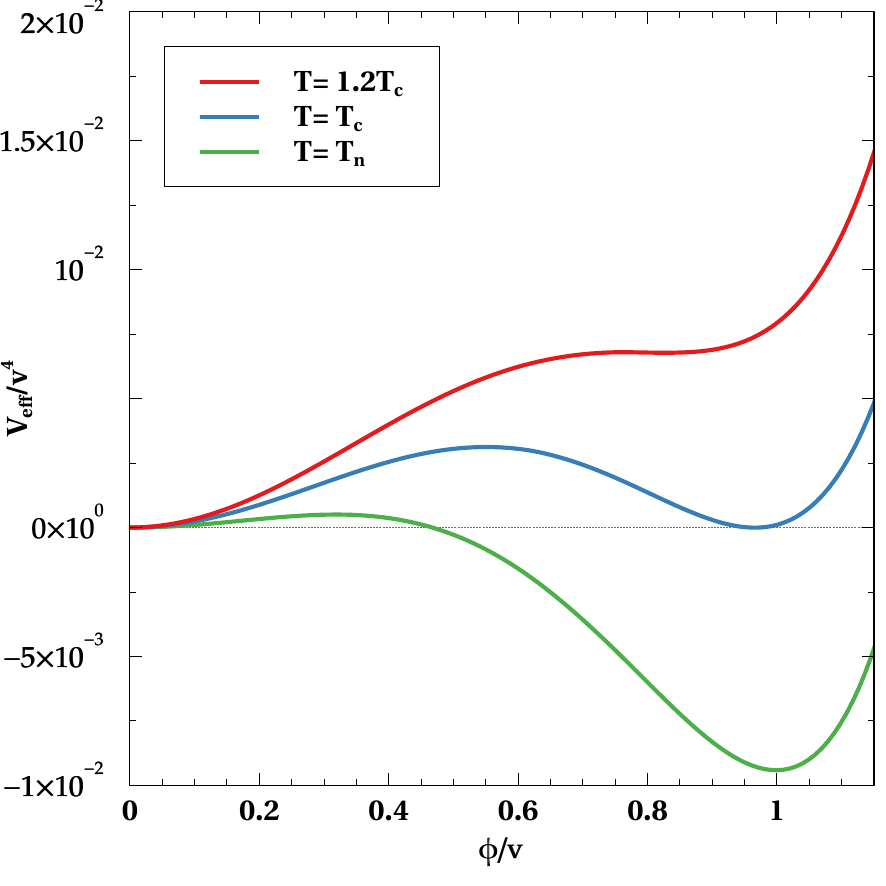}
    \caption{Variation of the potential at, above, and below the critical temperature $T_c$ for the BP1 listed in Table \ref{tab:bp}.}
    \label{fig:FO-EWPT}
\end{figure}
At finite temperature, IR divergences appear due to the Matsubara zero modes, which spoil the perturbative expansion and introduce significant uncertainties in the extraction of phase transition parameters \cite{Dolan:1973qd,Weinberg:1974hy,Linde:1978px,Linde:1980ts}. Daisy resummation addresses this challenge by systematically summing the dominant IR-sensitive ring diagrams \cite{Carrington:1991hz,Parwani:1991gq,Arnold:1992rz}. This process generates thermal Debye masses for the soft bosonic modes, thereby screening the IR divergences and restoring the perturbative control. Following the Arnold-Espinosa method, the ring (or daisy) diagram contributions \cite{Arnold:1992rz} to the effective potential
\begin{align}
V_T^{\rm ring}=\frac{T}{12\pi}\sum_{i= \varphi,W,Z}n_i \left( (M_i^2(\varphi, 0))^{3/2} - (M_i^2(\varphi, T))^{3/2}\right),
\label{eq:vring}
\end{align}
where $M_i^2(\varphi, T)=M_i^2(\varphi) + \Pi_i(T)$ and $\Pi_i(T)$ is the thermal self-energy defined as
\begin{align}
&\Pi_{\varphi}(T)\equiv  T^2\left(  \frac{ \lambda}{2} + \frac{ 3g^2}{16}  + \frac{{g'}^2}{16} +\frac{Y_t^2}{4} \right),
\\
&\Pi_{i}^{\rm (L,T)}(T)\equiv \frac{11T^2}{6}a_i^{\rm (L,T)}
\left(\begin{array}{cccccccc} 
 g^2  &0&0&0 \\
0& g^2  &0&0 \\
0&0& g^2 &0 \\
0&0&0& {g'}^2 
\end{array}\right),~~{\rm for}~i=W^1, W^2, W^3, B,
\end{align}
with $a^{\rm L}_i=1$, and $a^{\rm T}_i=0$.
\begin{figure}[t]
    \centering
    \includegraphics[width=0.45\textwidth]{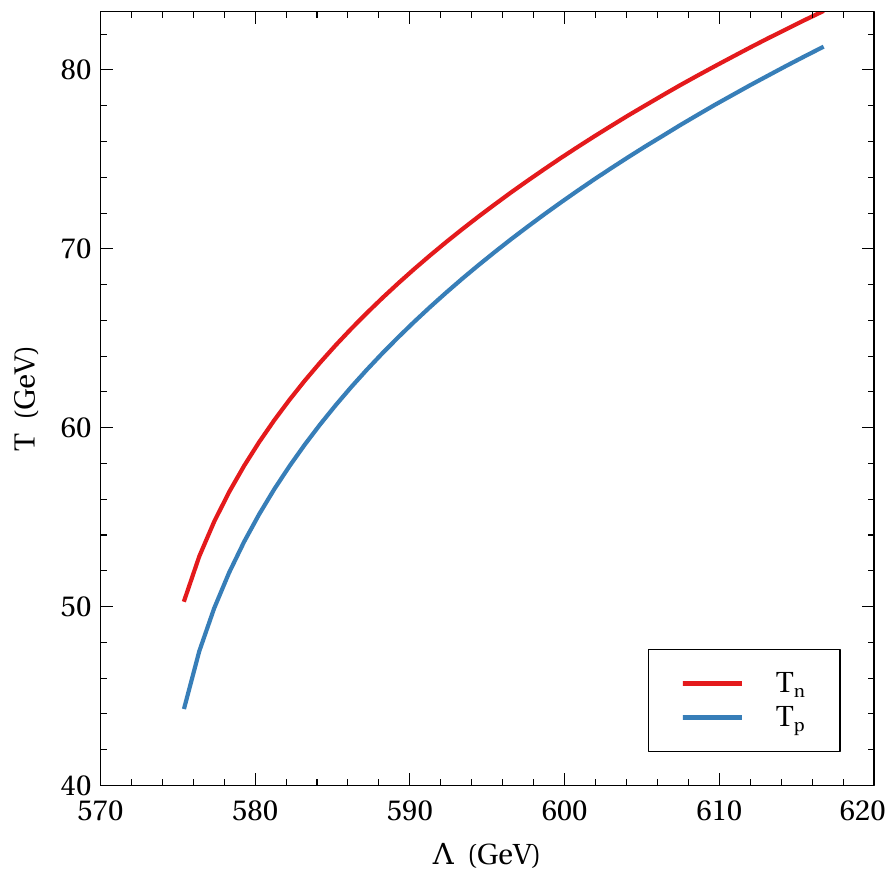}~~
     \includegraphics[width=0.45\textwidth]{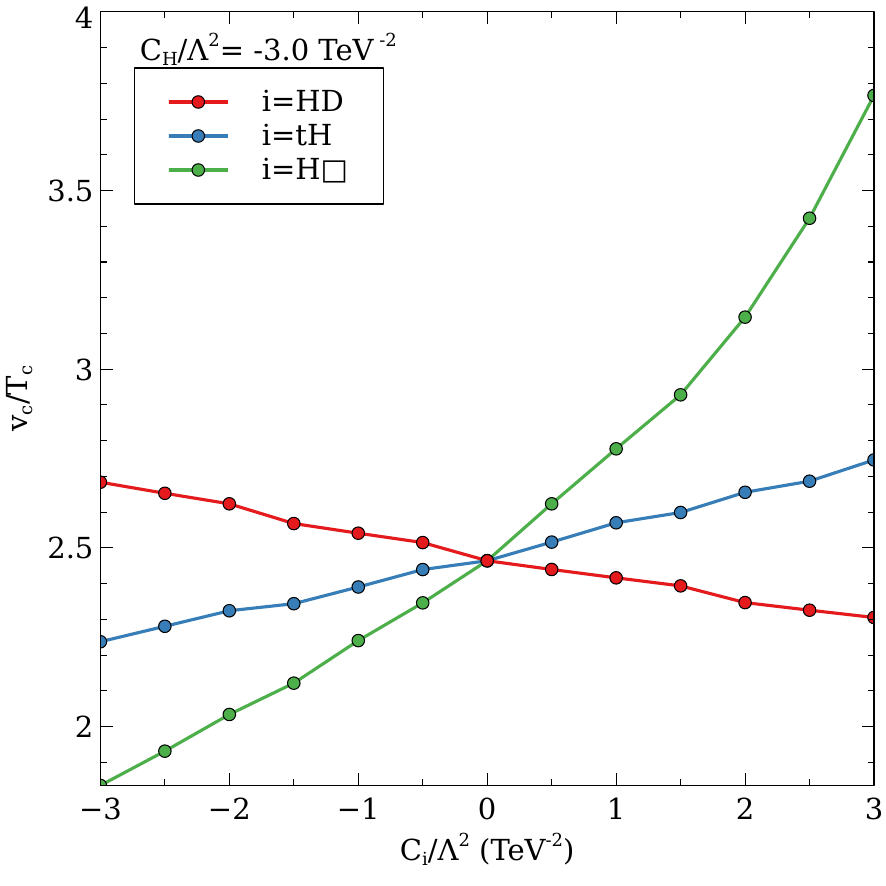}
    \caption{Left panel: Variation of the nucleation and percolation temperatures with $\Lambda$ for the operator $\mathcal{O}_H$ keeping the other three operators' contributions to zero. Right panel: Effect of the other three operators on the EWPT in the presence of the operator $\mathcal{O}_H$ with $C_H/\Lambda^2=-3.0$ TeV$^{-2}$. }
    \label{fig:ewpt-wc}
\end{figure}
\begin{table}[htb!]
    \centering
    {\footnotesize
    \begin{tabular}{ccccccccccc}
    \hline \hline
    \multirow{2}*{BPs} & $C_H/\Lambda^{2}$ & $C_{HD}/\Lambda^{2}$ & $C_{H\Box}/\Lambda^{2}$ &  $C_{tH}/\Lambda^{2}$ & $T_c$  & $v_c$ & $T_n$ & $T_p$ & \multirow{2}*{$\beta/\mathcal{H}$} & \multirow{2}*{$\alpha$} \\
    & $\rm (TeV^{-2})$ & $\rm (TeV^{-2})$ & $\rm (TeV^{-2})$  & $\rm (TeV^{-2})$ & $\rm (GeV)$ & $\rm (GeV)$ & $\rm (GeV)$ & $\rm (GeV)$ &  &  \\
    \hline \hline
    BP1 & -3.0 & 0 & 0 & 0 & 95.9 & 236.4 & 54.7 & 49.9 & 119.2 & 0.13 \\
    \hline
        BP2 & -3.0 & -0.3 & 0 & 0 & 95.3 & 238.8 & 48.5 & 41.9 & 48.9 & 0.22 \\
    \hline
        BP3 & -3.0 & 0 & -0.3 & 0 & 98.2 & 236.4 & 64.7 & 63.5 & 311.9 & 0.06 \\
    \hline
        BP4  & -3.0 & 0 & 0 & -0.3 & 96.5 & 236.4 & 61.6 & 58.1 & 237.1 & 0.08 \\
    \hline \hline
    \end{tabular}}
    \caption{Benchmark points consistent with FO-EWPT for different combinations of effective couplings. The numerical values of the EWPT parameters for different BPs tabulated in the table are also cross-checked using {\tt CosmoTransitions} \cite{Wainwright:2011kj}.}
    \label{tab:bp}
\end{table}
The strength of the strong FO-EWPT is characterized by the ratio $v_c/T_c$, where $T_c$ and $v_c$ are determined from the following two conditions:
\begin{align}
\begin{split}
    \partial_{\varphi} V_{\rm eff}(\varphi,T_c)|_{\varphi=v_c}=0,
    \\
    V_{\rm eff}(v_c,T_c)=V_{\rm eff}(0,T_c).\label{eq:cond2}
\end{split}
\end{align}
The evolution of this quartic potential with temperature is shown in figure~\ref{fig:FO-EWPT}. For temperatures larger than the critical temperature, $T>T_c$, the potential has a minimum at $\phi=0$, which corresponds to the ground (equilibrium) state of the system. At the critical temperature $T_c$, the effective potential develops two degenerate minima at $\varphi = 0$ and $\varphi = v_c$, which are separated by a potential barrier as seen in figure~\ref{fig:FO-EWPT} for BP1. Such degenerate minima give rise to bubble formation, with the true minimum realized inside the bubble and the false minimum persisting outside. The dynamics of these expanding bubbles subsequently generate a stochastic background of gravitational waves. For temperatures below the critical temperature, the minimum at a non-zero field value becomes the global minimum, corresponding to the true (stable) ground state. We note that other BPs tabulated in Table \ref{tab:bp} also show similar behavior as depicted in figure~\ref{fig:FO-EWPT}. Of the four operators, $\mathcal{O}_{H}$ provides the dominant contribution to the EWPT, whereas the remaining three mainly influence the strength of the EWPT. We show the variation of the nucleation and percolation temperature in the left panel of figure~\ref{fig:ewpt-wc}. We find that below $\Lambda \simeq 575$ GeV, percolation fails to occur\footnote{However, it was shown in Ref.~\cite{Ellis:2018mja} that percolation fails to occur below $\Lambda \simeq 545$ GeV.}. In the right panel of figure~\ref{fig:ewpt-wc}, we illustrate the effect of the remaining three operators on the EWPT alongside the operator $\mathcal{O}_H$, with $C_H/\Lambda^2 = -3.0$ TeV$^{-2}$. From the right panel of figure~\ref{fig:ewpt-wc}, it is evident that for positive values of $C_i/\Lambda^2$ (with $i = HD,~tH,~\text{and}~H\Box$), the operator $\mathcal{O}_{H \Box}$ provides the dominant enhancement to the EWPT, followed by $\mathcal{O}_{tH}$ and $\mathcal{O}_{H D}$. For negative values, the hierarchy of contributions is reversed. We note that the FO-EWPT scenario considered in this work is realized through the formation of a tree-level barrier in the effective potential. This requires relatively large values of $|C_H|/\Lambda^2$ and negative Higgs self-coupling. On the other hand, FO-EWPT can also arise from a radiatively generated barrier, which typically leads to a comparatively weaker phase transition. In this case, the required values of $|C_H|/\Lambda^2$ are comparatively smaller, and Higgs self-coupling is positive~\cite{Camargo-Molina:2021zgz,Chala:2025xlk}.

\section{Gravitational wave spectrum}
\label{sec:gw}
Stochastic gravitational wave background produced by FO-EWPT transition originates mainly from three phenomena: bubble collision \cite{Turner:1990rc,Kosowsky:1991ua,Kosowsky:1992rz,Kosowsky:1992vn,Turner:1992tz}, sound wave propagation in the plasma \cite{Hindmarsh:2013xza,Giblin:2014qia,Hindmarsh:2015qta,Hindmarsh:2017gnf}, and plasma turbulence \cite{Kamionkowski:1993fg,Kosowsky:2001xp,Caprini:2006jb,Gogoberidze:2007an,Caprini:2009yp,Niksa:2018ofa}. The GW spectrum is described by four key parameters, within which the effective couplings of the SMEFT are encoded. These four parameters, $T_n$, $\alpha$, $\beta/H$, and $v_w$, are collectively referred to as the transition parameters. $T_n$ represents the temperature corresponding to bubble nucleation, at which the bubble of true vacuum expands to cover the Universe, and is defined by
\begin{equation}
\frac{\Gamma}{\mathcal{H}^4}\Bigg|_{T=T_n}=1,
\label{nucleation}
\end{equation}
where $\mathcal{H}=8\pi^3 g_{\star}T^4/90M_{\text{Pl}}$ is the Hubble parameter during radiation domination, $M_{\text{Pl}}$  is the Planck mass, and $g_{\star}=106.75$ is the SM degrees of freedom. $\Gamma$ is the bubble nucleation rate per unit volume and time and is expressed as 
\begin{equation}
    \Gamma \simeq T^4 \left(\frac{S_3}{2\pi T}\right)^{3/2}e^{-S_3/T},
\end{equation}
where $S_3$ is a 3-dimensional Euclidean action, {\it i.e.}, symmetric bounce action. At finite temperature, the $O(3)$-symmetric bounce configuration \cite{Linde:1980tt} is determined by the radial field equation in Euclidean space, {\it i.e.}, bounce equation
\begin{align}
    \frac{d^2\varphi}{dr^2} + \frac{2}{r}\frac{d\varphi}{dr} 
    = \frac{\partial V_{\rm eff}}{\partial \varphi},
    \label{eq:bounce diff}
\end{align}
which governs the profile of the tunneling solution. To specify the bounce uniquely, one imposes the boundary conditions
\begin{align}
    \varphi(r\to\infty) = \varphi_{\rm false}, \qquad
    \left.\frac{d\varphi}{dr}\right|_{r=0} = 0,
    \label{eq:boundary condition}
\end{align}
where $\phi_{\rm false}$ denotes the value of the field at the false minimum of the potential. With the solution $\phi(r)$ satisfying the above conditions, the three-dimensional Euclidean action takes the form
\begin{align}
    S_3 = \int_{0}^{\infty} dr\, 4\pi r^2 
    \left[
        \frac{1}{2}\left( \frac{d\varphi}{dr} \right)^2 
        + V_{\rm eff}(\varphi, T)
    \right].
    \label{s3eq}
\end{align}
To evaluate the bounce action, we use the Mathematica-based package \texttt{FindBounce} \cite{Guada:2020xnz}. The nucleation temperature is subsequently obtained using eq.~\eqref{nucleation}.
The second parameter, $\alpha$, quantifies the strength of the phase transition and is defined as the ratio of the released latent heat, $\epsilon$, to the radiation energy density of the plasma, $\rho_{\rm rad}(T) = (\pi^2/30) g_* T^4$, evaluated at $T = T_n$, as follows: 
	\begin{align} 
	\alpha\equiv \frac{\epsilon(T_n)}{\rho_{\rm rad}(T_n)}.
	\end{align}
Here, the released latent heat, $\epsilon$, is expressed as 
 \begin{align}
 \label{latenth}
  \epsilon(T)
  =  \Delta V_{\rm eff} -\frac{T}{4}
  \frac{\partial  \Delta V_{\rm eff} }{\partial T},\quad  \Delta V_{\rm eff} =  V_{\rm eff}(\varphi_-(T),T) - V_{\rm eff}(\varphi_+(T),T),
\end{align}
where $V_{\rm eff}$ (defined in eq.~\eqref{eq:veff}) denotes the effective potential, and $\varphi_{+}$ and $\varphi_{-}$ stand for the order parameters corresponding to the broken and unbroken phases, respectively. The third parameter, $\beta/\mathcal{H}$, characterizes the inverse duration of the phase transition and is defined as
\begin{align} 
	\frac{\beta}{\mathcal{H}}\equiv T_n\left.\frac{d}{dT}\left(\frac{S_3}{T}\right)\right|_{T=T_n}.
	\end{align}
Following the prescription of Refs.~\cite{Ellis:2018mja,Ellis:2020nnr}, the percolation temperature $T_p$ is determined from the probability that a given point in space remains in the false vacuum,
\begin{equation}
P(T) = e^{-I(T)} ,
\end{equation}
where
\begin{equation}
I(T) = \frac{4\pi}{3} \int_{T}^{T_c} \frac{dT'}{T'^4} \,
\frac{\Gamma(T')}{\mathcal{H}(T')} 
\left( \int_{T}^{T'} \frac{d\tilde{T}}{\mathcal{H}(\tilde{T})} \right)^3 .
\end{equation}
The percolation temperature is obtained by solving
\begin{equation}
I(T_p) = 0.34 ,
\end{equation}
which corresponds to the condition that approximately $34\%$ of the comoving volume has transitioned to the true vacuum~\cite{Ellis:2018mja}.

During FO-EWPT, expanding bubbles of the true vacuum collide and interact with the surrounding plasma, leading to the generation of a stochastic background of GW. The total GW energy density spectrum is typically expressed as a sum of three main contributions: (i) collisions of scalar field bubble walls, (ii) sound waves in the plasma, and (iii) magnetohydrodynamic (MHD) turbulence generated in the plasma after bubble collisions. Each of these sources produces a characteristic spectral shape, peak frequency, and amplitude that depend on the thermodynamic and hydrodynamic parameters of the transition such as the transition strength $\alpha$, the inverse time duration of the transition $\beta/\mathcal{H}$, and the bubble wall velocity $v_w$.
The GW spectrum generated from the collision of bubble walls can be parameterized as \cite{Caprini:2015zlo,Caprini:2024hue,Athron:2023xlk}
\begin{equation}
    \Omega_{\text{col}} h^2 = 1.67 \times 10^{-5} 
    \left(\frac{100}{g_*}\right)^{1/3}
    \left(\frac{\mathcal{H}}{\beta}\right)^2
    \left(\frac{\kappa_\varphi \alpha}{1+\alpha}\right)^2
    \frac{0.11 v_w^3}{0.42+v_w^2}
    \frac{3.8(f/f_{\rm peak}^{\rm PT, \varphi})^{2.8}}
         {1+2.8 (f/f_{\rm peak}^{\rm PT, \varphi})^{3.8}} \, .
\end{equation}
Here, $\mathcal{H}=H(T_n)$ is the Hubble expansion rate at the nucleation temperature $T_n$, and $g_*$ denotes the number of relativistic degrees of freedom at that epoch. The corresponding peak frequency of this contribution is given by \cite{Caprini:2015zlo,Caprini:2024hue,Athron:2023xlk}
\begin{equation}
    f_{\rm peak}^{\rm PT, \text{col}} = 1.65 \times 10^{-5} \, {\rm Hz} 
    \left(\frac{g_*}{100}\right)^{1/6}
    \left(\frac{T_n}{100\,{\rm GeV}}\right)
    \frac{0.62}{1.8-0.1v_w+v_w^2}
    \left(\frac{\beta}{\mathcal{H}}\right).
\end{equation}
The estimation of the bubble wall velocity depends on the particle species present in the plasma and their interactions with the advancing wall. A precise determination, therefore, requires a rigorous computation accounting for the friction from the thermal plasma together with the associated theoretical uncertainties. In the present scenario, the expanding broken-phase bubbles satisfy the leading-order runaway criterion~\cite{Bodeker:2009qy}, and hence we expect ultra-relativistic bubble expansion with velocity $v_w \sim 1$, which we adopt in our subsequent calculations. However, this choice should be regarded as a representative value and not as a robust prediction. From the expressions for the peak GW energy density and peak frequency, one can see that the GW amplitude is proportional to $v_w$, whereas the peak frequency is inversely proportional to it as discussed later in this section.
The efficiency factor $\kappa_{\text{col}}$, which characterizes the fraction of vacuum energy converted into the scalar field's kinetic energy, is approximated by \cite{Kamionkowski:1993fg}
\begin{align}
    \kappa_{\text{col}} = 
    \frac{1}{1+0.715 \alpha}
    \left(0.715\alpha + \frac{4}{27}\sqrt{\frac{3\alpha}{2}}\right).
\end{align}

After the bubble walls collide, most of the released energy is transferred into the plasma in the form of acoustic waves, which act as a long-lasting and efficient GW source. The corresponding GW energy density spectrum can be written as \cite{Caprini:2019egz,Guo:2020grp,Caprini:2024hue,Athron:2023xlk}
\begin{equation}
    \Omega_{\rm sw} h^2 = 2.65 \times 10^{-6} 
    \left(\frac{100}{g_*}\right)^{1/3}
    \left(\frac{\mathcal{H}}{\beta}\right)
    \left(\frac{\kappa_{\rm sw} \alpha}{1+\alpha}\right)^2
    v_w
    \left(\frac{f}{f_{\rm peak}^{\rm PT, sw}}\right)^3
    \left[\frac{7}{4+3 (f/f_{\rm peak}^{\rm PT, sw})^{2}}\right]^{7/2}
    \Upsilon.
    \label{eq:gw_sound}
\end{equation}
The peak frequency for this component is given by \cite{Athron:2023xlk}
\begin{equation}
    f_{\rm peak}^{\rm PT, sw} = 1.65 \times 10^{-5} \, {\rm Hz} \frac{1}{v_w}
    \left(\frac{g_*}{100}\right)^{1/6}
    \left(\frac{T_n}{100\,{\rm GeV}}\right)
    \left(\frac{\beta}{\mathcal{H}}\right)
    \frac{2}{\sqrt{3}}.
\end{equation}
The efficiency factor $\kappa_{\rm sw}$, quantifying the fraction of released vacuum energy converted into bulk fluid motion, is approximated by \cite{Espinosa:2010hh}
\begin{align}
    \kappa_{\rm sw} = 
    \frac{\sqrt{\alpha}}{0.135 + \sqrt{0.98 + \alpha}}.
\end{align}
A suppression factor $\Upsilon = 1 - 1/\sqrt{1 + 2\tau_{\rm sw}\mathcal{H}}$ accounts for the finite lifetime of the acoustic source, where the sound wave lifetime is given approximately by $\tau_{\rm sw} \sim R_*/\bar{U}_f$ \cite{Guo:2020grp}. Here, $R_* = (8\pi)^{1/3}v_w/\beta$ is the mean bubble separation, and the rms fluid velocity is $\bar{U}_f = \sqrt{3\kappa_{\rm sw}\alpha/4}$.

Following bubble collisions, a fraction of the plasma kinetic energy can cascade into MHD turbulence, producing an additional but typically subdominant GW signal. The GW spectrum from turbulence is parameterized as \cite{Caprini:2015zlo,Caprini:2024hue,Athron:2023xlk}
\begin{equation}
    \Omega_{\rm turb} h^2 = 3.35 \times 10^{-4} 
    \left(\frac{100}{g_*}\right)^{1/3}
    \left(\frac{\mathcal{H}}{\beta}\right)
    \left(\frac{\kappa_{\rm turb} \alpha}{1+\alpha}\right)^{3/2}
    \frac{v_w(f/f_{\rm peak}^{\rm PT, turb})^3}
         {(1+ f/f_{\rm peak}^{\rm PT, turb})^{11/3}(1+8\pi f/h_*)}.
\end{equation}
The corresponding peak frequency is \cite{Caprini:2015zlo}
\begin{equation}
    f_{\rm peak}^{\rm PT, turb} = 1.65 \times 10^{-5} \, {\rm Hz}\frac{1}{v_w}
    \left(\frac{g_*}{100}\right)^{1/6}
    \left(\frac{T_n}{100\,{\rm GeV}}\right)
    \frac{3.5}{2}
    \left(\frac{\beta}{\mathcal{H}}\right),
    \label{eq:peak_freak}
\end{equation}
with the efficiency factor approximated as $\kappa_{\rm turb} \simeq 0.1 \kappa_{\rm sw}$ \cite{Caprini:2015zlo}.  
The inverse Hubble time at the epoch of GW emission, redshifted to the present, is
\begin{equation}
   h_* = 1.65\times 10^{-5}
   \frac{T_n}{100\,{\rm GeV}}
   \left(\frac{g_*}{100}\right)^{1/6}.
\end{equation}

\begin{figure}[t]
    \centering
    \includegraphics[width=0.4\textwidth]{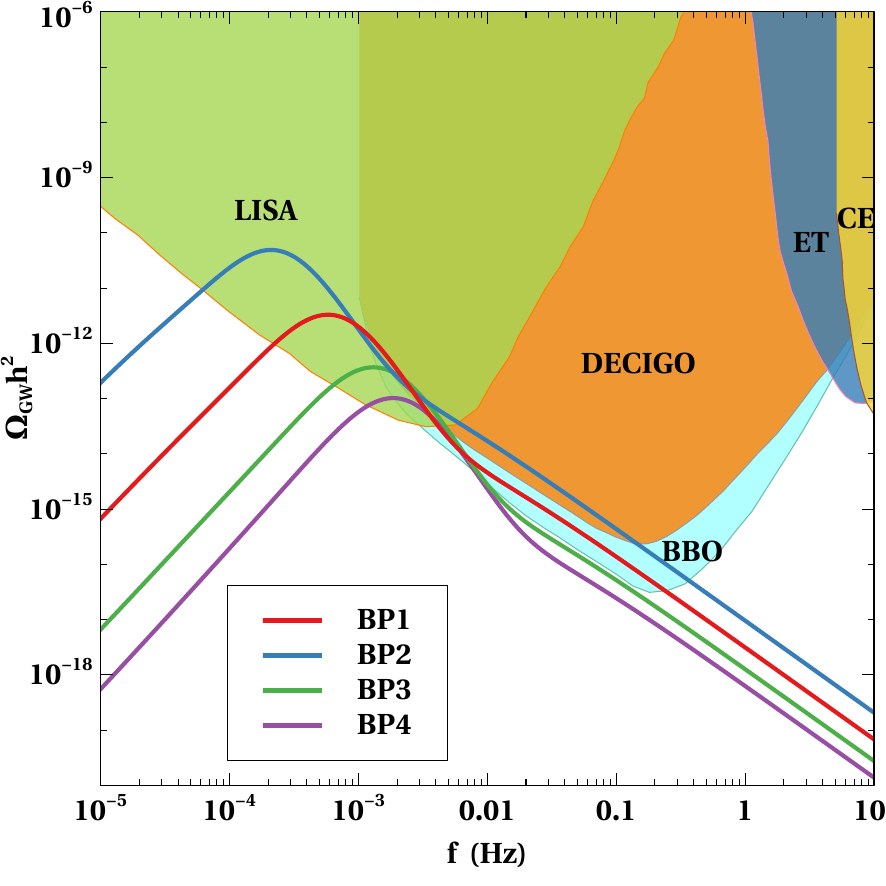}~~
    \includegraphics[width=0.4\textwidth]{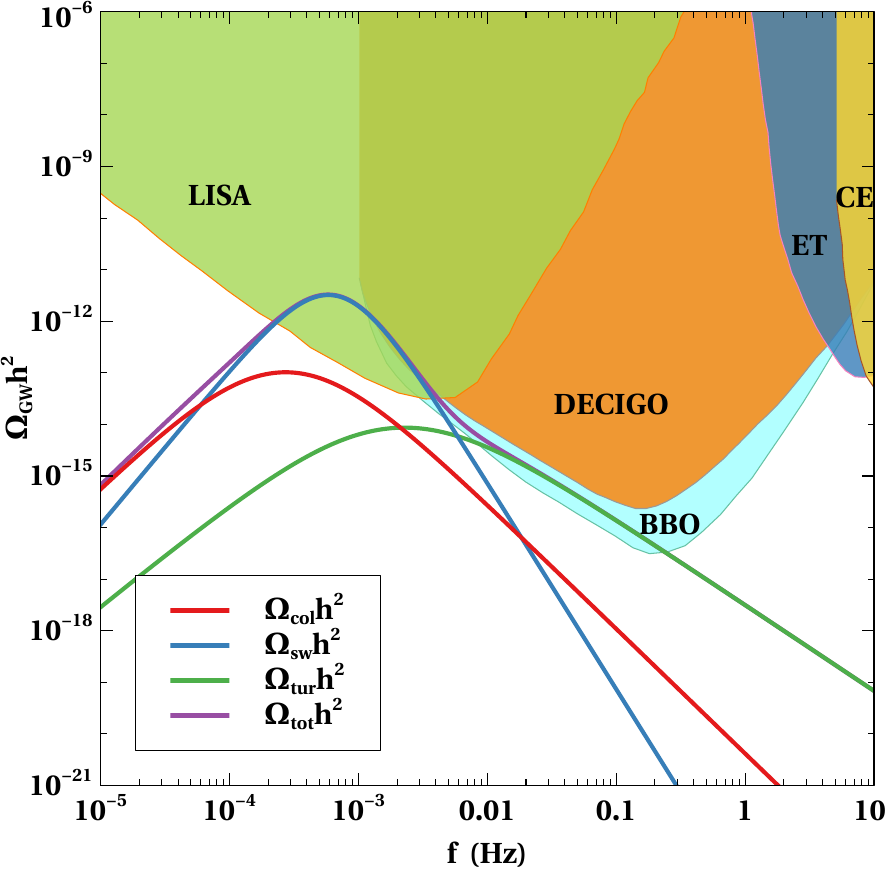}
    \caption{Left panel: Resultant GW spectra from the FO-EWPT for different BPs listed in Table \ref{tab:bp}. Right panel: Individual contribution to the GW spectra from bubble collision, sound wave, and plasma turbulence for BP1.}
    \label{fig:gw.spectrum}
\end{figure}

From the above expressions, it is evident that among the three contributions, the sound-wave component dominates the GW spectrum. This is because the acoustic phase of the plasma persists longer than the brief period of bubble collision and generates a coherent source of stress-energy perturbations. Consequently, the overall GW peak amplitude and frequency are primarily determined by the sound-wave contribution, and the total GW spectrum typically peaks around $f_{\rm peak}^{\rm PT, sw}$. 

GW produced during FO-EWPT generally exhibits a broad frequency spectrum. The precise form of the GW energy density spectrum, $\Omega_{\mathrm{GW}}(f)$, is determined by the underlying dynamics of the generating processes, including bubble collisions, sound waves, and plasma turbulence. For many GW sources, the spectrum can be approximated by a power-law dependence on frequency, $\Omega_{\mathrm{GW}}(f) \propto f^{n}$, where $n$ is the spectral index. Typically, such spectra feature a characteristic peak whose position is governed by the temperature and dynamics of the phase transition. The GW energy density rises at low frequencies, reaches a maximum near the peak frequency, and subsequently decreases at higher frequencies, as illustrated in the left panel of figure~\ref{fig:gw.spectrum} for different BPs tabulated in Table \ref{tab:bp}. The contribution from sound waves is found to be the dominant component of the GW spectrum as shown in the right panel of figure~\ref{fig:gw.spectrum}. The corresponding peak frequency of the total signal lies well within the sensitivity ranges of future detectors such as LISA, DECIGO, and BBO. In the left panel of figure~\ref{fig:gw.spectrum}, the red solid line shows the variation of the GW energy density with frequency for the operator $\mathcal{O}_H$ (BP1). We then include the contributions from the remaining operators, one at a time, to examine their impact on the GW energy density. We find that the inclusion of the operator $\mathcal{O}_{HD}$ (BP2) enhances the GW energy density. In contrast, the inclusion of the operators $\mathcal{O}_{H\Box}$ (BP3) and $\mathcal{O}_{tH}$ (BP4) leads to a suppression of the GW energy density, consistent with the change in the strength of the EWPT, $v_c/T_c$, shown in the right panel of figure~\ref{fig:ewpt-wc} for negative values of effective couplings.

The sensitivity of upcoming GW detectors sensitive to the effective couplings can be effectively quantified through the signal-to-noise ratio (SNR). The SNR corresponding to the detection of the GW spectrum is expressed as \cite{Seto:2005qy,Hashino:2018wee}
\begin{equation}
    \text{SNR}=\sqrt{\delta \times T_{\text{obs}}\int^{\infty}_{0}df\left[\frac{\Omega_{\text{GW}}(f)}{\Omega_{\text{sen}}(f)}\right]^2},
\end{equation}
where $T_{\text{obs}}$ is the observation period, $\Omega_{\text{sen}}=2 \pi^2 f^3/2\mathcal{H}_0$, $\delta$ is the number of independent channels for different GW experiments, i.e., $\delta=1$ for LISA and $\delta=2$ for DECIGO and BBO. In future space-based GW experiments, the threshold SNR for detection is typically taken as 5 for LISA and 25 for both DECIGO and BBO, reflecting their respective sensitivities and detection criteria.

\begin{figure}[t]
    \centering
    \includegraphics[width = 0.3\textwidth]{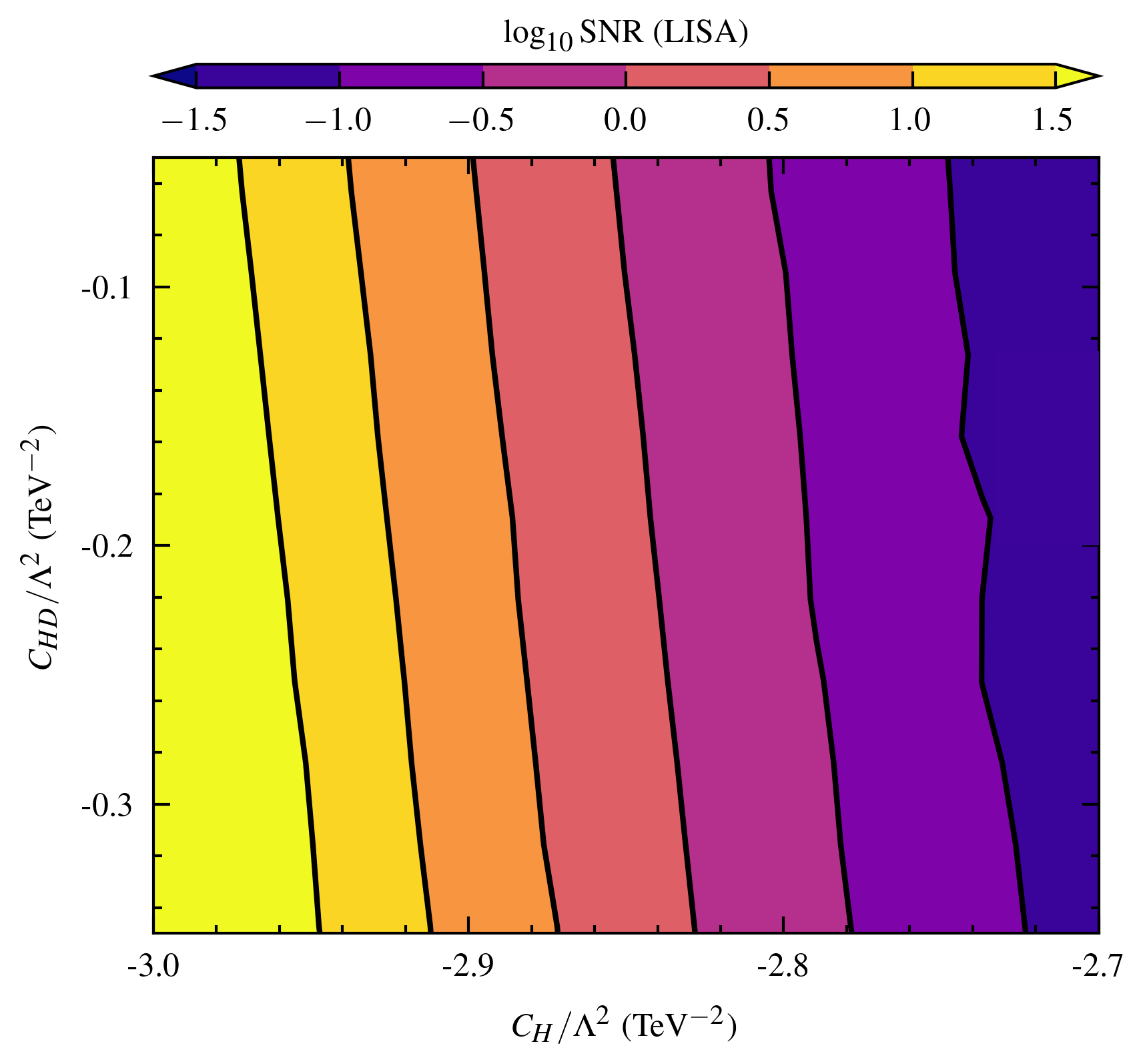} \quad
    \includegraphics[width = 0.3\textwidth]{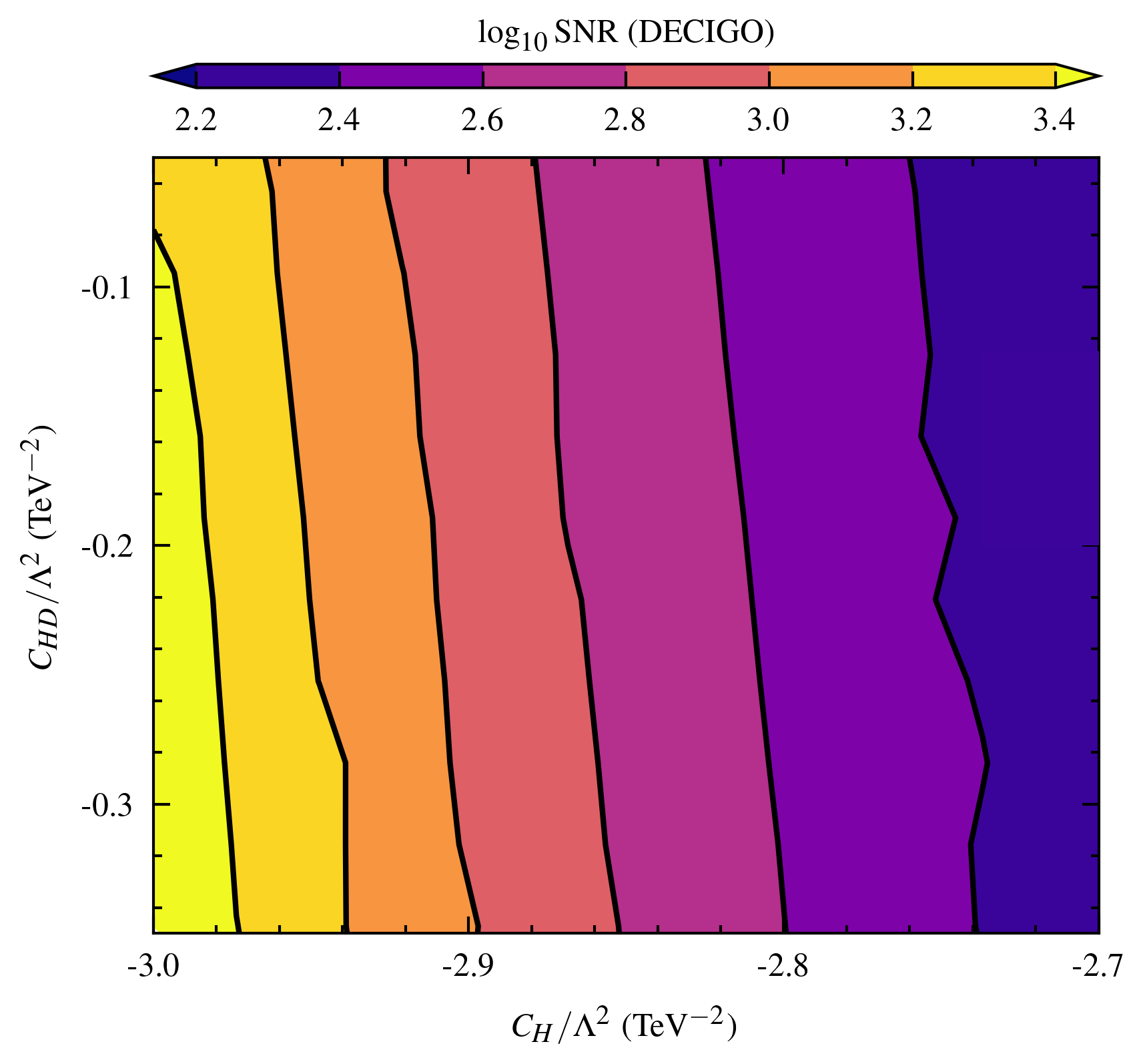} \quad
    \includegraphics[width = 0.3\textwidth]{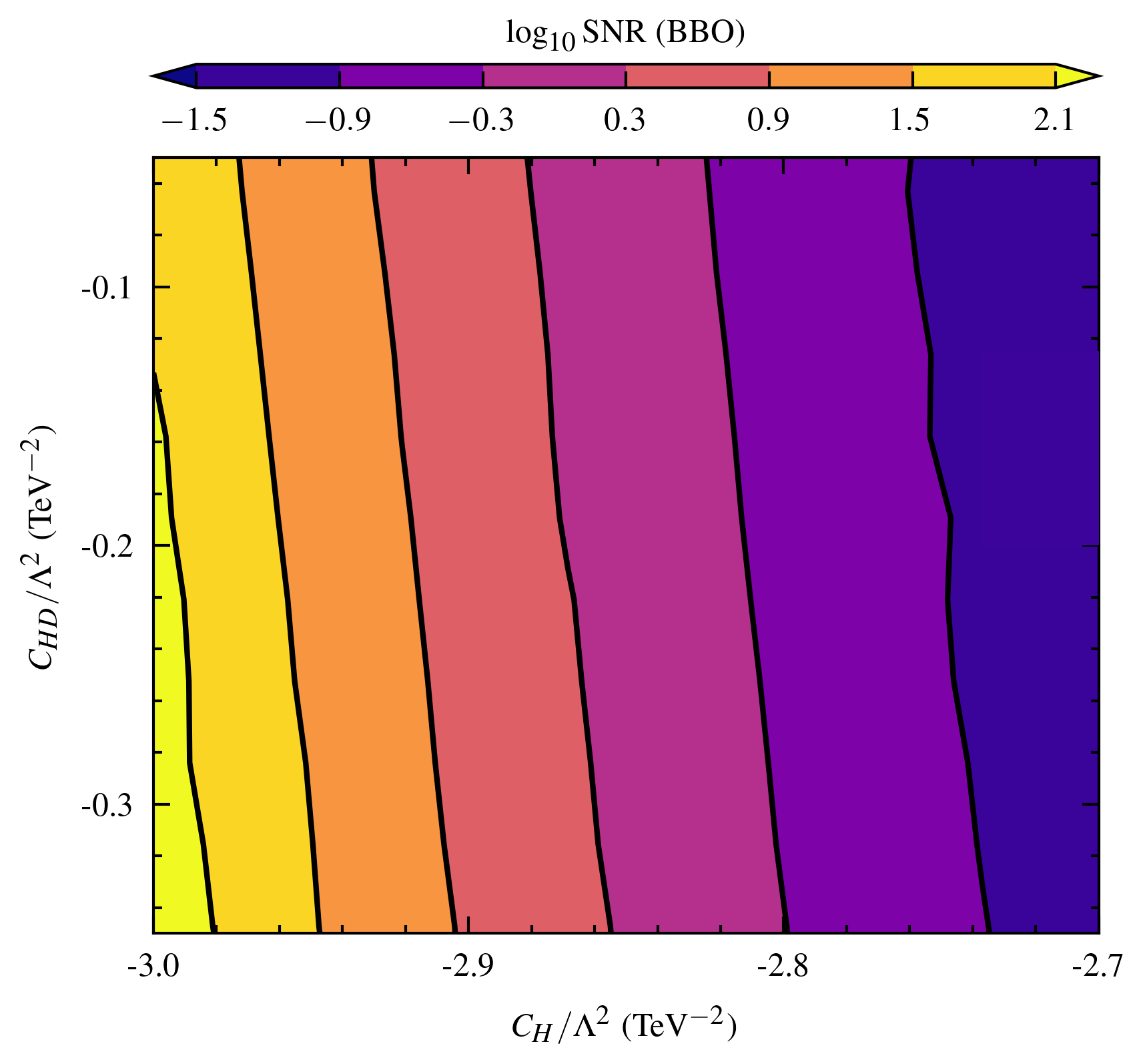} \\
    \includegraphics[width = 0.3\textwidth]{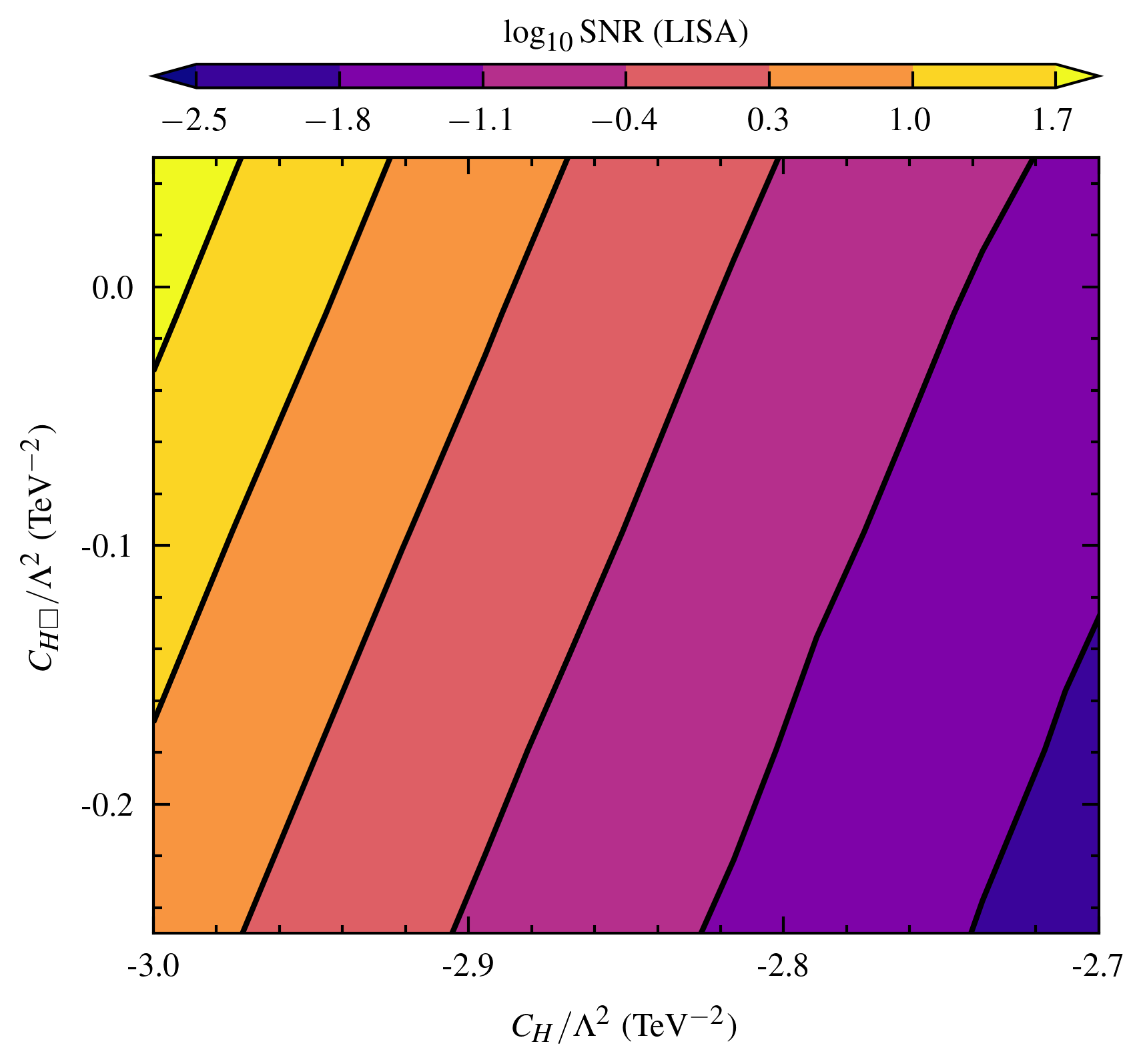} \quad
    \includegraphics[width = 0.3\textwidth]{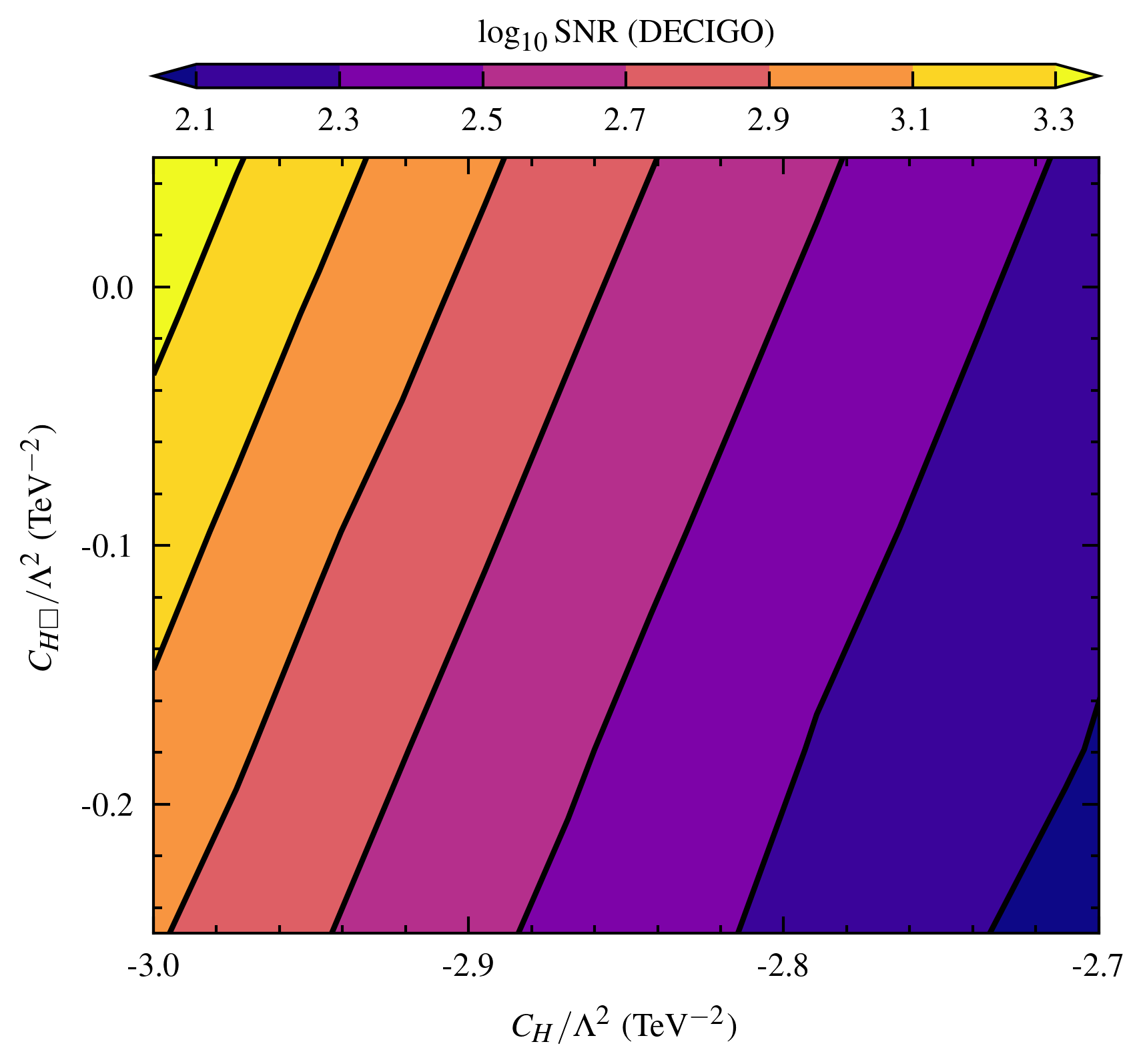} \quad
    \includegraphics[width = 0.3\textwidth]{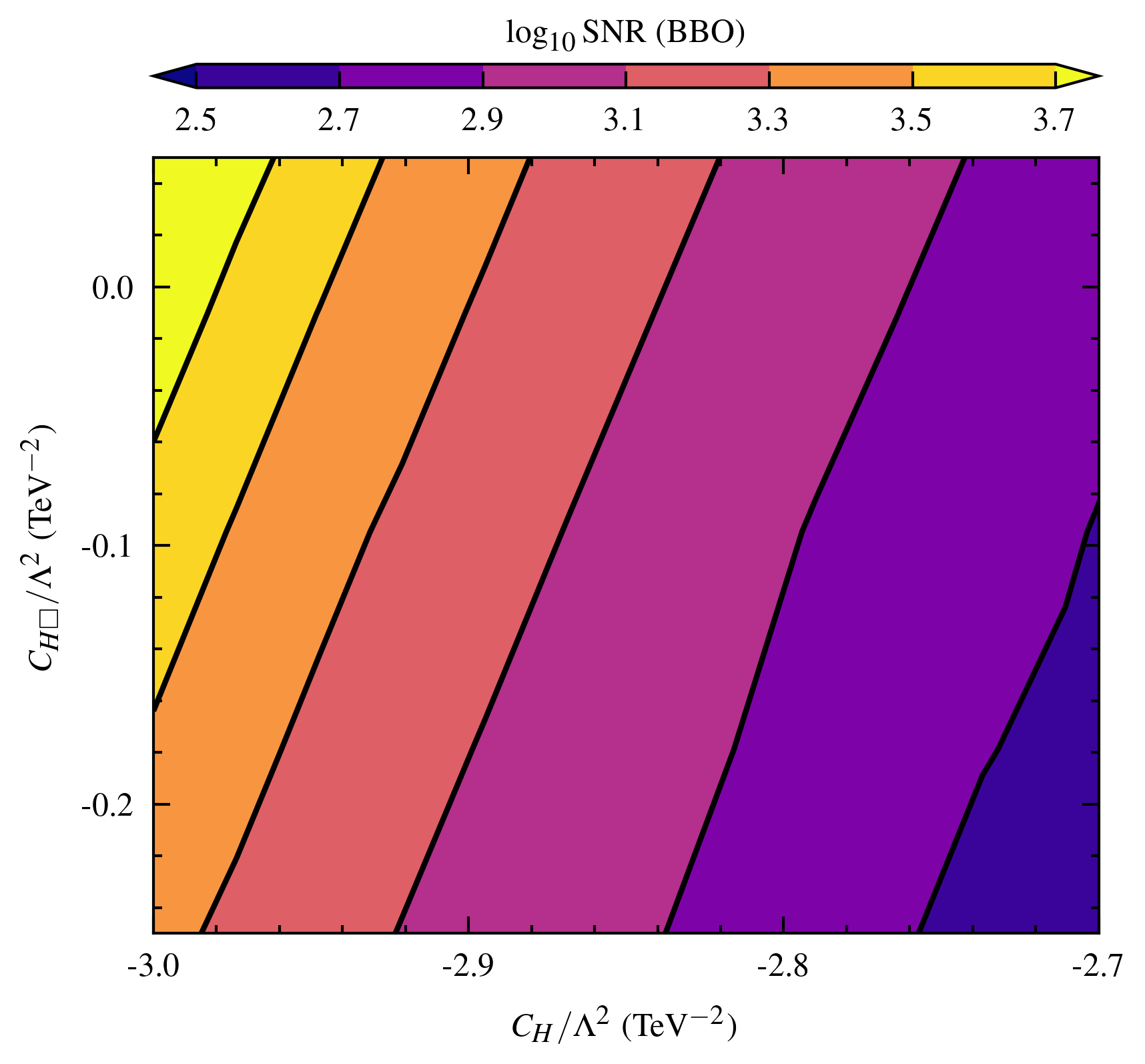} \\
    \includegraphics[width = 0.3\textwidth]{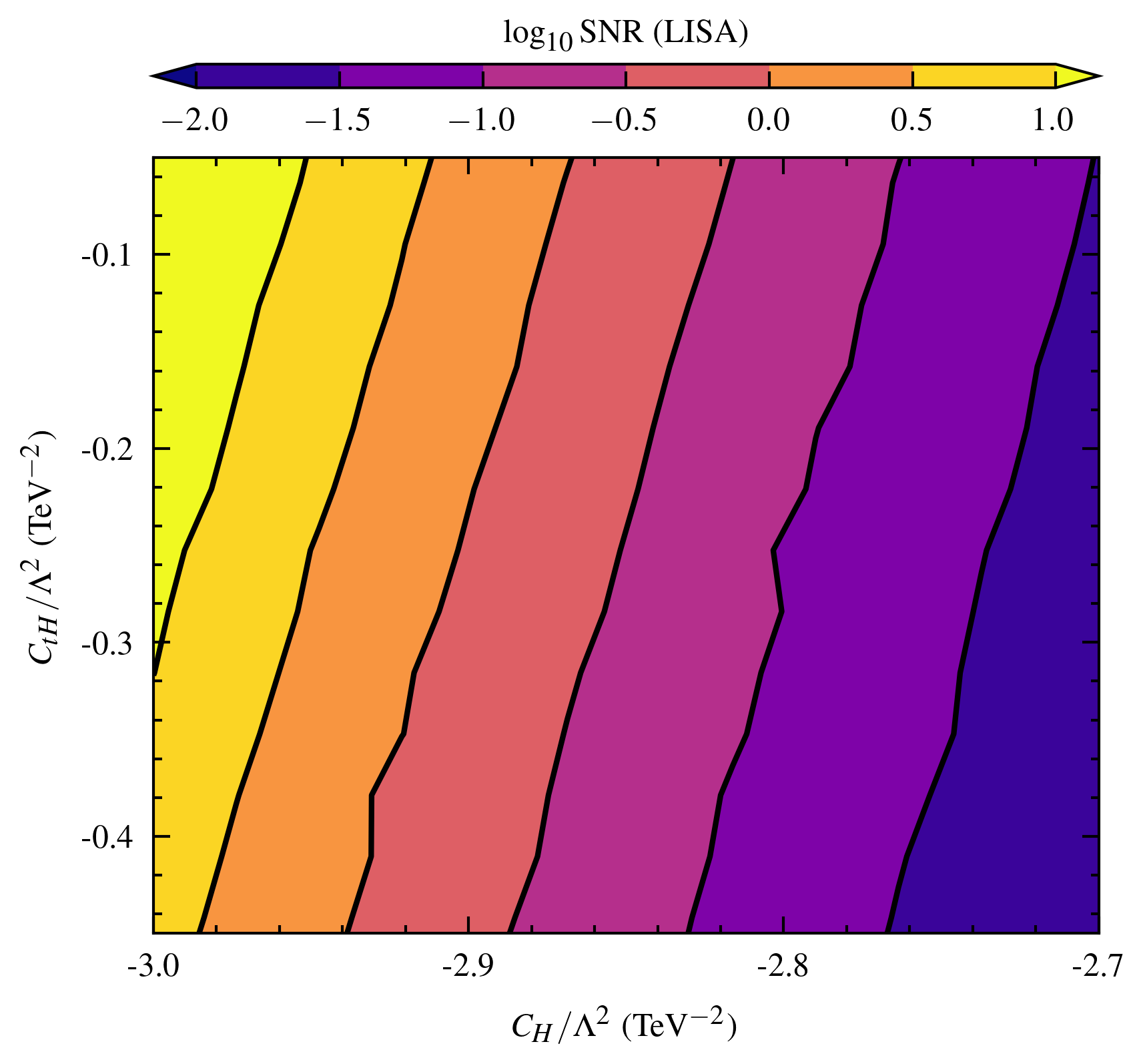} \quad
    \includegraphics[width = 0.3\textwidth]{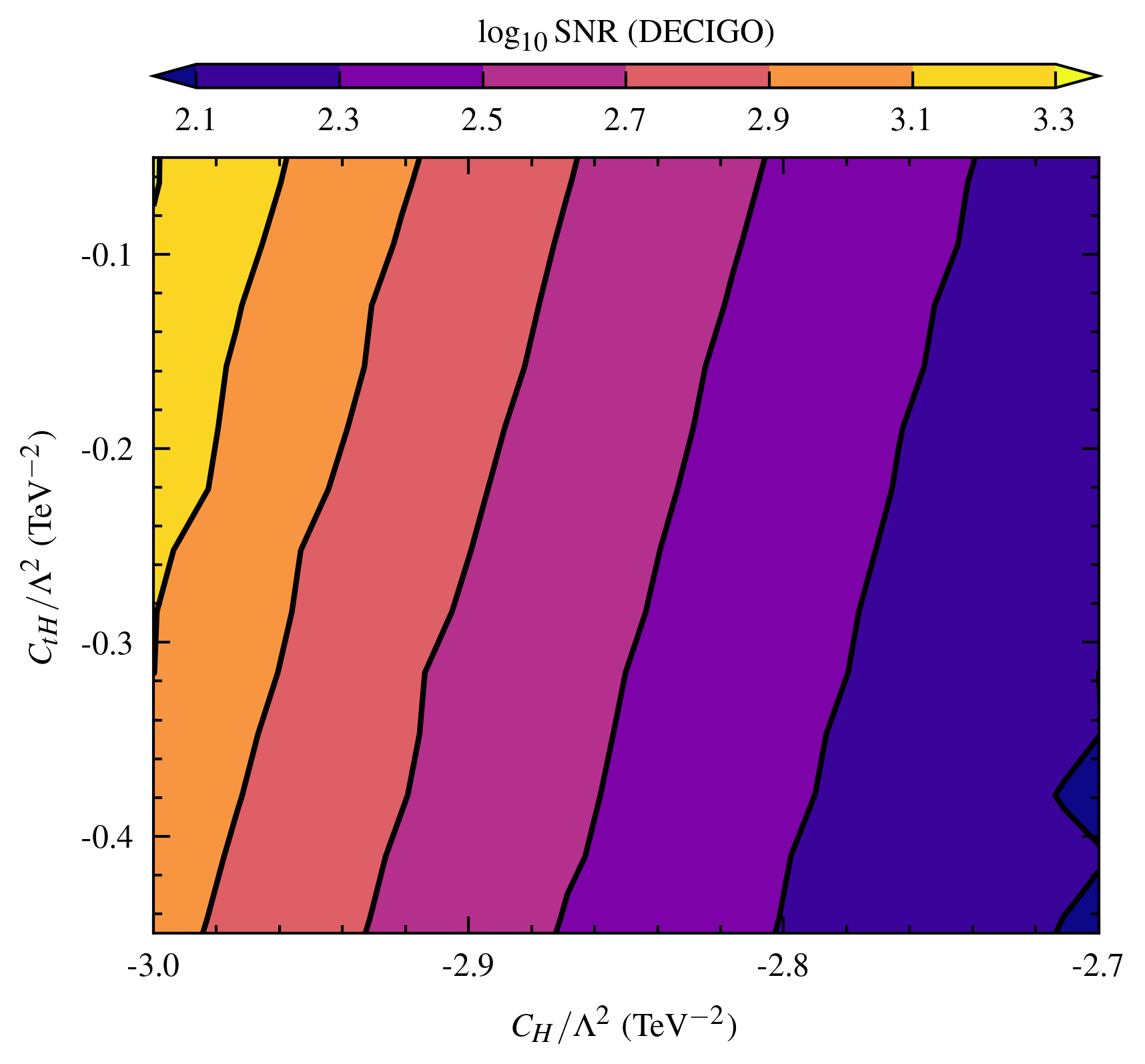} \quad
    \includegraphics[width = 0.3\textwidth]{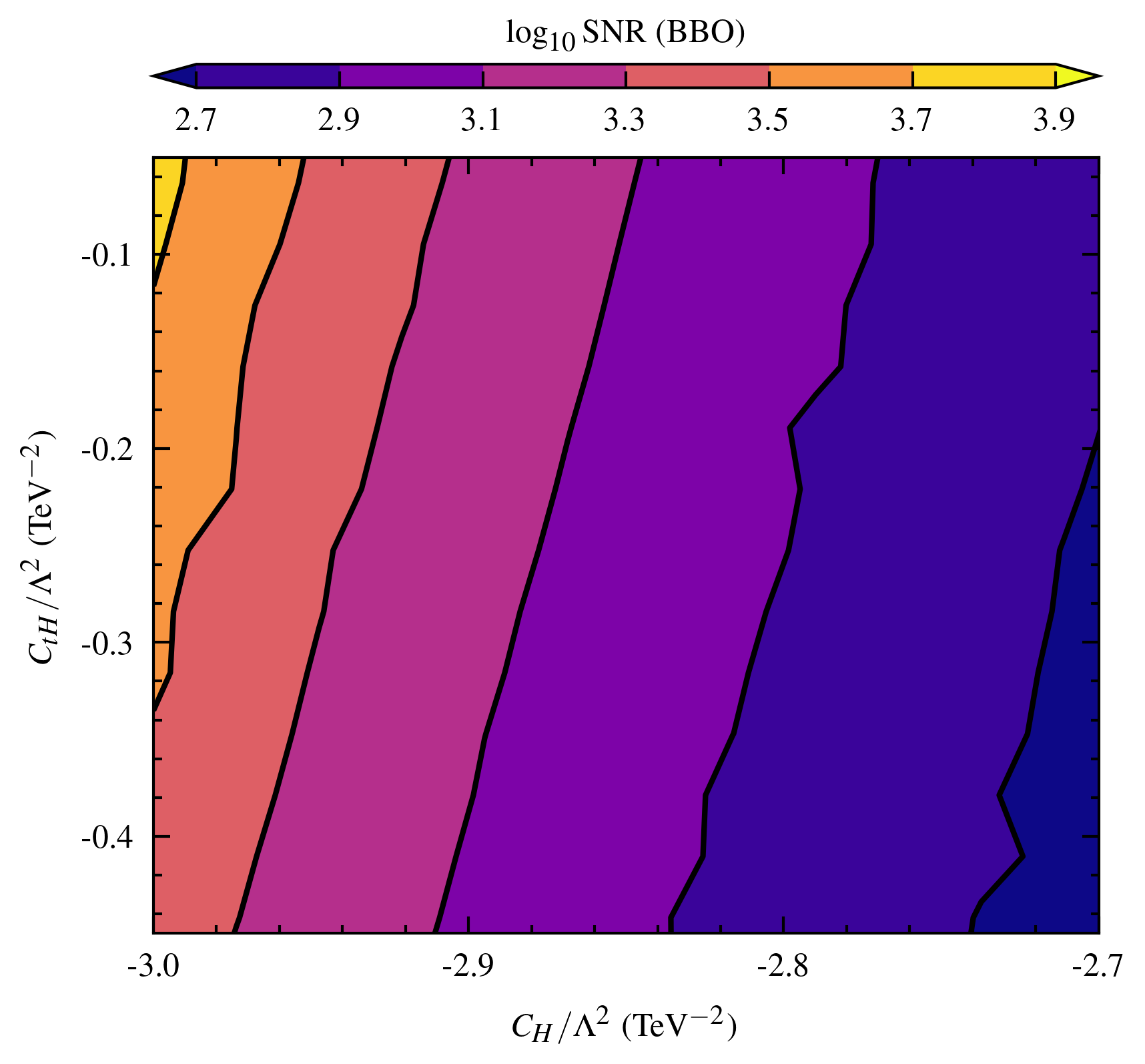} \\
    \caption{Expected SNR in future GW experiments such as LISA (\textit{left}), DECIGO (\textit{middle}) and BBO (\textit{right}). Top panel: $C_H/\Lambda^2-C_{HD}/\Lambda^2$ plane; middle panel: $C_H/\Lambda^2-C_{H\Box}/\Lambda^2$ plane; bottom panel: $C_H/\Lambda^2-C_{tH}/\Lambda^2$ plane.}
    \label{fig:snr}
\end{figure}

To quantitatively assess the sensitivity of future GW experiments to these SMEFT operators effect, we compute the SNR\footnote{To estimate the SNR for different experiments, we include all three contributions to the GW spectrum discussed above. However, we have verified that neglecting bubble collisions does not significantly alter the SNR.} for LISA, DECIGO, and BBO on the $C_{H}/\Lambda^2-C_{HD}/\Lambda^2$, $C_{H}/\Lambda^2-C_{H\Box}/\Lambda^2$ and $C_{H}/\Lambda^2-C_{tH}/\Lambda^2$ planes, keeping the others WC contributions equal to zero. Since $C_{H}/\Lambda^{2}$ yields the dominant contribution among the four NP couplings, we consistently include it together with the other three operators in the correlation analysis.
The analysis is performed with an observation time of $T_{\text{obs}} = 4~\text{year}$, as illustrated in figure~\ref{fig:snr}. It's worthwhile to mention that the regions in figure~\ref{fig:snr} satisfy $v_c/T_c>1$, implying the EWPT to be first order. It is clear from figure~\ref{fig:snr} that the SNR exhibits a clear dependence on the correlation of three WCs with $C_H/\Lambda^2$ that control the strength of the EWPT. For the operator $\mathcal{O}_{HD}$ (top row), the SNR increases toward more negative values of $C_{HD}/\Lambda^2$, indicating a stronger GW signal in the region where the EWPT becomes stronger. In contrast, for the operators $\mathcal{O}_{H\Box}$ (middle row) and $\mathcal{O}_{tH}$ (bottom row), the SNR gradually decreases across the negative values of the parameter space, reflecting the weakening of the strength of the EWPT for the corresponding ranges of WCs. Comparing the detectors, DECIGO and BBO generally predict larger SNR values and therefore exhibit higher sensitivity to the GW signal, while LISA shows comparatively smaller SNR in the same parameter regions. These results demonstrate that future space-based GW detectors, particularly DECIGO and BBO, can probe sizable regions of the SMEFT parameter space associated with EWPT. 

\begin{figure}[t]
    \centering
    \includegraphics[width=0.4\textwidth]{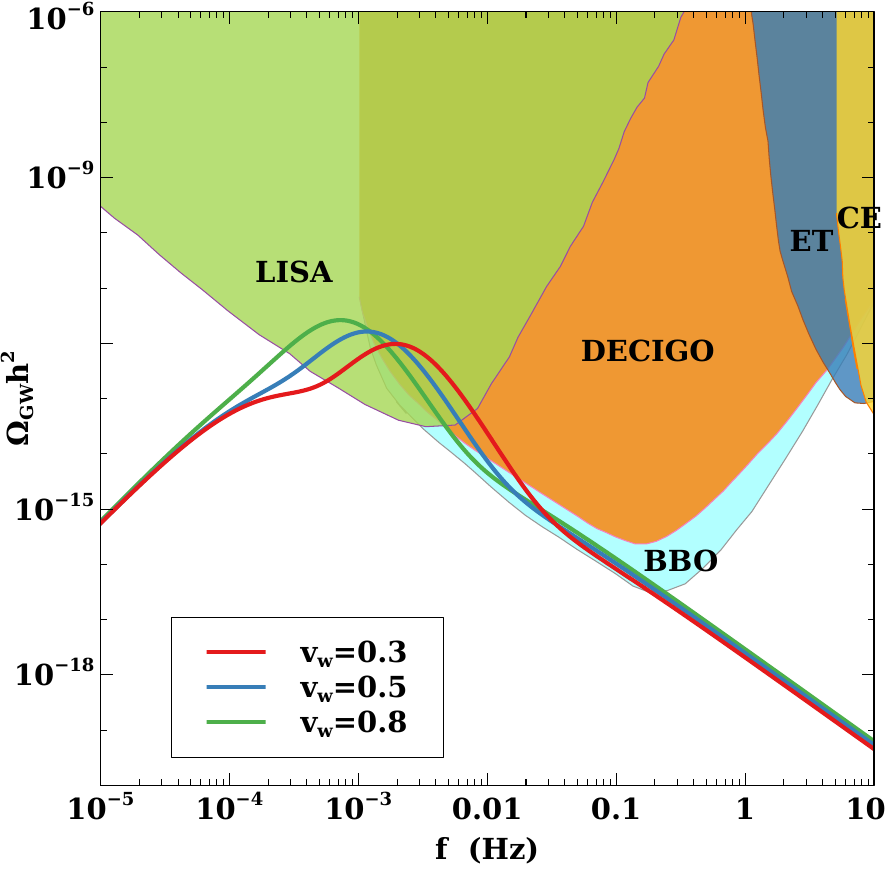}~~
    \includegraphics[width=0.4\textwidth]{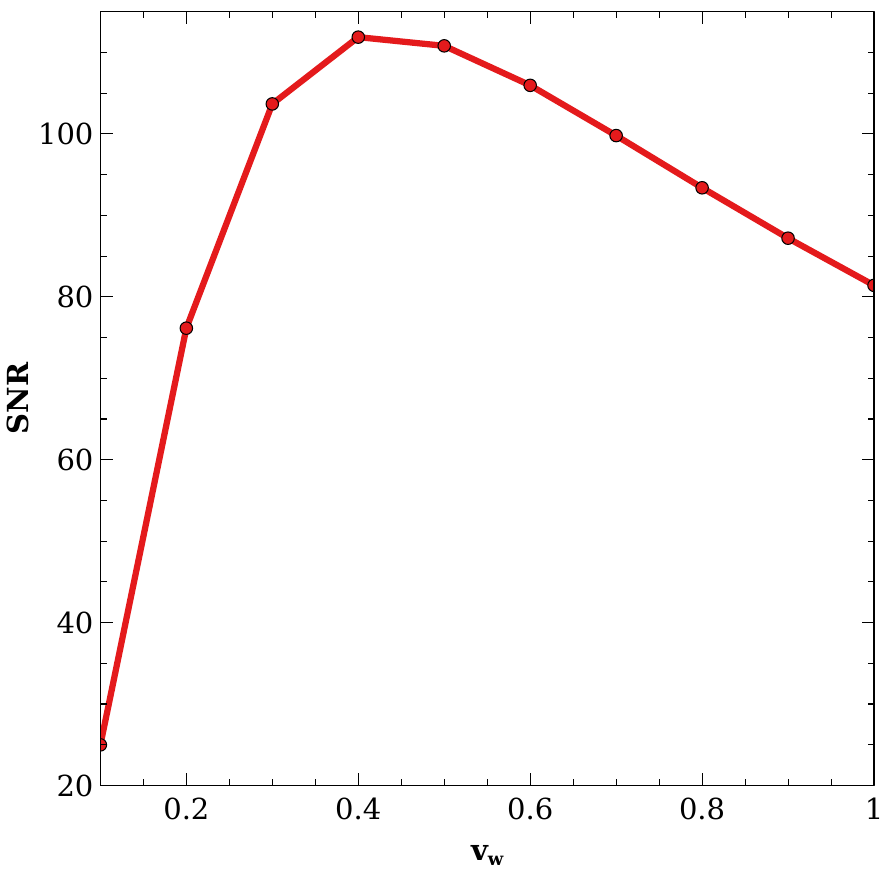}
    \caption{Left panel: Resultant GW spectra from the FO-EWPT for different $v_w$ for BP1. Right panel: Variation of the SNR with $v_w$.}
    \label{fig:vw_SNR}
\end{figure}

In the left panel of figure~\ref{fig:vw_SNR}, we present the GW spectra for different values of $v_w$ for BP1. We note that the amplitude of the GW spectrum increases with increasing $v_w$, consistent with eq.~\eqref{eq:gw_sound}, as the dominant contribution to the spectrum arises from sound waves in the plasma. On the other hand, the peak frequency decreases, in accordance with eq.~\eqref{eq:peak_freak}. In the right panel, we show the variation of the SNR with the $v_w$. We find that the SNR is maximized for $v_w \sim 0.4$. Several studies have explored methods to estimate the bubble wall velocity \cite{Ai:2023see,Ekstedt:2024fyq,vandeVis:2025plm}. However, these approaches are subject to significant uncertainties, which propagate into the predictions of the GW spectra and, consequently, into the corresponding SNR. The choice of $v_w \sim1$ considered here is rather conservative and simplified. This ultra-relativistic limit of the wall velocity can arise in scenarios where friction is insufficient to balance the vacuum pressure, though this is not a generic outcome. 

We would like to point out that the phase transition parameters and the SNR depend on the choice of gauge and renormalization scale of the effective potential~\cite{Zhu:2025pht,Croon:2020cgk}. The gauge dependence of the FOPT parameters with the cutoff scale for the operator $\mathcal{O}_H$ is discussed in Ref.~\cite{Zhu:2025pht}. Efforts to eliminate the gauge dependence of the effective potential have also been explored in Ref.~\cite{Patel:2011th}.  In addition, the choice of renormalization scale can also affect the FOPT parameters and corresponding SNR, which is elaborately discussed in Ref.~\cite{Croon:2020cgk}. This scale dependence can be mitigated by incorporating the renormalization group equations improved treatment of the effective potential ~\cite{Andreassen:2014gha,Andreassen:2014eha}.
\section{Constraints from existing collider studies}
\label{sec:collider}
Collider experiments, particularly the LHC, play a crucial role in constraining the EFT operator coefficients that parameterize possible NP effects beyond the Standard Model. By comparing measured observables with SM predictions extended by higher-dimensional operators, one can extract upper limits on the deviations induced by the NP. These constraints arise primarily from precision measurements of processes sensitive to Higgs couplings, electroweak interactions, and rare decay modes. Electroweak precision observables, Higgs signal strengths, and differential cross-section measurements collectively narrow the allowed parameter space of the EFT coefficients. As the LHC continues to accumulate data and improve measurement precision, these bounds are expected to tighten further, offering an increasingly stringent test of the SM and its possible extensions.

\subsection{Trilinear Higgs self-coupling modifier}
NP effects in the Higgs sector are commonly analyzed at the collider experiments within the $\kappa$-framework. In this approach, deviations from the SM are parameterized as multiplicative modifiers of the Higgs couplings, assuming that NP contributes as an anomalous correction with the same Lorentz structure as the SM interaction. The trilinear Higgs self-coupling modifier is defined as
\begin{equation}
    \kappa_{\lambda} = \frac{\lambda_{\rm obs}}{\lambda_{\rm SM}}\,,
\end{equation}
where $\lambda_{\rm SM}$ is the Higgs self-coupling in the SM. A deviation of $\kappa_{\lambda}$ from unity signals the presence of NP in the Higgs potential. In the SMEFT framework, such a deviation can originate from the operator $\mathcal{O}_{H}$, which modifies the Higgs potential without altering its Lorentz structure. The corresponding parameterization reads
\begin{equation}
    \kappa_{\lambda} = 1 - \frac{5v^{2}}{2 \lambda_{\rm SM}}\left(\frac{C_{H}}{\Lambda^{2}}\right)\,.
\end{equation}
The most stringent current constraints on $\kappa_{\lambda}$ arise from the ATLAS measurements of single- and double-Higgs production at a center-of-mass (CM) energy of 13 TeV with an integrated luminosity of 126-139 fb$^{-1}$~\cite{ATLAS:2022jtk}. The corresponding bounds on $\kappa_{\lambda}$ and the inferred limits on $C_H/\Lambda^2$ are summarized in Table~\ref{tab:c1}.

\begin{table}[h!]
    \centering
    \begin{tabular}{ccc}
    \hline \hline
     Production Channel & 95\% C.L. Bound on $\kappa_{\lambda}$ & Bound on $C_{H}/\Lambda^{2}$ (TeV$^{-2}$) \\ \hline
     $hh$ & $[-0.6, +6.6]$ & $[-4.77, +1.36]$ \\
     $h + hh$ & $[-0.4, +6.3]$ & $[-4.52, +1.19]$ \\ \hline \hline
    \end{tabular}
    \caption{Current 95\% C.L. bounds on $\kappa_{\lambda}$ and $C_{H}/\Lambda^{2}$ from ATLAS analyses of single and double Higgs production at $\sqrt{s} = 13$ TeV and $\mathfrak{L}_{\rm int} = 126$-$139$ fb$^{-1}$. Both gluon-gluon fusion (ggF) and vector boson fusion (VBF) production modes are included. All other couplings are assumed to be SM-like.}
    \label{tab:c1}
\end{table}
\subsection{Top Yukawa coupling modifier}
Similar to the Higgs self-coupling modifier case, the operator, $\mathcal{O}_{tH}$, shifts the top quark Yukawa coupling, post EWSB, and, hence, is sensitive to the top Yukawa coupling modifier measurements at the LHC. The matching equation is
\begin{equation}
    \kappa_{t} = 1 + \frac{v^{2}}{2 Y_{t}} \left(\frac{C_{tH}}{\Lambda^{2}}\right)
\end{equation}
The most stringent limit comes from measurements in the combined study of Higgs couplings at the ATLAS experiment~\cite{ATLAS:2022vkf}. The 95\% C.L. limit of $\kappa_{t}$ and the corresponding translated bound on $C_{tH}/\Lambda^{2}$ is
\begin{equation}
    \kappa_{t}: [0.88, 1.02] \;\;\; \longrightarrow \;\;\; C_{tH}/\Lambda^{2}\;(\text{TeV}^{-2}): [-3.94,+0.66]\,.
\end{equation}
\subsection{Constraints on other SMEFT operators}
Beyond the trilinear Higgs coupling, other dimension-6 SMEFT operators also affect Higgs production and decay processes. In particular, $\mathcal{O}_{H\Box}$ and $\mathcal{O}_{HD}$ modify the Higgs kinetic terms and its couplings to electroweak gauge bosons, respectively. Consequently, they can be constrained through combined fits to Higgs and electroweak data. The ATLAS collaboration has performed a global interpretation of EFT coefficients using combined datasets from LHC Higgs and weak boson measurements, supplemented with Electroweak Precision Data (EWPD) from the Large Electron-Positron collider (LEP)~\cite{ATL-PHYS-PUB-2022-037}. The resulting 95\% confidence level (C.L.) bounds on $C_{H\Box}/\Lambda^{2}$ are:
\begin{equation}
\begin{split}
    C_{H\Box}/\Lambda^{2}\; ({\rm TeV}^{-2}): & \quad [-1.23,\, +3.62] \quad {\rm (LHC)} \,, \\
    C_{H\Box}/\Lambda^{2}\; ({\rm TeV}^{-2}): & \quad [-2.82,\, +2.75] \quad {\rm (LHC+EWPO)} \,.
\end{split}
\end{equation}

Complementary results have also been provided by the CMS collaboration from analyses of Higgs decays into electroweak gauge bosons. The constraints on $C_{H\Box}/\Lambda^{2}$ and $C_{HD}/\Lambda^{2}$ obtained from $h \rightarrow ZZ^{*} \rightarrow 4\ell$~\cite{CMS:2021nnc} and $h \rightarrow WW^{*}$~\cite{CMS:2024bua} channels are as follows:
\begin{equation}
\begin{split}
    C_{H\Box}/\Lambda^{2}\; ({\rm TeV}^{-2}): & \quad [-0.41,\, +0.47] \quad (h \rightarrow ZZ \rightarrow 4\ell)\,, \\
    C_{H\Box}/\Lambda^{2}\; ({\rm TeV}^{-2}): & \quad [-4.19,\, +0.67] \quad (h \rightarrow WW)\,, \\
    C_{HD}/\Lambda^{2}\; ({\rm TeV}^{-2}): & \quad [-4.94,\, +0.24] \quad (h \rightarrow ZZ \rightarrow 4\ell)\,, \\
    C_{HD}/\Lambda^{2}\; ({\rm TeV}^{-2}): & \quad [-0.44,\, +0.81] \quad (h \rightarrow WW)\,.
\end{split}
\end{equation}
These results collectively indicate that current LHC measurements already impose meaningful constraints on the effective couplings of dimension-6 operators in the Higgs sector. However, future runs of the HL-LHC and possible next-generation colliders will be essential to further refine these limits and probe smaller deviations from the SM expectations.

\section{\textit{Di}-Higgs production at future LHC runs}
\label{sec:di-higgs}
A comprehensive understanding of the Higgs sector is essential for uncovering the mechanism of electroweak symmetry breaking and testing the structure of the Higgs potential. In this context, the measurement of the Higgs self-coupling, accessible through Higgs boson pair production (\textit{di}-Higgs production), plays a central role. The \textit{di}-Higgs process provides a direct probe of the Higgs potential and serves as a sensitive window to possible NP effects that could modify the Higgs self-interactions. At the HL-LHC, operating at a CM energy of 14 TeV with an expected integrated luminosity of 3 ab$^{-1}$, and at the HE-LHC with 27 TeV and 15 ab$^{-1}$, the sensitivity to \textit{di}-Higgs production is expected to improve substantially. In the SM, this process proceeds predominantly via gluon-gluon fusion (ggF), mediated by top-quark loop. However, the small production cross section and large backgrounds render its observation extremely challenging. In this study, we investigate the prospects of \textit{di}-Higgs production at the HL- and HE-LHC in the presence of above mentioned dimension-6 effective operators, which can significantly alter both the production rate and the kinematic properties of the Higgs pair system.

\begin{figure}[htb!]
	\centering
	\begin{tikzpicture}[baseline={(current bounding box.center)},style={scale=0.85, transform shape}]
			\begin{feynman}
				\vertex(a);
				\vertex[left = 1cm of a] (a1) {$g$};
				\vertex[below = 2cm of a] (b);
				\vertex[left = 1cm of b] (b1) {$g$};
				\vertex[below right = 1 cm and 1 cm of a] (c);
				\vertex[right = 1cm of c] (d);
				\vertex[above right = 1 cm and 1 cm of d] (d1) {$h$};
				\vertex[below right = 1 cm and 1 cm of d] (d2) {$h$};
				\diagram*{
					(a1) -- [ gluon, arrow size=0.7pt] (a),
					(a) -- [ fermion, arrow size=0.7pt] (b),
					(b1) -- [ gluon, arrow size=0.7pt] (b),
					(c) -- [ fermion, arrow size=0.7pt] (a),
					(b) -- [ fermion, arrow size=0.7pt] (c),
					(c) -- [ scalar, arrow size=0.7pt] (d),
					(d) -- [ scalar, arrow size=0.7pt] (d1),
					(d) -- [ scalar, arrow size=0.7pt] (d2),
					};
			\end{feynman}
		\end{tikzpicture}
		\hspace{1.00cm}
		\begin{tikzpicture}[baseline={(current bounding box.center)},style={scale=0.85, transform shape}]
			\begin{feynman}
				\vertex(a);
				\vertex[left = 1cm of a] (a1) {$g$};
				\vertex[below = 2cm of a] (b);
				\vertex[left = 1cm of b] (b1) {$g$};
				\vertex[below right = 1 cm and 1 cm of a] (c);
				\vertex[right = 1cm of c] (d);
				\vertex[above right = 1 cm and 1 cm of d] (d1) {$h$};
				\vertex[below right = 1 cm and 1 cm of d] (d2) {$h$};
				\diagram*{
					(a1) -- [ gluon, arrow size=0.7pt] (a),
					(a) -- [ fermion, arrow size=0.7pt] (b),
					(b1) -- [ gluon, arrow size=0.7pt] (b),
					(c) -- [ fermion, arrow size=0.7pt] (a),
					(b) -- [ fermion, arrow size=0.7pt] (c),
					(c) -- [ scalar, arrow size=0.7pt] (d),
					(d) -- [ scalar, arrow size=0.7pt] (d1),
					(d) -- [ scalar, arrow size=0.7pt] (d2),
				};
			\end{feynman}
			\node at (d)[black,fill,style=black,inner sep=3pt]{};
		\end{tikzpicture}
		\begin{tikzpicture}[baseline={(current bounding box.center)},style={scale=0.85, transform shape}]
			\begin{feynman}
				\vertex(a);
				\vertex[left = 1cm of a] (a1) {$g$};
				\vertex[below = 2cm of a] (b);
				\vertex[left = 1cm of b] (b1) {$g$};
				\vertex[below right = 1 cm and 1 cm of a] (c);
				\vertex[right = 1cm of c] (d);
				\vertex[above right = 1 cm and 1 cm of d] (d1) {$h$};
				\vertex[below right = 1 cm and 1 cm of d] (d2) {$h$};
				\diagram*{
					(a1) -- [ gluon, arrow size=0.7pt] (a),
					(a) -- [ fermion, arrow size=0.7pt] (b),
					(b1) -- [ gluon, arrow size=0.7pt] (b),
					(c) -- [ fermion, arrow size=0.7pt] (a),
					(b) -- [ fermion, arrow size=0.7pt] (c),
					(c) -- [ scalar, arrow size=0.7pt] (d),
					(d) -- [ scalar, arrow size=0.7pt] (d1),
					(d) -- [ scalar, arrow size=0.7pt] (d2),
				};
			\end{feynman}
			\node at (c)[black,fill,style=black,inner sep=3pt]{};
		\end{tikzpicture} \\
		\begin{tikzpicture}[baseline={(current bounding box.center)},style={scale=0.85, transform shape}]
			\begin{feynman}
				\vertex(a);
				\vertex[left = 1cm of a] (a1) {$g$};
				\vertex[below = 2cm of a] (b);
				\vertex[left = 1cm of b] (b1) {$g$};
				\vertex[ right = 2 cm of a] (c1);
				\vertex[ right = 2 cm of b] (c2);
				\vertex[right = 1 cm of c1] (d1) {$h$};
				\vertex[right = 1 cm of c2] (d2) {$h$};
				\diagram*{
					(a1) -- [ gluon, arrow size=0.7pt] (a),
					(a) -- [ fermion, arrow size=0.7pt] (b),
					(b1) -- [ gluon, arrow size=0.7pt] (b),
					(a) -- [ fermion, arrow size=0.7pt] (c1),
					(b) -- [ fermion, arrow size=0.7pt] (c2),
					(c1) -- [ fermion, arrow size=0.7pt] (c2),
					(c1) -- [ scalar, arrow size=0.7pt] (d1),
					(c2) -- [ scalar, arrow size=0.7pt] (d2),
				};
			\end{feynman}
		\end{tikzpicture}
            \hspace{1.10cm}
		\begin{tikzpicture}[baseline={(current bounding box.center)},style={scale=0.85, transform shape}]
			\begin{feynman}
				\vertex(a);
				\vertex[left = 1cm of a] (a1) {$g$};
				\vertex[below = 2cm of a] (b);
				\vertex[left = 1cm of b] (b1) {$g$};
				\vertex[ right = 2 cm of a] (c1);
				\vertex[ right = 2 cm of b] (c2);
				\vertex[right = 1 cm of c1] (d1) {$h$};
				\vertex[right = 1 cm of c2] (d2) {$h$};
				\diagram*{
					(a1) -- [ gluon, arrow size=0.7pt] (a),
					(a) -- [ fermion, arrow size=0.7pt] (b),
					(b1) -- [ gluon, arrow size=0.7pt] (b),
					(a) -- [ fermion, arrow size=0.7pt] (c1),
					(b) -- [ fermion, arrow size=0.7pt] (c2),
					(c1) -- [ fermion, arrow size=0.7pt] (c2),
					(c1) -- [ scalar, arrow size=0.7pt] (d1),
					(c2) -- [ scalar, arrow size=0.7pt] (d2),
				};
			\end{feynman}
                \node at (c1)[black,fill,style=black,inner sep=3pt]{};
		\end{tikzpicture}
            \begin{tikzpicture}[baseline={(current bounding box.center)},style={scale=0.85, transform shape}]
			\begin{feynman}
				\vertex(a);
				\vertex[left = 1cm of a] (a1) {$g$};
				\vertex[below = 2cm of a] (b);
				\vertex[left = 1cm of b] (b1) {$g$};
				\vertex[below right = 1 cm and 1.5 cm of a] (c);
				\vertex[above right = 1 cm and 1.5 cm of c] (d1) {$h$};
				\vertex[below right = 1 cm and 1.5 cm of c] (d2) {$h$};
				\diagram*{
					(a1) -- [ gluon, arrow size=0.7pt] (a),
					(a) -- [ fermion, arrow size=0.7pt] (b),
					(b1) -- [ gluon, arrow size=0.7pt] (b),
					(c) -- [ fermion, arrow size=0.7pt] (a),
					(b) -- [ fermion, arrow size=0.7pt] (c),
					(c) -- [ scalar, arrow size=0.7pt] (d1),
					(c) -- [ scalar, arrow size=0.7pt] (d2),
				};
			\end{feynman}
			\node at (c)[black,fill,style=black,inner sep=3pt]{};
		\end{tikzpicture}
	\caption{Feynman graphs concerning Higgs pair production: SM (\textit{left}) and EFT (\textit{right}) contributions, via gluon-gluon fusion (ggF) production mode at the LHC. The black node represents the effective vertex.}
	\label{fig:hh_lhc}
\end{figure}

\subsection{Signal-background simulation}
The final state signal of our interest is $2b2\tau$. The EFT model is implemented using \texttt{FeynRules}~\cite{Alloul:2013bka}. The corresponding QCD ultraviolet (UV) counterterms and $R_2$ terms are generated with \texttt{NLOCT}~\cite{Degrande:2014vpa}. The signal process, $pp \rightarrow h(b\bar{b})h(\tau^+\tau^-)$, shown in figure~\ref{fig:hh_lhc}, is simulated in \texttt{MG5\_aMC@NLO}~\cite{Alwall:2011uj}, with top-quark loop contribution evaluated using \texttt{MadLoop}~\cite{Hirschi:2011pa}. Additional contributions to the \textit{di}-Higgs production can come from the VBF mode, however, we only consider the dominant ggF mode for our study. Event generation is performed for CM energies of 14 TeV and 27 TeV, corresponding to the HL-LHC and HE-LHC runs, respectively. The dominant Higgs decay mode is $h \rightarrow b\overline{b}$, which accounts for approximately 58.2\% of the total decay width. However, allowing both Higgs bosons to decay into $b\overline{b}$ leads to substantial QCD multijet background contamination. To mitigate this, one Higgs boson is decayed to $b\overline{b}$ while the other is decayed to $\tau^{+}\tau^{-}$ using \texttt{MadSpin}. Although the $\tau^{+}\tau^{-}$ decay has a smaller branching ratio, it benefits from significantly reduced background contamination and offers a cleaner final state for reconstruction. Although the $2b2\gamma$ channel suffers the least from background contamination, the Higgs diphoton branching ratio is extremely small, and extracting such a rare signal reliably requires highly specialized reconstruction techniques; therefore, we do not include this channel in our analysis. In this analysis, we focus on the final state consisting of two $b$-tagged jets and two hadronically decaying $\tau$ leptons. The hadronic $\tau$ decays, which produce narrow jets with distinctive signatures, can be efficiently identified at collider detectors through specialized $\tau$-tagging algorithms. Consequently, the $2b2\tau$ final state provides an optimal balance between signal yield and background suppression. However, combining all channels would yield the highest overall sensitivity, but doing so demands substantial simulation resources and is therefore left for future work.

The dominant SM background processes considered in this analysis are $pp \rightarrow Z + {\rm jets}$ and $pp \rightarrow t\overline{t}$. Additional backgrounds include multijet fake backgrounds which are largely reducible~\cite{ATLAS:2024pov}. The dominant backgrounds are generated at leading order (LO) using \texttt{MG5\_aMC@NLO}. Hadronic processes at the LHC are subject to sizable QCD corrections, which can significantly affect both signal and background rates. To approximately incorporate these effects, we rescale the LO cross sections using appropriate next-to-leading order (NLO) $K$-factors. For the signal process, the $K_{\rm NLO}$ factors are 1.869 at 14 TeV and 1.863 at 27 TeV~\cite{AH:2022elh}. The corresponding background $K$-factors, obtained from \texttt{MG5\_aMC@NLO}, are:
\begin{equation}
\begin{split}
    K^{Zj}_{\rm NLO} &= 1.280, \quad K^{t\overline{t}}_{\rm NLO} = 1.485 \quad \text{at } 14~\text{TeV}, \\
    K^{Zj}_{\rm NLO} &= 1.304, \quad K^{t\overline{t}}_{\rm NLO} = 1.490 \quad \text{at } 27~\text{TeV}.
\end{split}
\end{equation}

The generated events are subsequently processed through \texttt{Pythia8}~\cite{Bierlich:2022pfr} for parton showering and hadronization, followed by detector simulation using \texttt{Delphes3}~\cite{deFavereau:2013fsa} with the \texttt{delphes\_card\_CMS} configuration. The detector card accounts for realistic reconstruction efficiencies, energy resolutions, and $b$- and $\tau$-jet tagging performances. Jet reconstruction is performed within \texttt{Delphes3} using the anti-$k_T$ algorithm, as implemented in \texttt{FastJet3}~\cite{Cacciari:2011ma}.
\subsection{ANN-based classifier}
\begin{figure}[htb!]
    \centering
    \includegraphics[width = 0.4\textwidth]{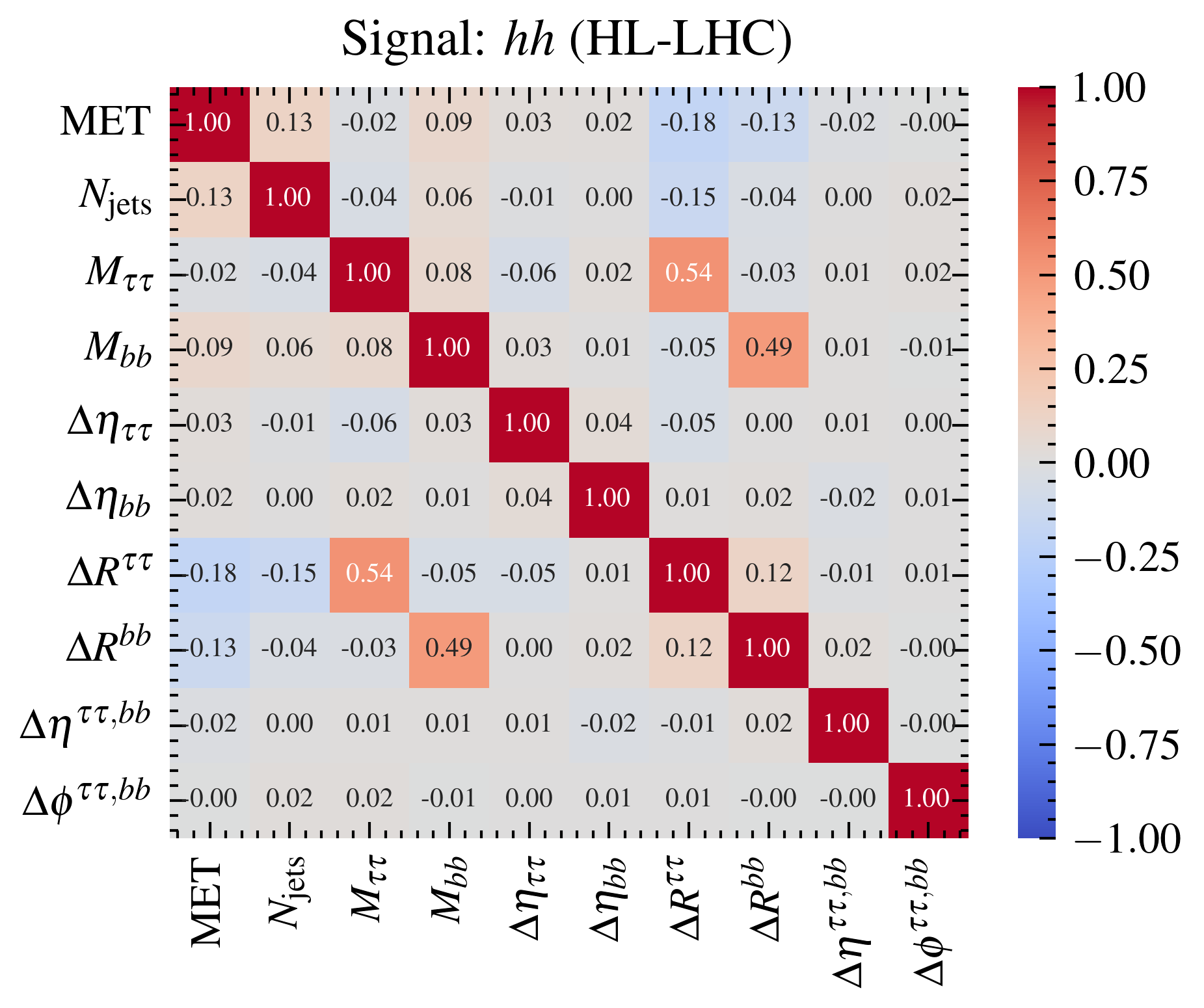} \qquad
    \includegraphics[width = 0.4\textwidth]{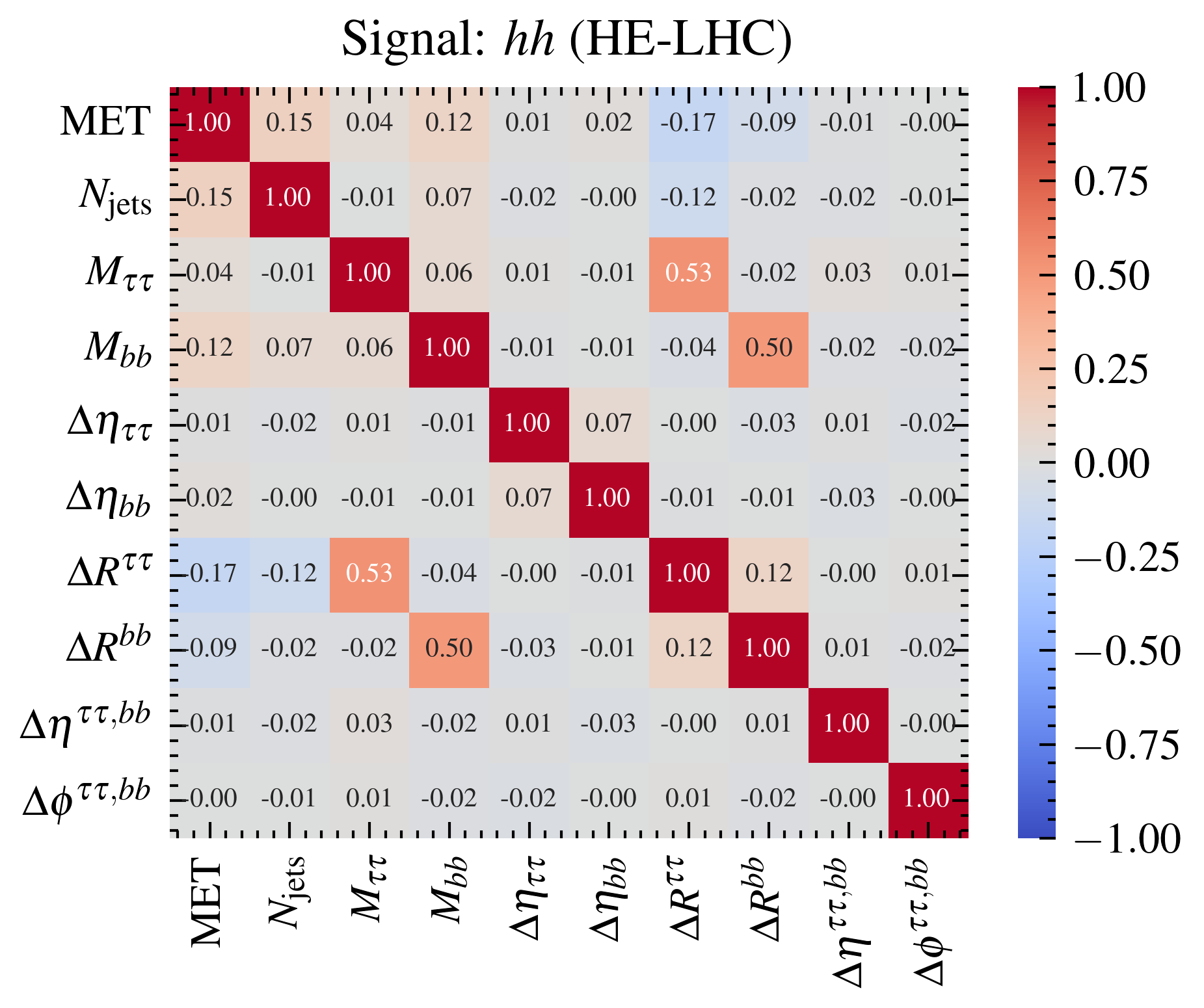} \\
    \includegraphics[width = 0.4\textwidth]{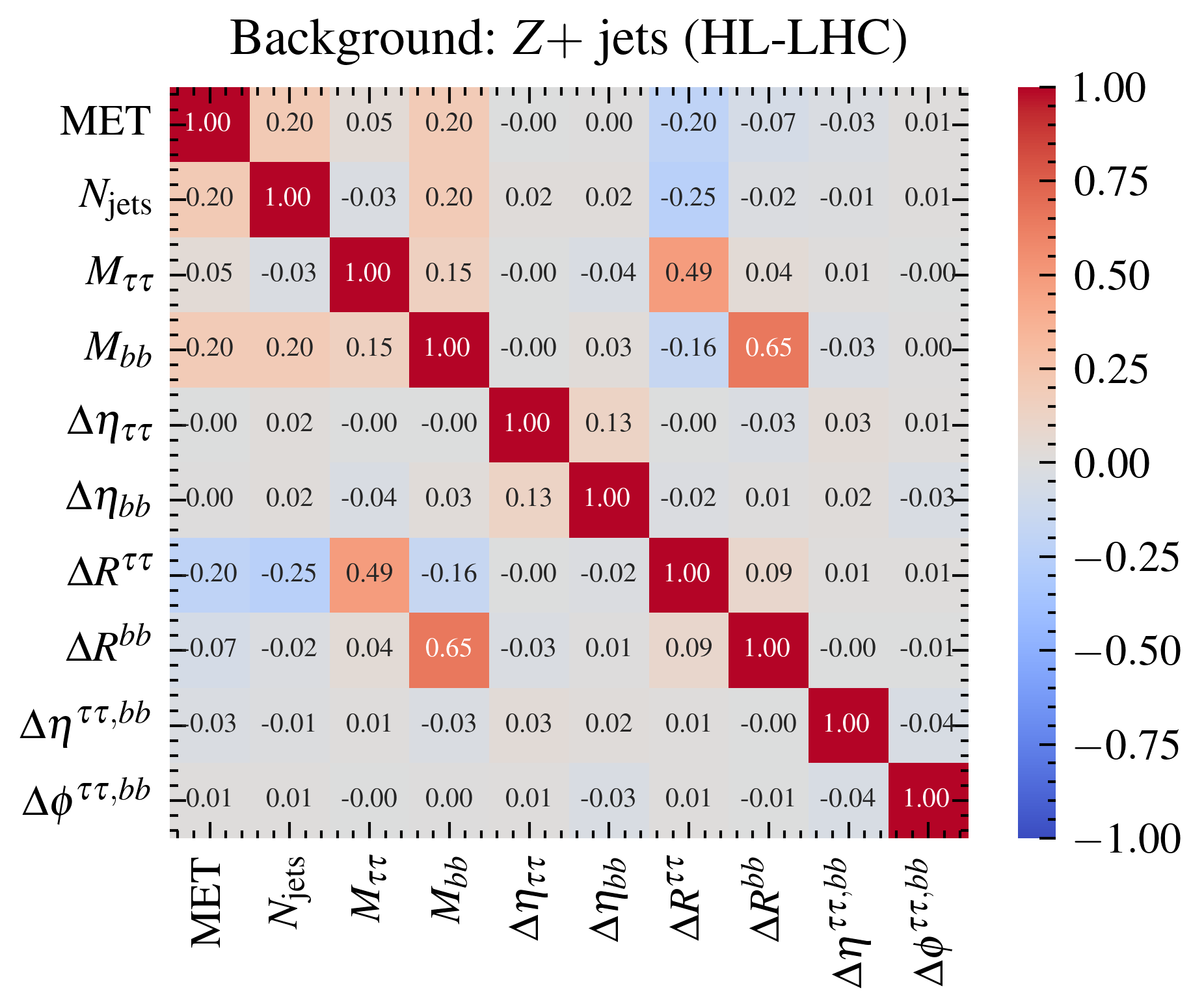}
    \qquad
    \includegraphics[width = 0.4\textwidth]{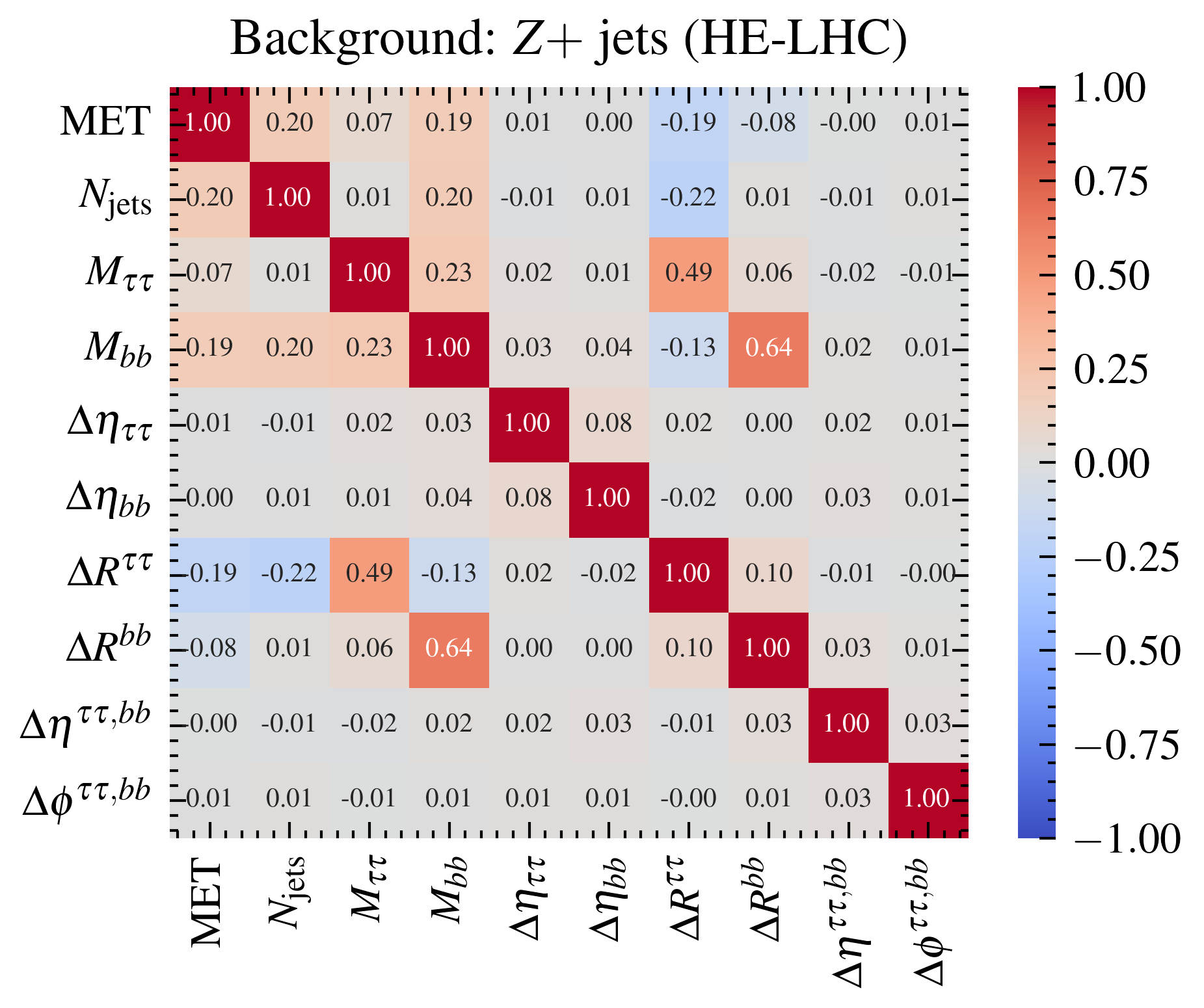} \\
    \includegraphics[width = 0.4\textwidth]{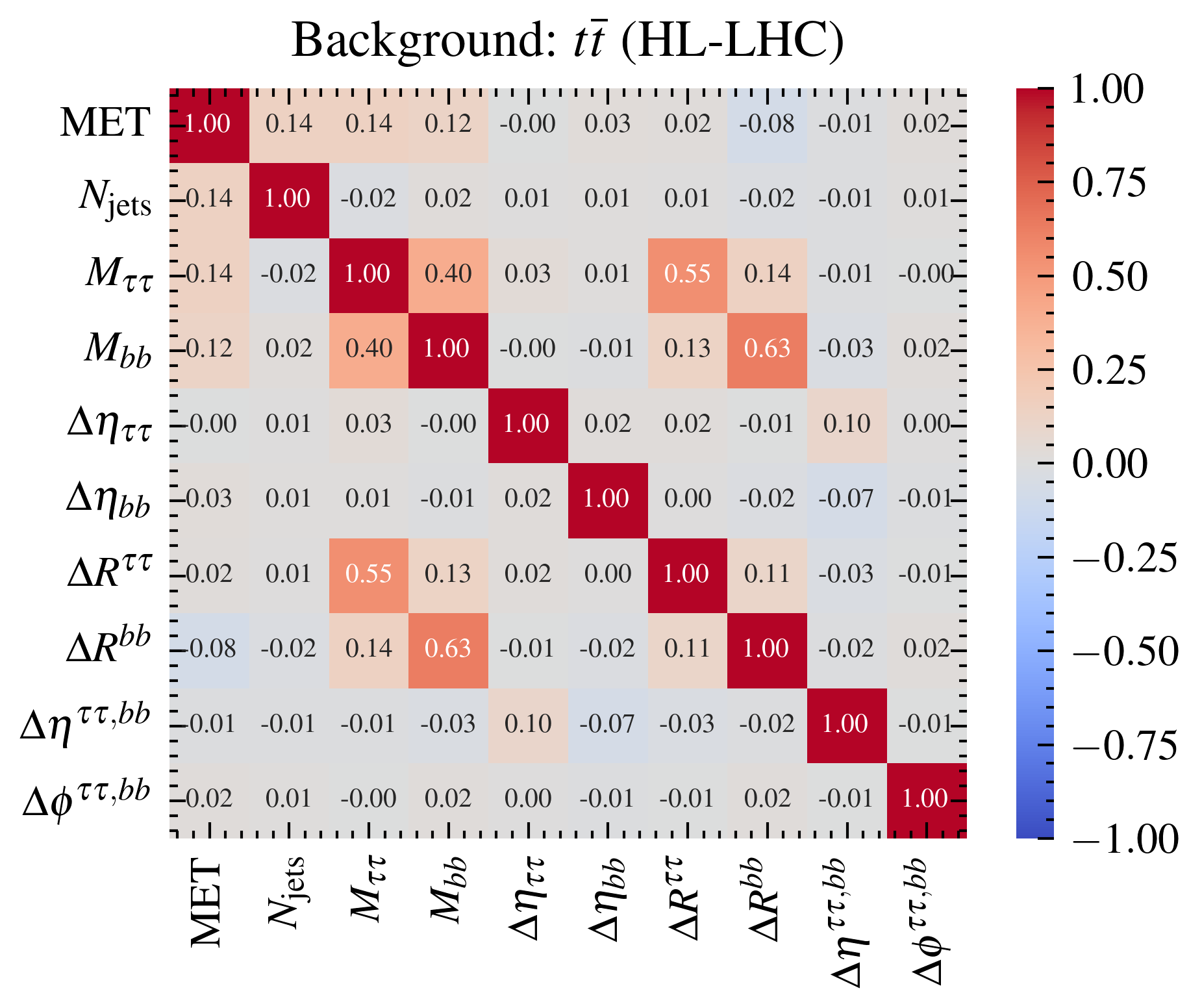} 
    \qquad    
    \includegraphics[width = 0.4\textwidth]{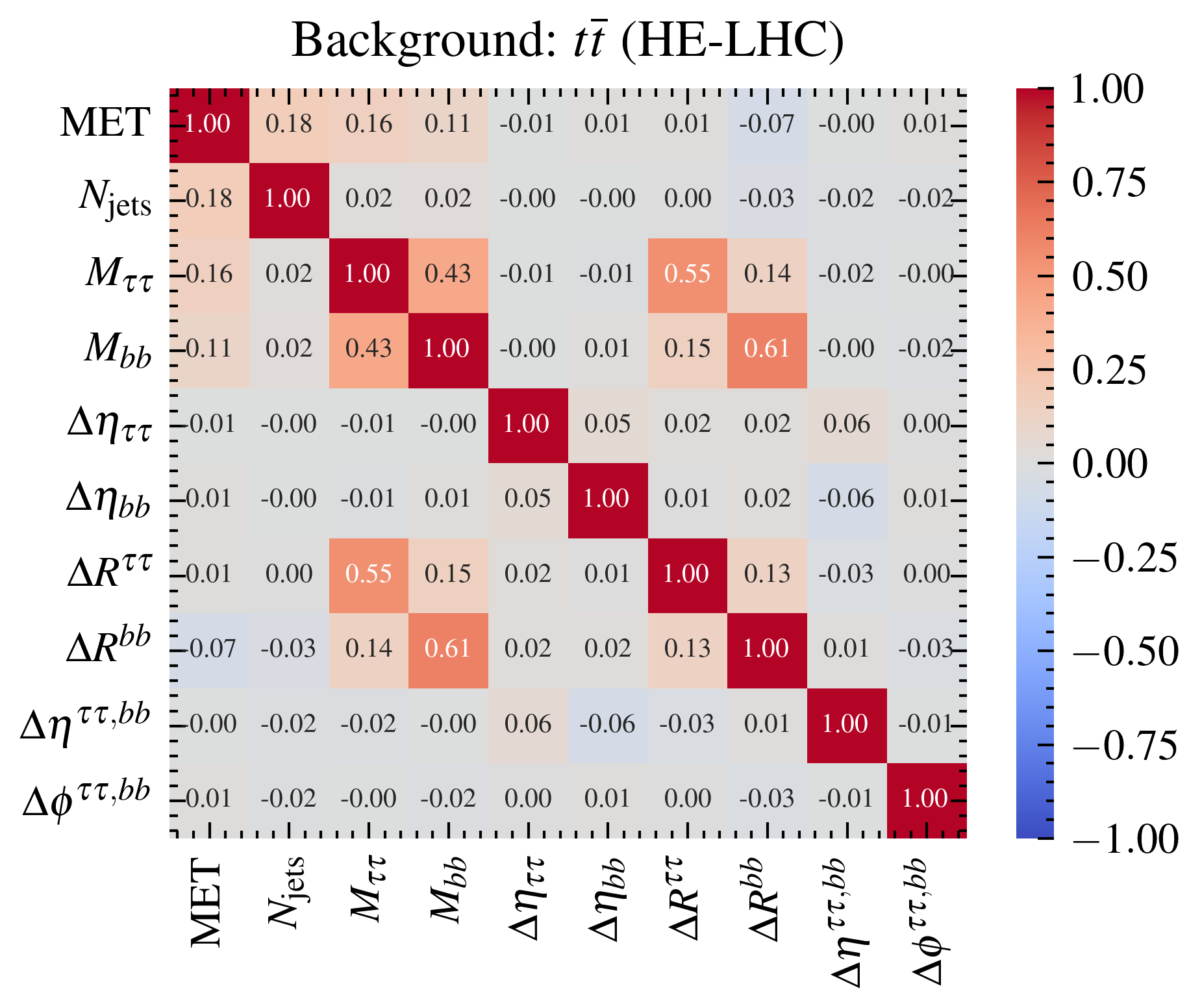} \\
    \caption{The kinematic variable correlations for signal and background processes for HL-LHC (14 TeV 3 ab$^{-1}$) and HE-LHC (27 TeV 15 ab$^{-1}$).}
    \label{fig:corr}
\end{figure}
For signal selection, events with at least four jets in the final state are considered, with exactly two $b$-tagged and two $\tau$-tagged jets. In cases where a jet is simultaneously identified as both $b$- and $\tau$-tagged, the $b$-tagging is given priority. Events containing hard leptons or photons with transverse momentum, $p_{T} > 10~\text{GeV}$, are rejected. The four-object final state enables the construction of a large set of angular observables, which can effectively discriminate the signal process from the background. Such discrimination can be efficiently achieved using machine learning-based classification techniques. A particularly powerful class of such methods is the ANNs, which have recently become popular in collider event analyses due to their strong non-linear learning capability and robustness against correlations among input observables. ANNs are well-suited for handling multi-dimensional feature spaces where conventional cut-based or likelihood approaches often fail to capture complex kinematic dependencies. A comparison with the traditional cut-based approach is provided in Appendix~\ref{app:1}. They learn directly from data distributions, optimizing separation between signal and background through iterative weight updates that minimize a chosen loss function. In this work, we employ a binary classification ANN to distinguish the SM \textit{di}-Higgs signal from other SM background processes. It is worth noting that since the dominant contribution to the signal originates from the SM itself, the training of the model is performed using \textit{SM-only} datasets. The robustness of the model across different EFT working points is validated in Appendix~\ref{app:2}. We construct the following kinematic observables as input features for discriminating the signal from the background:
\begin{itemize}
    \itemsep-0em
    \item \textbf{MET}: The missing transverse energy of the event.
    \item $\boldsymbol{N_{\rm jets}}$: The number of reconstructed jets in the final state.
    \item $\boldsymbol{M_{\tau\tau}}$: The invariant mass of the $\tau$-jet pair.
    \item $\boldsymbol{M_{bb}}$: The invariant mass of the $b$-jet pair.
    \item $\boldsymbol{\Delta \eta_{\tau\tau}}$: The pseudorapidity difference between the $\tau$-jets.
    \item $\boldsymbol{\Delta \eta_{bb}}$: The pseudorapidity difference between the $b$-jets.
    \item $\boldsymbol{\Delta R_{\tau\tau}}$: The separation between the $\tau$-jets in the $\eta$--$\phi$ plane.
    \item $\boldsymbol{\Delta R_{bb}}$: The separation between the $b$-jets in the $\eta$--$\phi$ plane.
    \item $\boldsymbol{\Delta \eta_{\tau\tau,bb}}$: The pseudorapidity difference between the $\tau$ and $b$ jet systems.
    \item $\boldsymbol{\Delta \phi_{\tau\tau,bb}}$: The azimuthal angle difference between the $\tau$ and $b$ jet systems.
\end{itemize}
\begin{figure}[htb!]
    \centering
    \includegraphics[width = 0.4\textwidth]{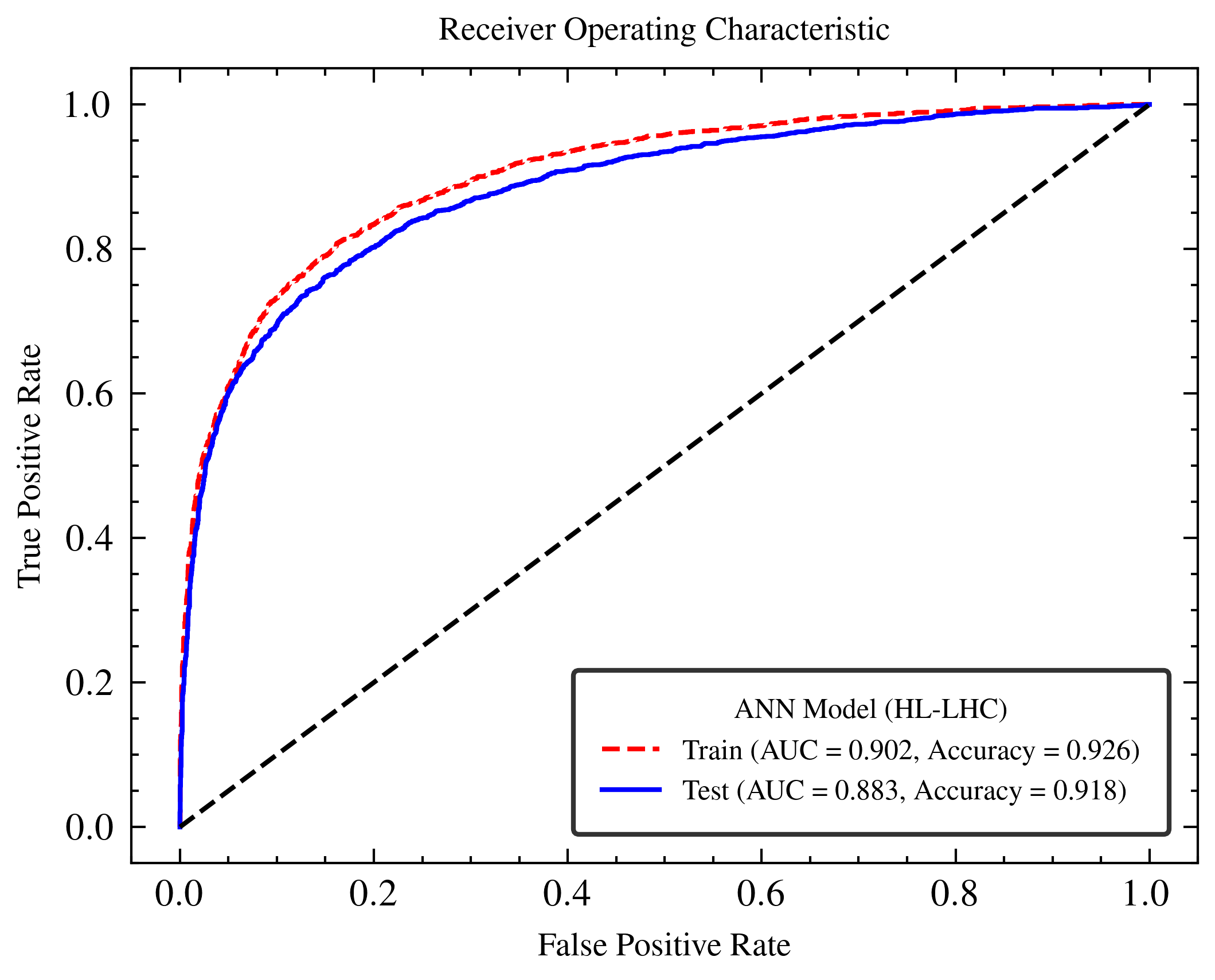} \qquad
    \includegraphics[width = 0.4\textwidth]{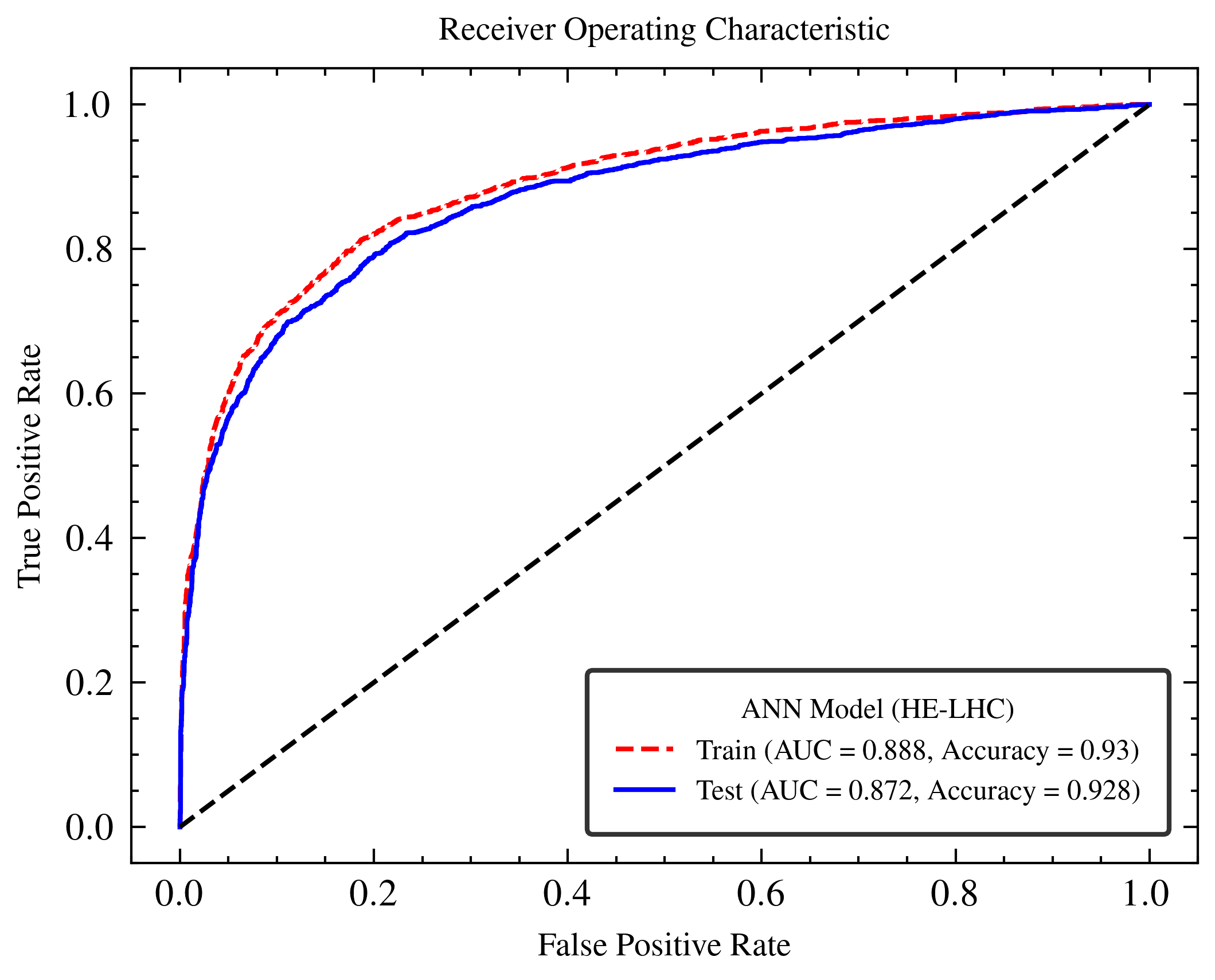} \\
    \includegraphics[width = 0.4\textwidth]{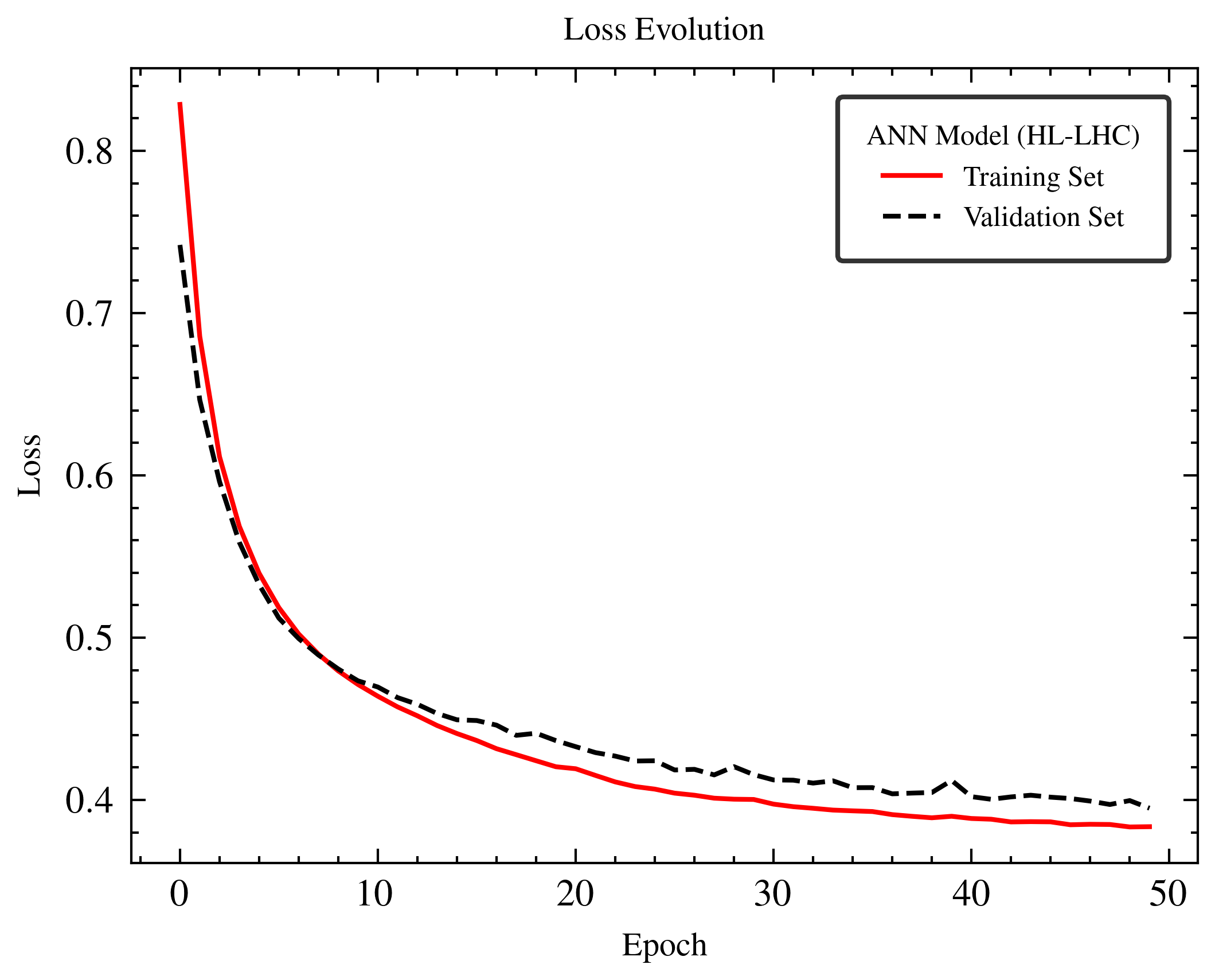} \qquad
    \includegraphics[width = 0.4\textwidth]{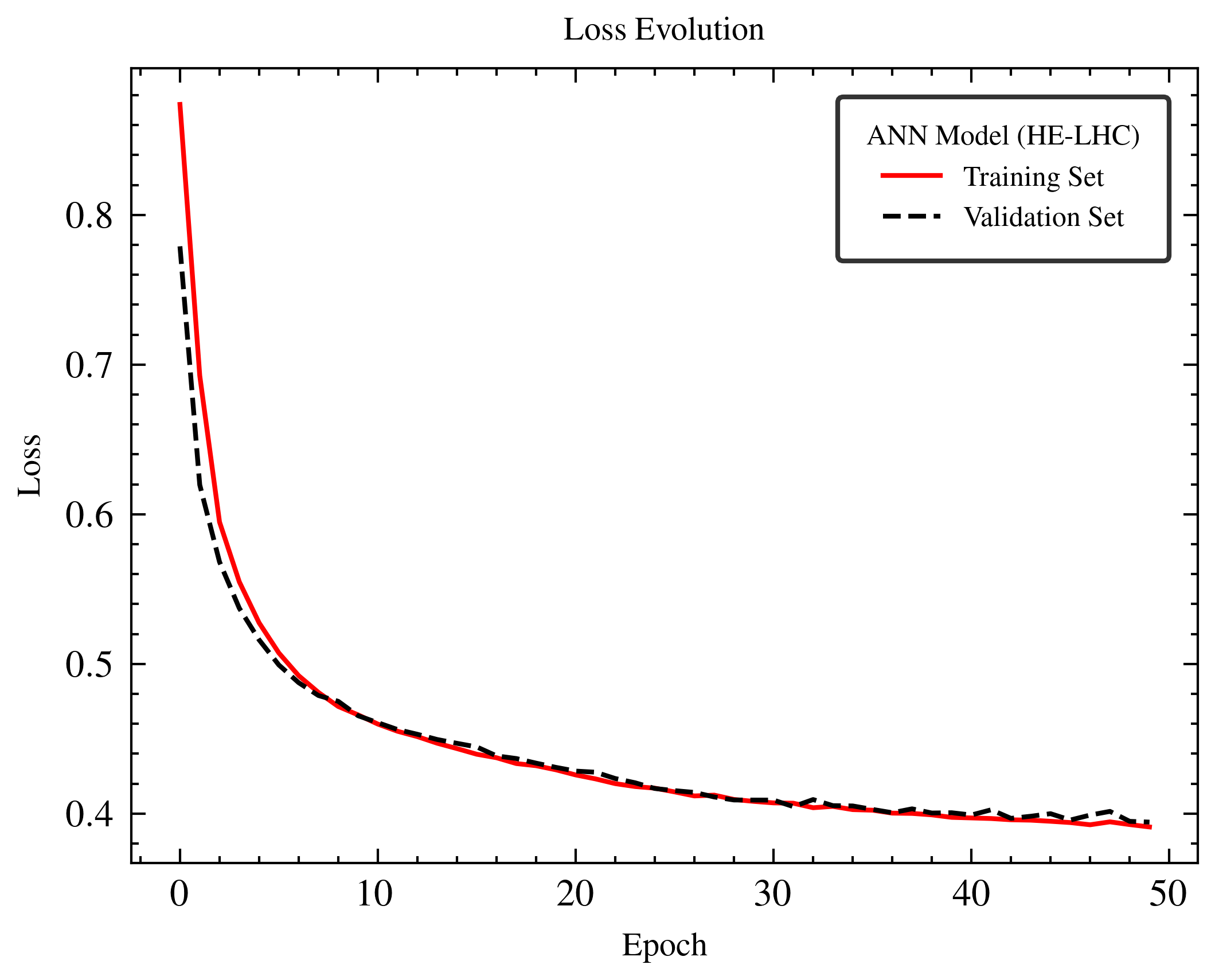} \\
    \includegraphics[width = 0.4\textwidth]{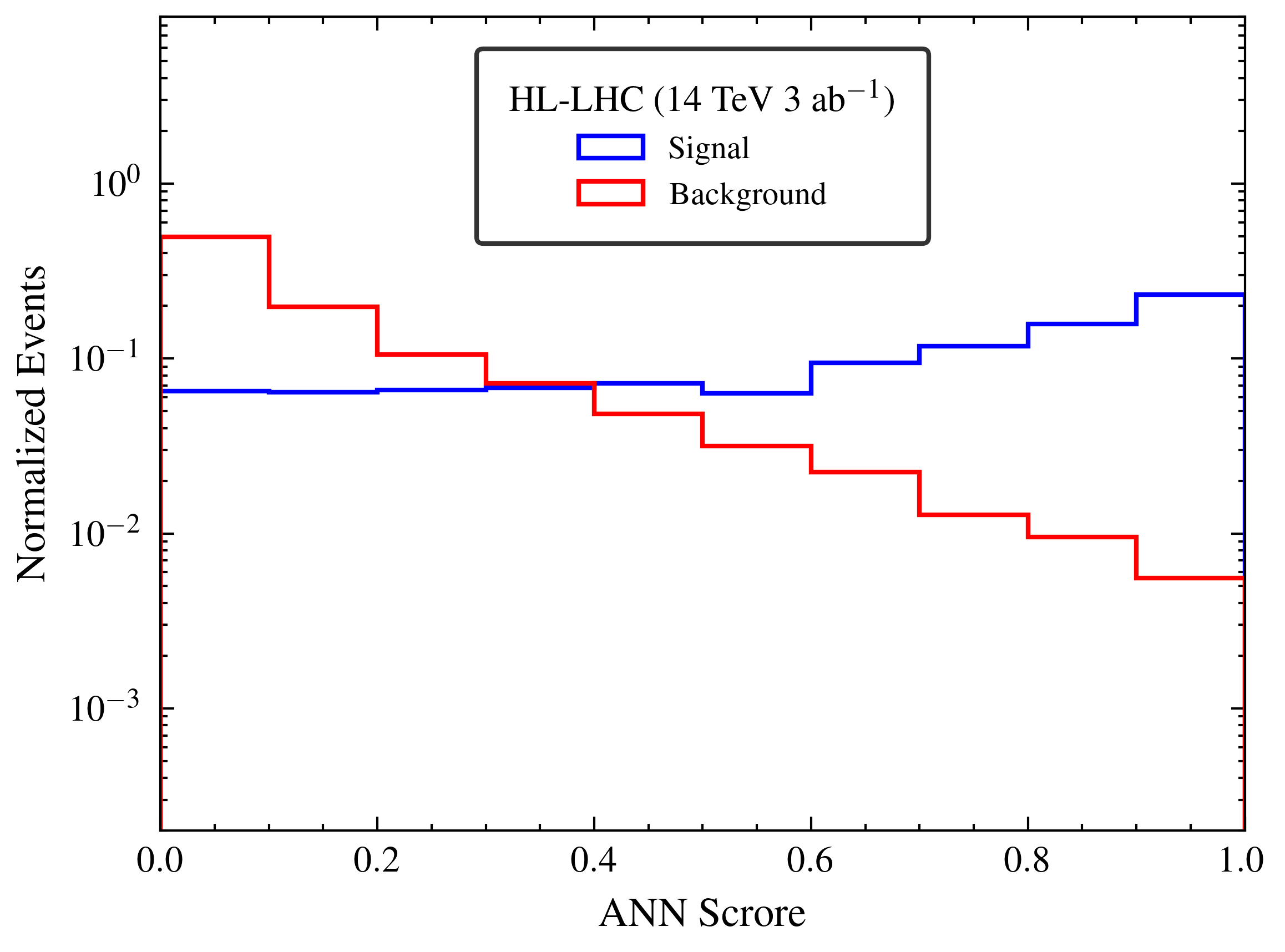} \qquad 
    \includegraphics[width = 0.4\textwidth]{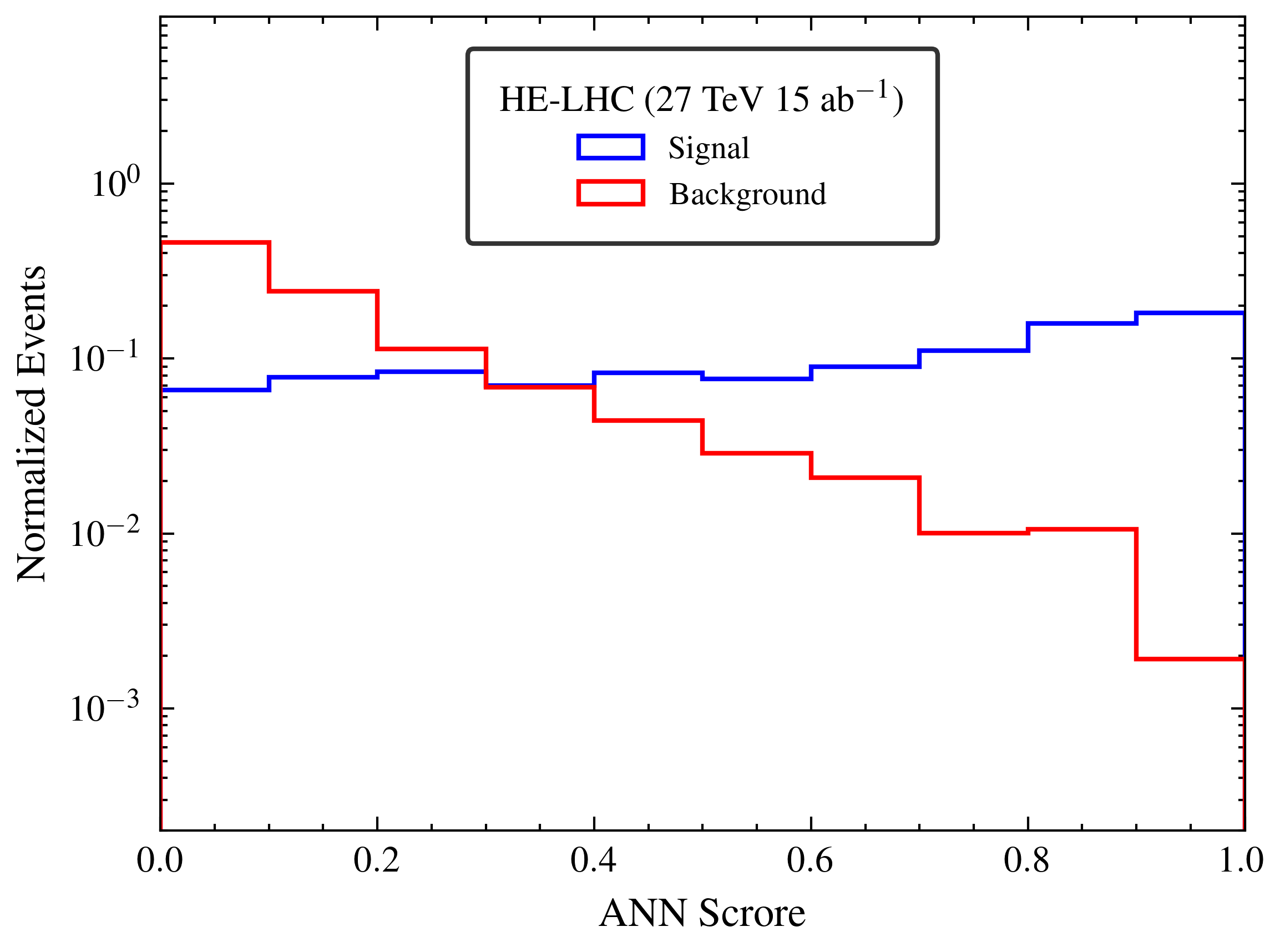} \\
    \caption{ROC curve (\textit{upper row}), loss evolution (\textit{center row}) and ANN score (\textit{lower row}) for the ANN models for HL-LHC (14 TeV 3 ab$^{-1}$, \textit{left} panel) and HE-LHC (27 TeV 15 ab$^{-1}$, \textit{right} panel).}
    \label{fig:loss}
\end{figure}
The definitions of all kinematic variables follow the standard conventions used in collider analyses. The normalized signal-background event distributions are shown in figure~\ref{fig:dist} of Appendix~\ref{app:1}. The correlations between different variables for the signal and backgrounds are shown in figure~\ref{fig:corr}. Each event, for both signal and background samples, is assigned a weight proportional to its corresponding production cross section. The full dataset is partitioned into three mutually exclusive subsets for machine learning: training, validation, and test samples, in the ratio $5\!:\!1\!:\!4$. To avoid class imbalance and ensure unbiased learning, equal class weights are applied for signal and background events. The data preprocessing is performed using the \texttt{StandardScaler} function from the \texttt{scikit-learn} module to normalize all input features around their mean values and scale them to unit variance. A sequential neural network model is implemented using the \texttt{Keras} API within \texttt{TensorFlow}. During training, equal class weights are assigned to the signal and background datasets to prevent bias, while for the validation and test samples, the true event weights corresponding to their cross sections are applied. The sequential model comprises three fully connected hidden layers with depths of 32, 16, and 8 neurons, respectively, each activated using the \texttt{ReLU} function. An \texttt{L2} regularization with a coefficient of 0.005 is applied at every layer to prevent overfitting. The output layer employs a \texttt{Sigmoid} activation function, suitable for binary classification. The model is trained using the \texttt{binary crossentropy} loss function and optimized with the \texttt{Adam} optimizer, using a learning rate of 0.001. The performance metric used during training is the area under the Receiver Operating Characteristic (ROC) curve (\texttt{AUC}). The model is trained for 50 epochs with an early stopping criterion applied to prevent overfitting, using a patience parameter of 5 epochs based on the validation loss. The ROC curves and the loss evolution for the ANN models corresponding to the HL-LHC (14 TeV, 3 ab$^{-1}$) and HE-LHC (27 TeV, 15 ab$^{-1}$) scenarios are shown in figure~\ref{fig:loss}. For the HL-LHC model, we obtain a test sample accuracy of 0.918 and an AUC score of 0.883, while for the HE-LHC model, the corresponding values are 0.928 and 0.872, respectively. The close agreement between the training and validation loss curves indicates stable training behavior and the absence of overfitting throughout the training process.
\begin{table}[htb!]
    \centering
    \begin{tabular}{cccc}
    \hline \hline
        ANN Model & Class & Events (before cut) & Events (after cut) \\ \hline
        \multirow{3}*{HL-LHC (14 TeV, 3 ab$^{-1}$)} 
        & $S$ & $340$  & $38$ \\
        & $B$ & $1047095$ & $1562$ \\ \cline{2-4}
        & $S:B$ & $3.247\times 10^{-4}$ & $2.433 \times 10^{-2}$ \\ \hline
        \multirow{3}*{HE-LHC (27 TeV, 15 ab$^{-1}$)} 
        & $S$ & $6941$  & $396$ \\
        & $B$ & $30909360$ & $19390$ \\ \cline{2-4}
        & $S:B$ & $2.246 \times 10^{-4}$ & $2.042 \times 10^{-2}$ \\ \hline \hline
    \end{tabular}
    \caption{Signal ($S$) and background ($B$) event counts before and after the ANN threshold cut for HL-LHC (14 TeV, 3 ab$^{-1}$) and HE-LHC (27 TeV, 15 ab$^{-1}$) scenarios.}
    \label{tab:sb}
\end{table}
The distributions of the ANN scores for the two models are shown in figure~\ref{fig:loss}. A threshold score value of 0.95 is chosen to retain signal-like events, effectively suppressing background contamination. The corresponding signal and background event counts before and after applying the threshold cut are summarized in Table~\ref{tab:sb}. It is evident that the application of the ANN threshold improves the signal-to-background ratio ($S:B$) by approximately one order of magnitude for both collider scenarios after employing the kinematical cuts.

\subsection{Signal significance}
\begin{figure}[htb!]
    \centering
    \includegraphics[width = 0.4\textwidth]{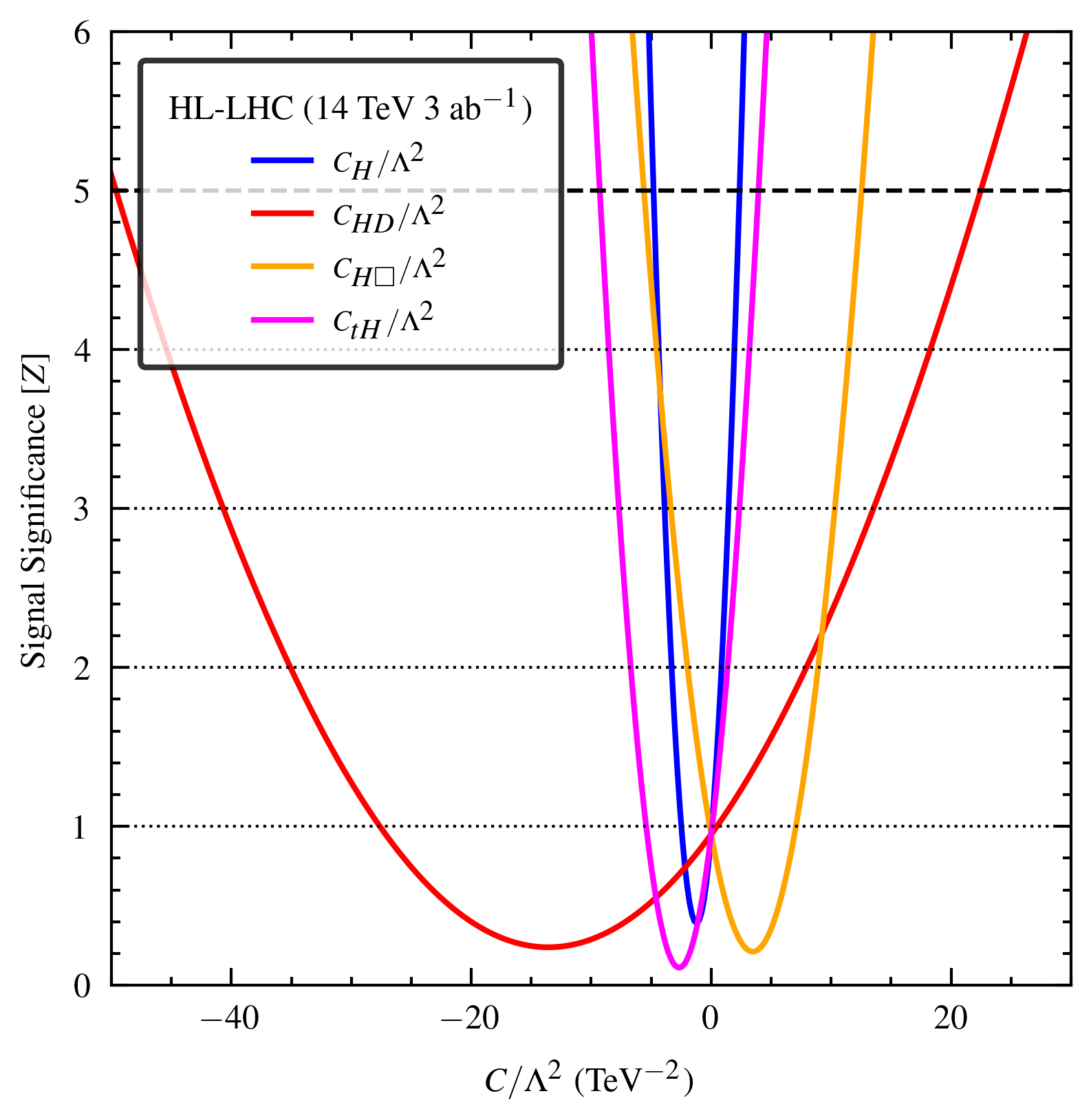} \qquad
    \includegraphics[width = 0.4\textwidth]{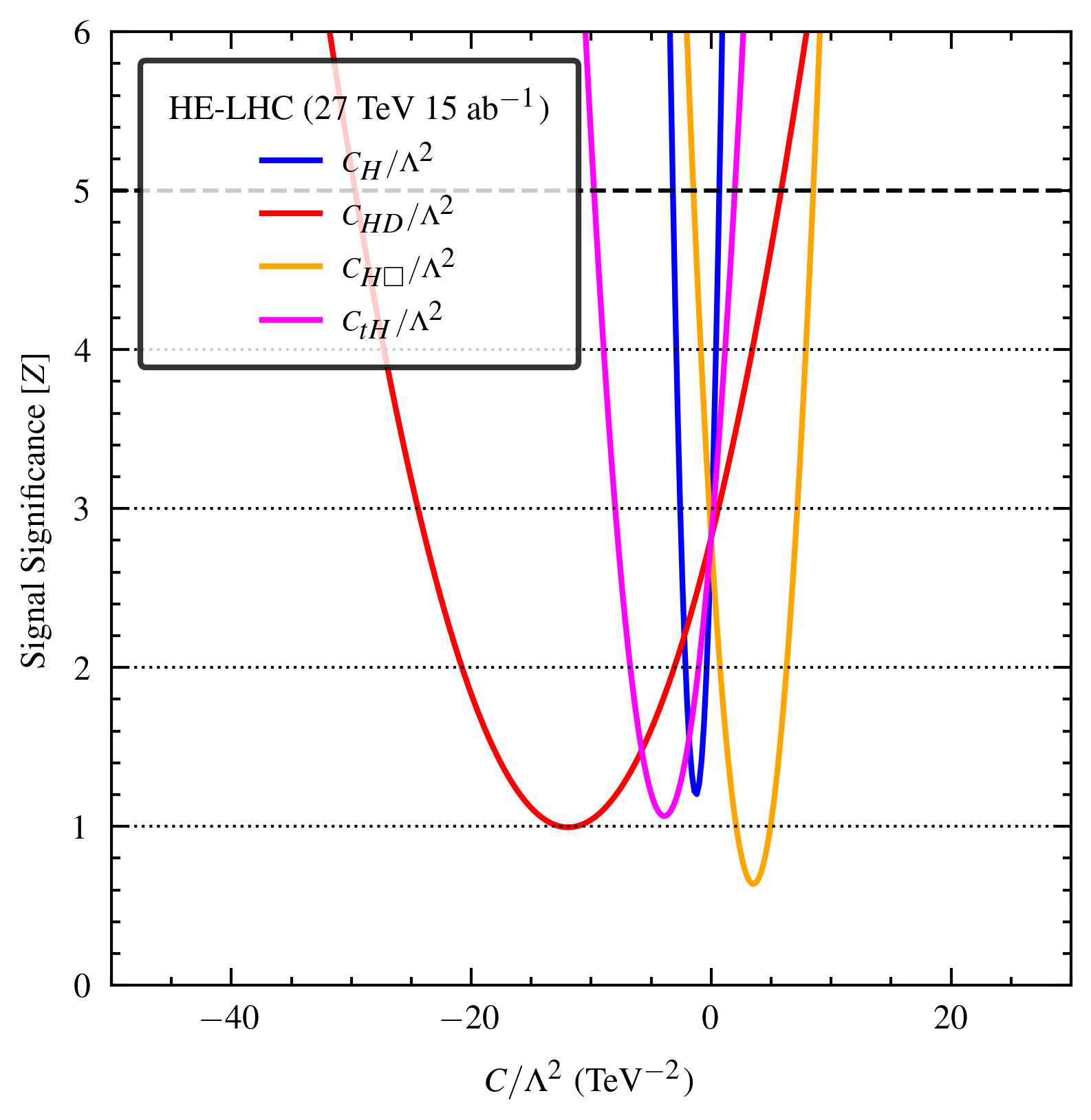} \\
    \caption{Signal Significance, $Z$ as a function of the effective couplings. Left panel: HL-LHC (14 TeV 3 ab$^{-1}$), right panel: HE-LHC (27 TeV 15 ab$^{-1}$). The \textit{black dotted} and \textit{black dashed} horizontal lines represent the $1\sigma-4\sigma$ and $5\sigma$ limits, respectively.}
    \label{fig:1d.sens}
\end{figure}
\noindent
To quantify the impact of the SMEFT operators, we evaluate the signal significance based on the event yields obtained after applying the ANN classifiers. For both the HL-LHC and HE-LHC setups, the signal event counts corresponding to different values of the effective couplings are processed through the trained models. The signal significance ($Z$) is defined as
\begin{equation}
    Z\!\left({C_{i}}/{\Lambda^{2}}\right)
    = \frac{S\!\left({C_{i}}/{\Lambda^{2}}\right)}
    {\sqrt{S\!\left({C_{i}}/{\Lambda^{2}}\right) + B}},
\end{equation}
where $S\left(C_{i}/\Lambda^{2}\right)$ is the number of signal events for a given EFT benchmark after classification, $S(0)$ corresponds to the SM prediction, and $B$ denotes the non-interfering SM background. 

\begin{figure}[t]
    \centering
    \includegraphics[width = 0.32\textwidth]{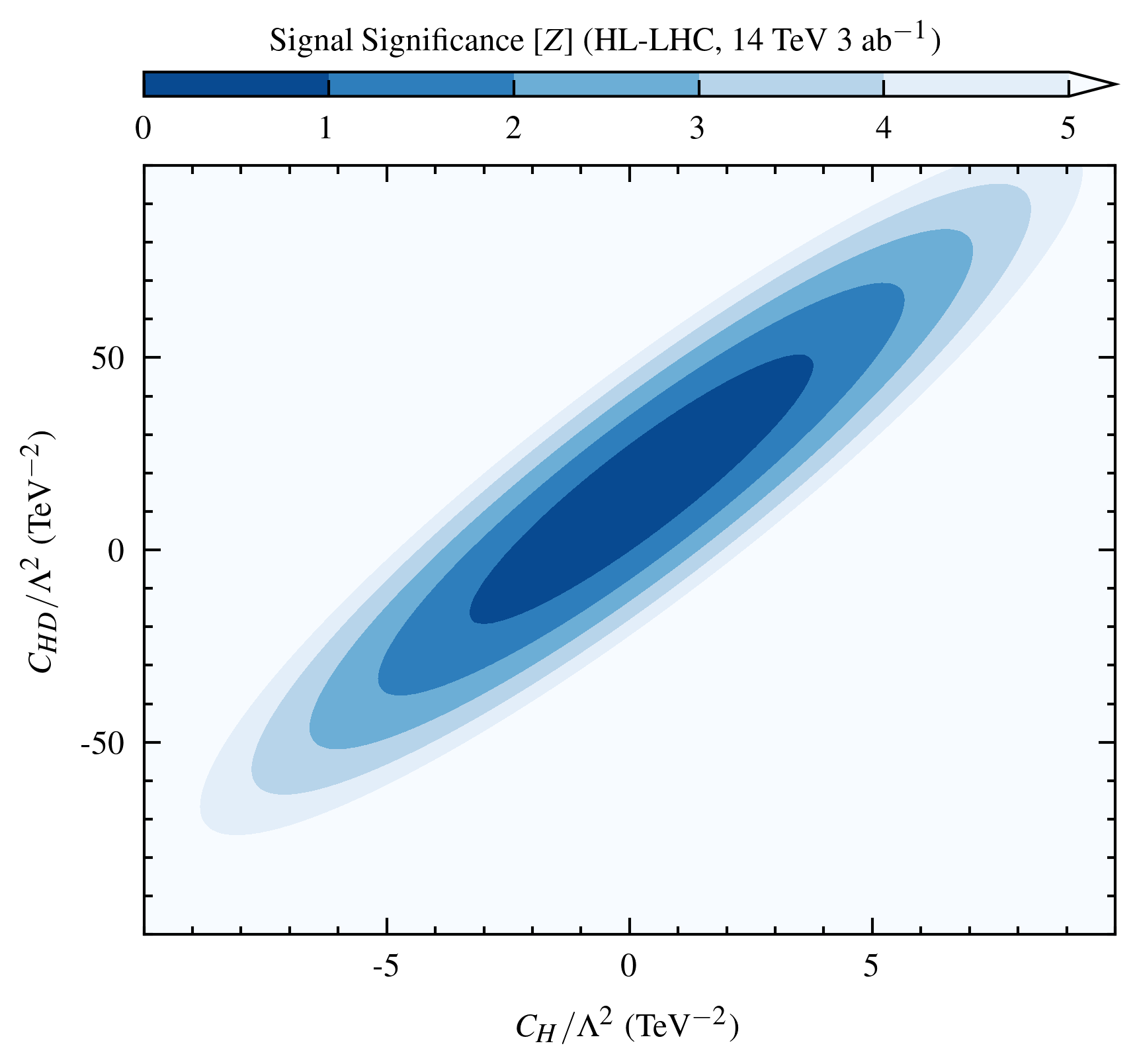}
    \includegraphics[width = 0.32\textwidth]{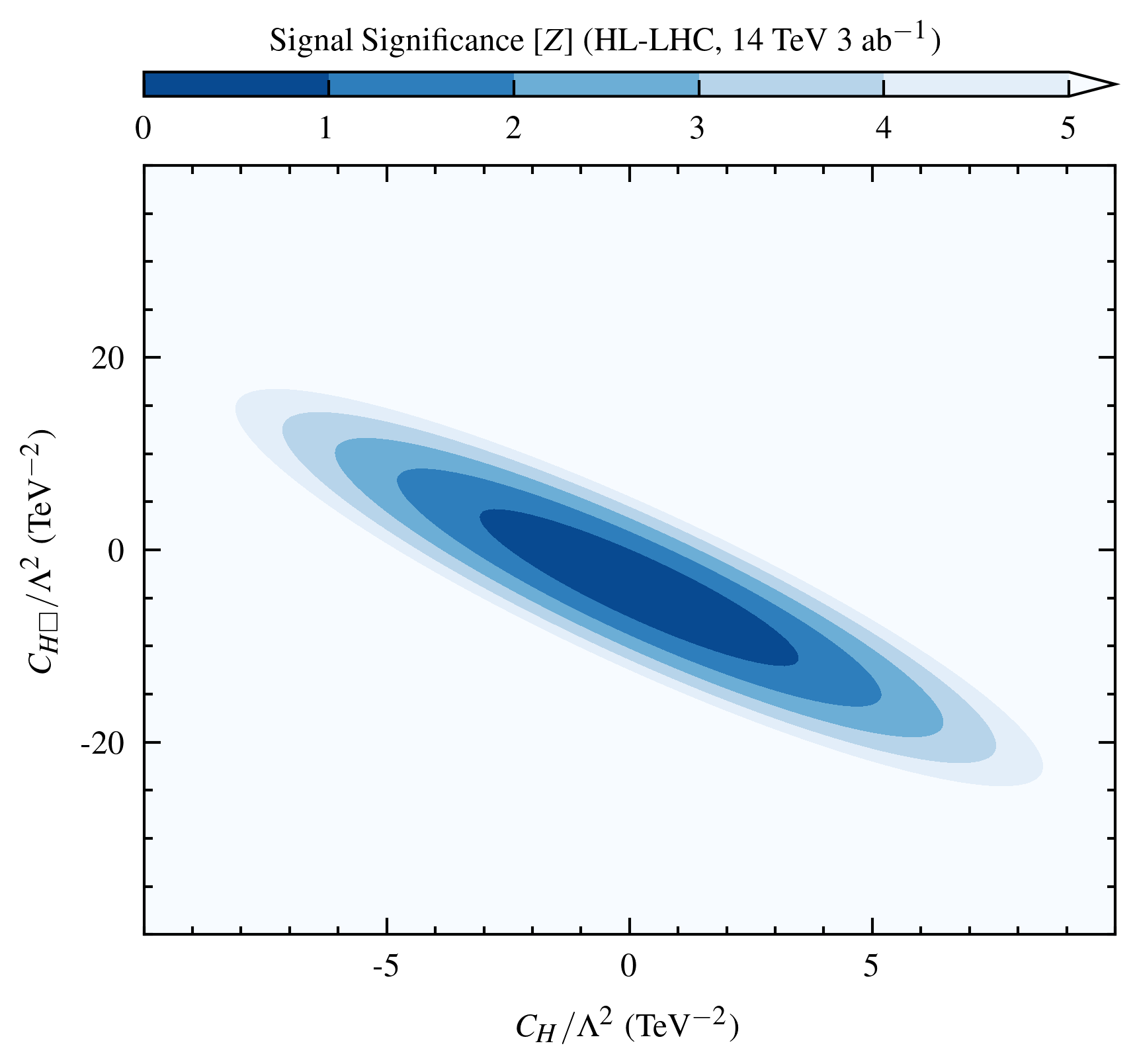}
    \includegraphics[width = 0.32\textwidth]{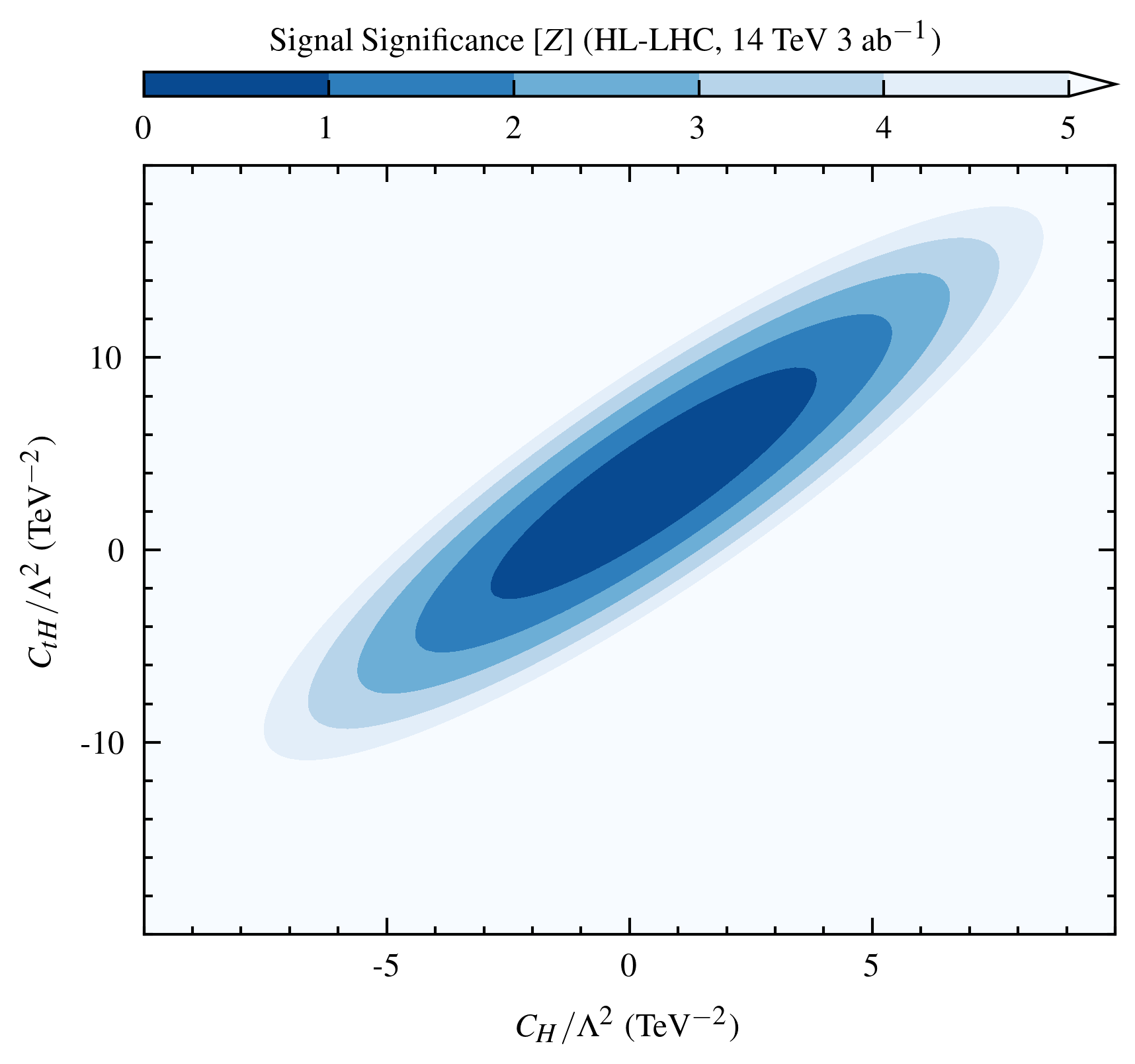} \\
    \includegraphics[width = 0.32\textwidth]{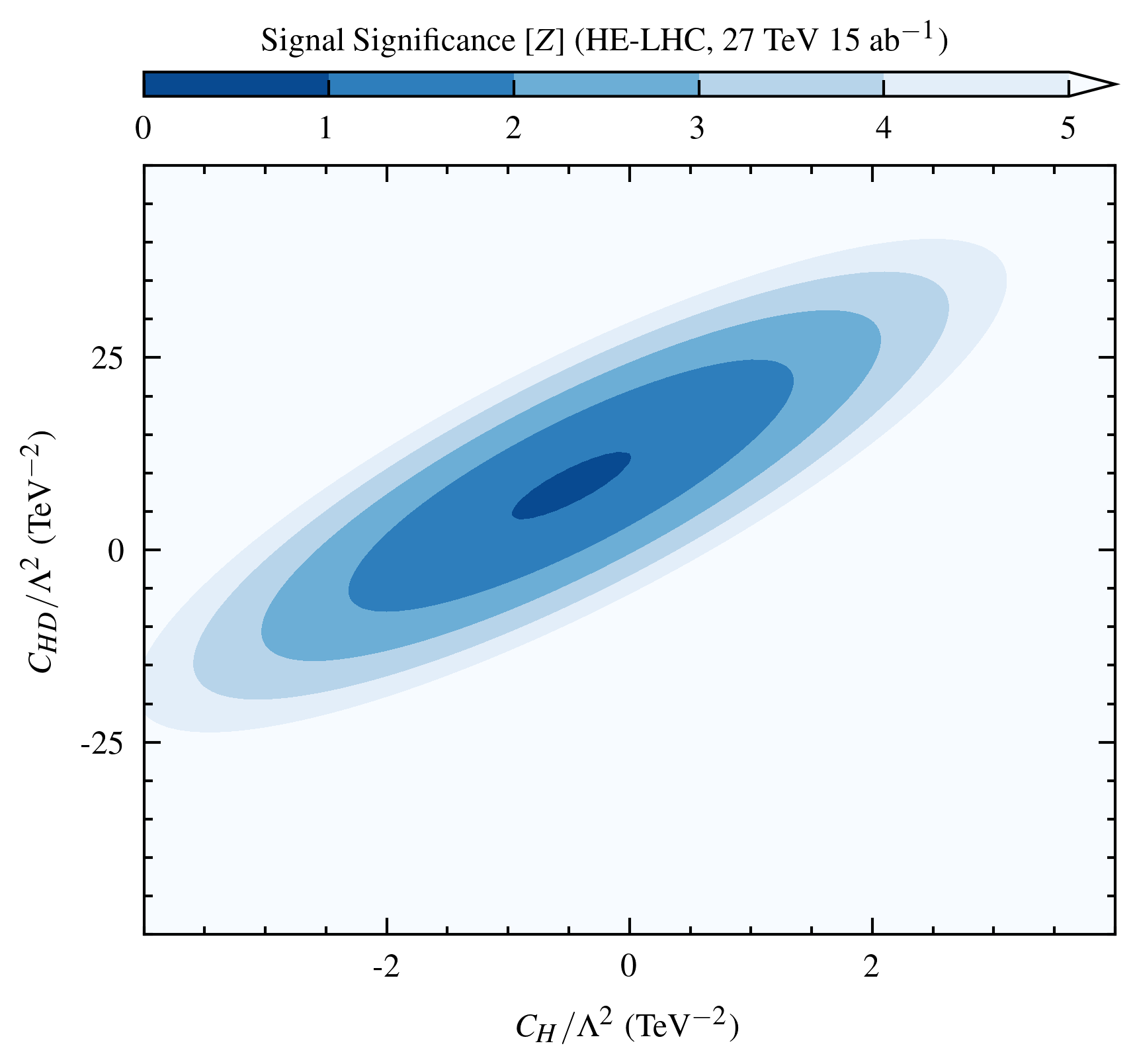}
    \includegraphics[width = 0.32\textwidth]{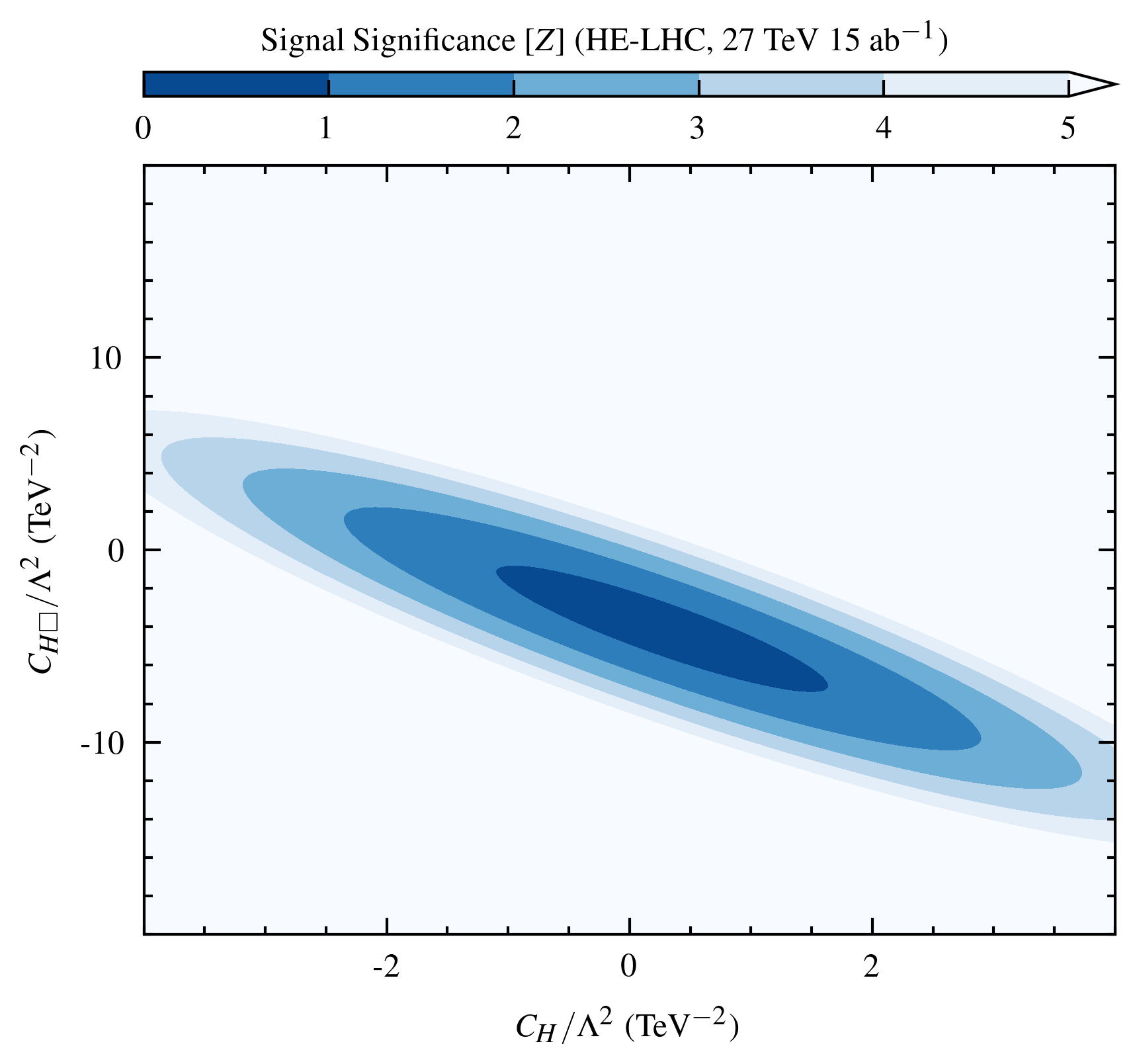}
    \includegraphics[width = 0.32\textwidth]{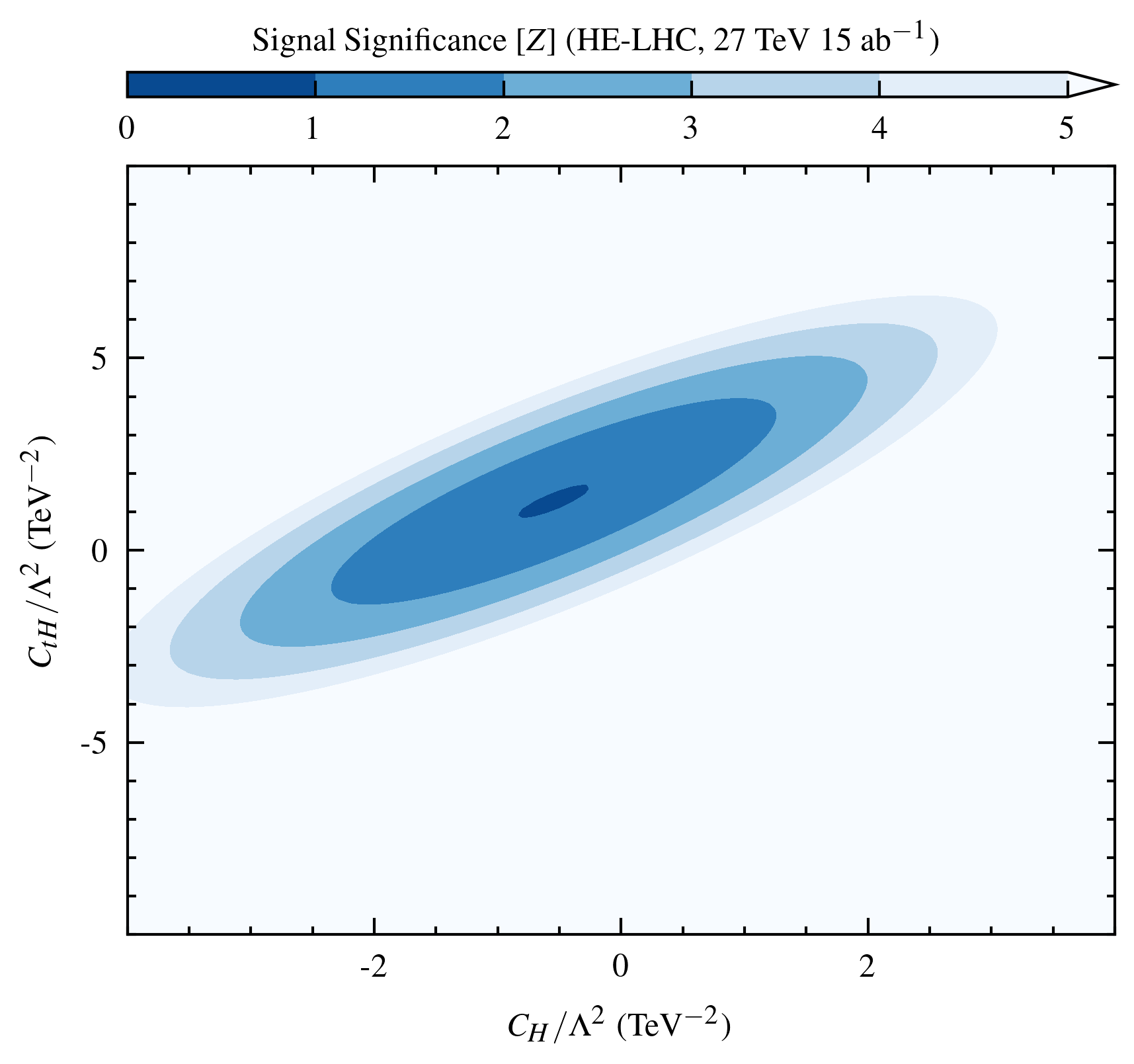}
    \caption{Two parameter signal significance $(Z)$ plots. Top panel: HL-LHC (14 TeV 3 ab$^{-1}$), bottom panel: HE-LHC (27 TeV 15 ab$^{-1}$).}
    \label{fig:2d.sens}
\end{figure}

The dependence of the signal significance on the three WCs is displayed in figure~\ref{fig:1d.sens} for both the HL-LHC and HE-LHC. Amongst them, $C_{H}/\Lambda^{2}$ shows the largest impact on the signal significance, followed by $C_{tH}/\Lambda^{2}$, $C_{H\Box}/\Lambda^{2}$, and $C_{HD}/\Lambda^{2}$. The corresponding two-dimensional signal significance contours are shown in figure~\ref{fig:2d.sens}. Motivated by FO-EWPT considerations, we fix $C_{H}/\Lambda^{2}$ along one axis and examine the correlated reach in $C_{H\Box}/\Lambda^{2}$, $C_{HD}/\Lambda^{2}$, and $C_{tH}/\Lambda^{2}$ along the other. The resulting contours show that the couplings $C_{HD}/\Lambda^{2}$ and $C_{tH}/\Lambda^{2}$ exhibit similar correlations with $C_{H}/\Lambda^{2}$, whereas $C_{H\Box}/\Lambda^{2}$ displays an opposite correlation. Moreover, the enhancement in signal significance induced by $C_{H}/\Lambda^{2}$ is considerably stronger in the presence of $C_{tH}/\Lambda^{2}$ than in the presence of $C_{H\Box}/\Lambda^{2}$ and $C_{HD}/\Lambda^{2}$.

\begin{figure}[t]
    \centering
    \includegraphics[width = 0.32\textwidth]{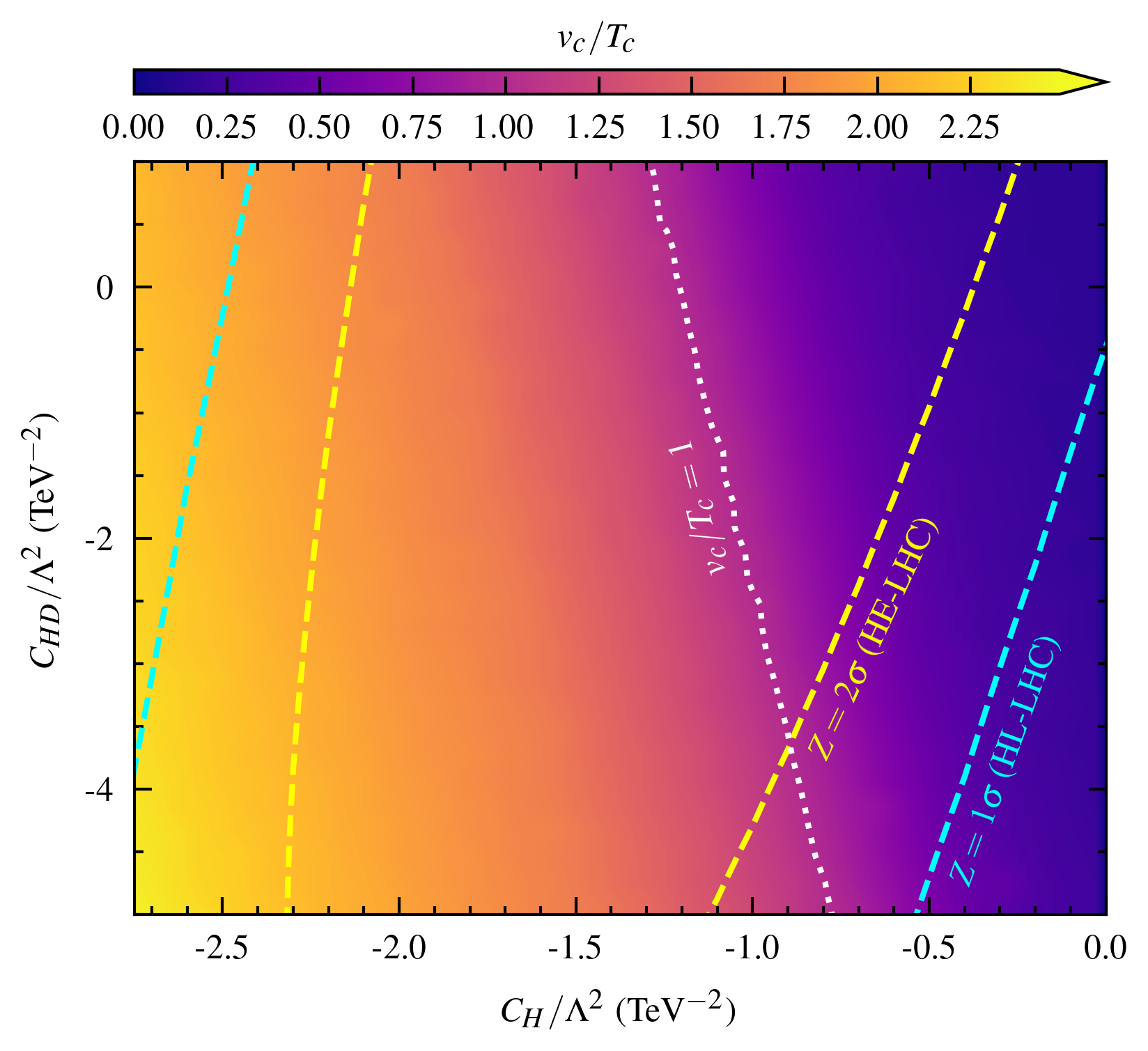}
    \includegraphics[width = 0.32\textwidth]{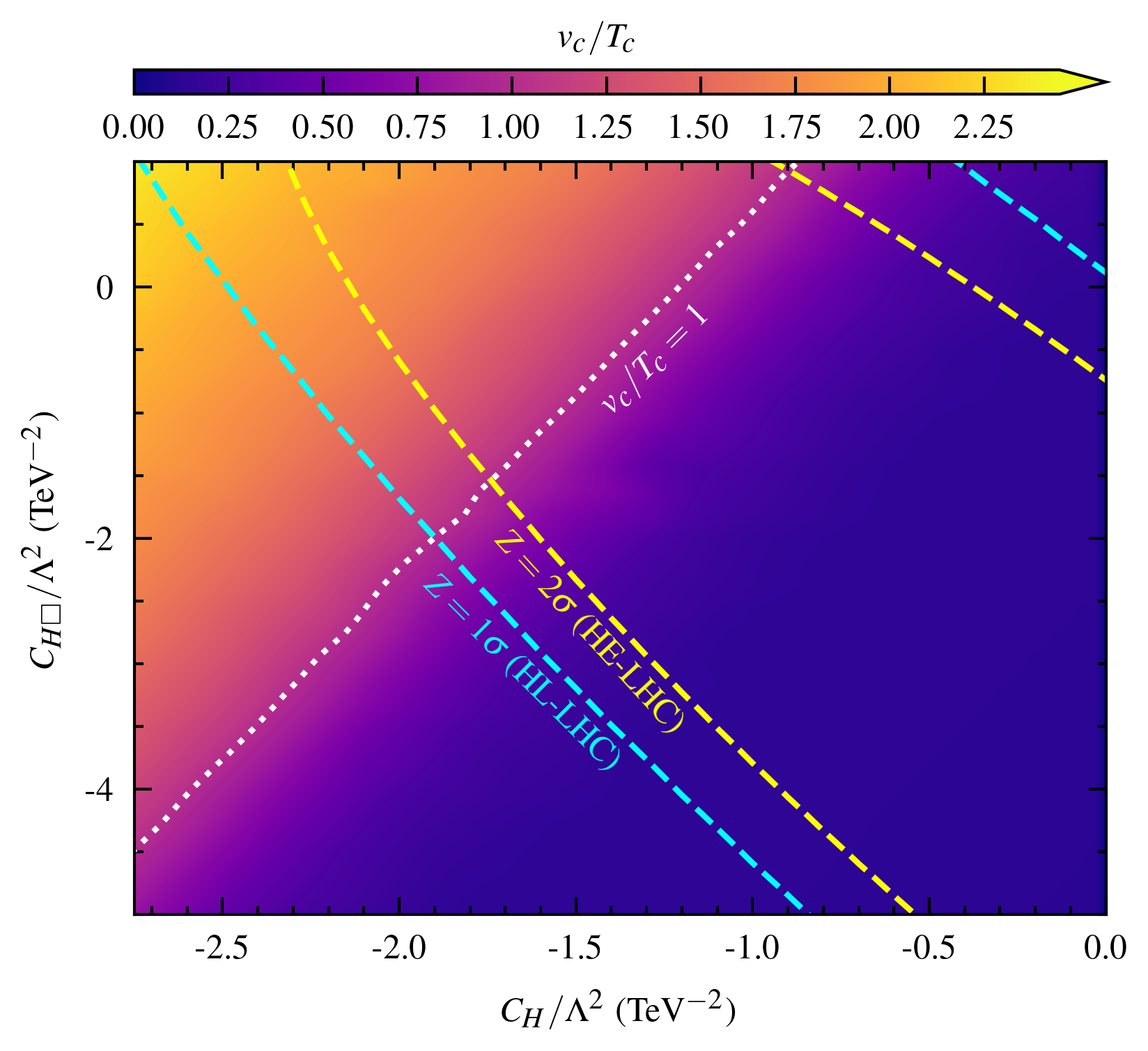}
    \includegraphics[width = 0.32\textwidth]{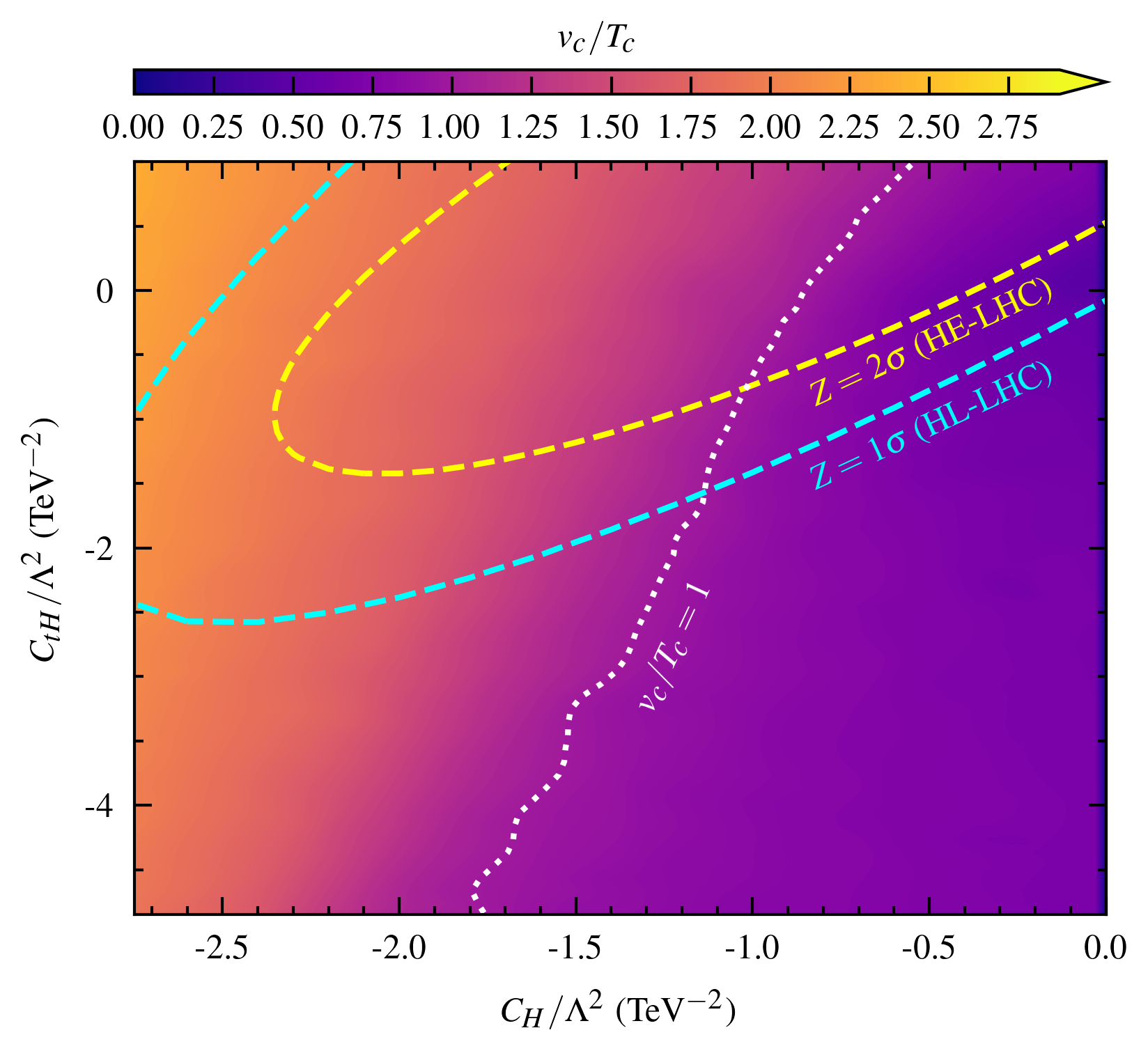}
    \caption{Signal significance levels: $Z=1\sigma$ for HL-LHC (14 TeV 3 ab$^{-1}$) and $Z=2\sigma$ for HE-LHC (27 TeV 15 ab$^{-1}$), overlaid with the FO-EWPT allowed parameter space. The color hue represents the $v_{c}/T_{c}$ values across the parameter space.}
    \label{fig:vtvalue}
\end{figure}

Figure~\ref{fig:vtvalue} shows the constant-significance contours corresponding to $Z = 1\sigma$ for the HL-LHC (14~TeV, 3~ab$^{-1}$) and $Z = 2\sigma$ for the HE-LHC (27~TeV, 15~ab$^{-1}$), overlaid with regions consistent with the FO-EWPT. The colour gradient represents the value of $v_{c}/T_{c}$ across the parameter space, where $v_{c}/T_{c} > 1$ denotes a strong FO-EWPT. Only negative values of $C_{H}/\Lambda^{2}$ are shown, since a strong FO-EWPT is realised exclusively in that region. From the summary plots, it is evident that the parameter space capable of generating a strong FO-EWPT lies well within the significance reach of both the HL-LHC and HE-LHC. This demonstrates that the di-Higgs production channel at future hadron colliders can serve as a powerful indirect probe of the FO-EWPT. The higher CM energy and luminosity of the HE-LHC further extend the reach in $C_{tH}/\Lambda^{2}$, $C_{H\Box}/\Lambda^{2}$ and $C_{HD}/\Lambda^{2}$, where the momentum-dependent operators induce increasingly pronounced effects. Overall, these results indicate that precision measurements of the Higgs self-interactions and kinematic observables at future colliders provide a crucial pathway to explore the dynamics of the FO-EWPT and test NP within the SMEFT framework.

\section{Conclusion}
\label{sec:con}
In this work, we have explored the phenomenological implications of the Standard Model Effective Field Theory (SMEFT) framework to investigate first-order electroweak phase transitions (FO-EWPTs), complemented by gravitational wave (GW) and collider probes. We have analyzed three tree level and one 1-loop level SMEFT operators that induce modifications to the Higgs potential and examined their impact on the thermal evolution of the effective potential at finite temperature. Subsequently, we have outlined the fitting formulas relevant to the GW production and discussed the individual contributions from the different sources. Considering an observation period of four year, we have explored the two-dimensional parameter space of the effective couplings to identify regions yielding a sufficient signal-to-noise ratio for GW detection at future GW observatories.

In the next part of our analysis, we have examined the capability of the HL-LHC and HE-LHC to probe the same four dimension-6 operators via \textit{di}-Higgs production. The \textit{di}-Higgs channel at hadron colliders offers a direct handle on the trilinear Higgs self-coupling and consequently provides a unique window into the structure of the Higgs potential and possible manifestations of new physics. In particular, modifications induced by $C_{H}/\Lambda^{2}$, $C_{H\Box}/\Lambda^{2}$, $C_{HD}/\Lambda^{2}$ and $C_{tH}/\Lambda^{2}$, alter both the total production rate and the kinematic distributions of \textit{di}-Higgs events, motivating a detailed signal–background analysis. To quantify the discrimination power, we focus on the $2b2\tau$ final state, which balances a large signal yield with manageable SM backgrounds. We employ state-of-the-art machine learning techniques based on an Artificial Neural Network (ANN) trained on a set of optimised kinematic observables. After applying an optimal threshold selection on the ANN output, the signal–background ratio improves approximately by one order for both collider setups, demonstrating a substantial enhancement in sensitivity compared to traditional cut-based approaches. The increased centre-of-mass energy and integrated luminosity of the HE-LHC significantly boost the statistical power of the measurement. In particular, the resulting signal significance improves by approximately a factor of three relative to the HL-LHC. This clearly highlights the ability of the HE-LHC to probe smaller deviations in the Higgs sector and to explore regions of parameter space that remain inaccessible at the HL-LHC.

Although future GW observations, specially the DECIGO and BBO, allow a substantial region of parameter space consistent with FO-EWPT, the HL and HE upgrades of the LHC provide a far more sharply defined and experimentally accessible window for testing this scenario. In particular, \textit{di}-Higgs production offers a direct collider-based probe of the scalar potential, enabling us to identify the regions of the SMEFT parameter space that are not only capable of realising a strong FO-EWPT but are also within reach of precision measurements at hadron colliders. This complementarity between collider measurements and gravitational-wave observations to test EWPT highlights the importance of pursuing both approaches simultaneously: while GWs survey the cosmological imprint of the EWPT, collider data determine the microscopic interactions responsible for it. 

\acknowledgments
We sincerely thank the anonymous referee for constructive comments that significantly improved the quality and clarity of the manuscript. We acknowledge the use of {\tt TikZ-Feynman} \cite{Ellis:2016jkw} to draw Feynman diagrams in this work.  {We are deeply grateful to Professor Mikael Chala for valuable discussions.}

\appendix
\section{Comparison: Cuts vs ANN}
\label{app:1}
\begin{figure}[htb!]
    \centering
    \includegraphics[width=0.325\linewidth]{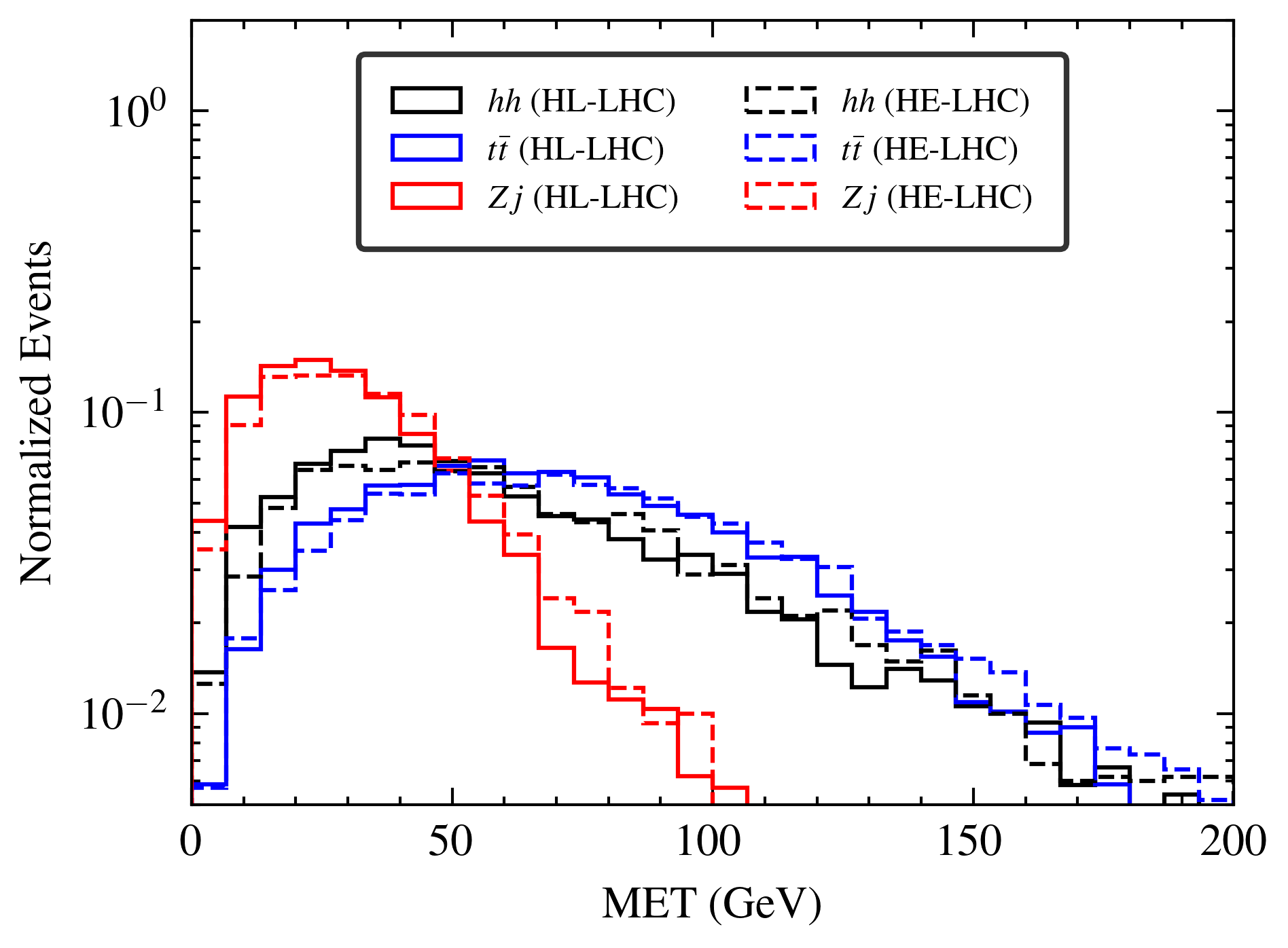}
    \includegraphics[width=0.325\linewidth]{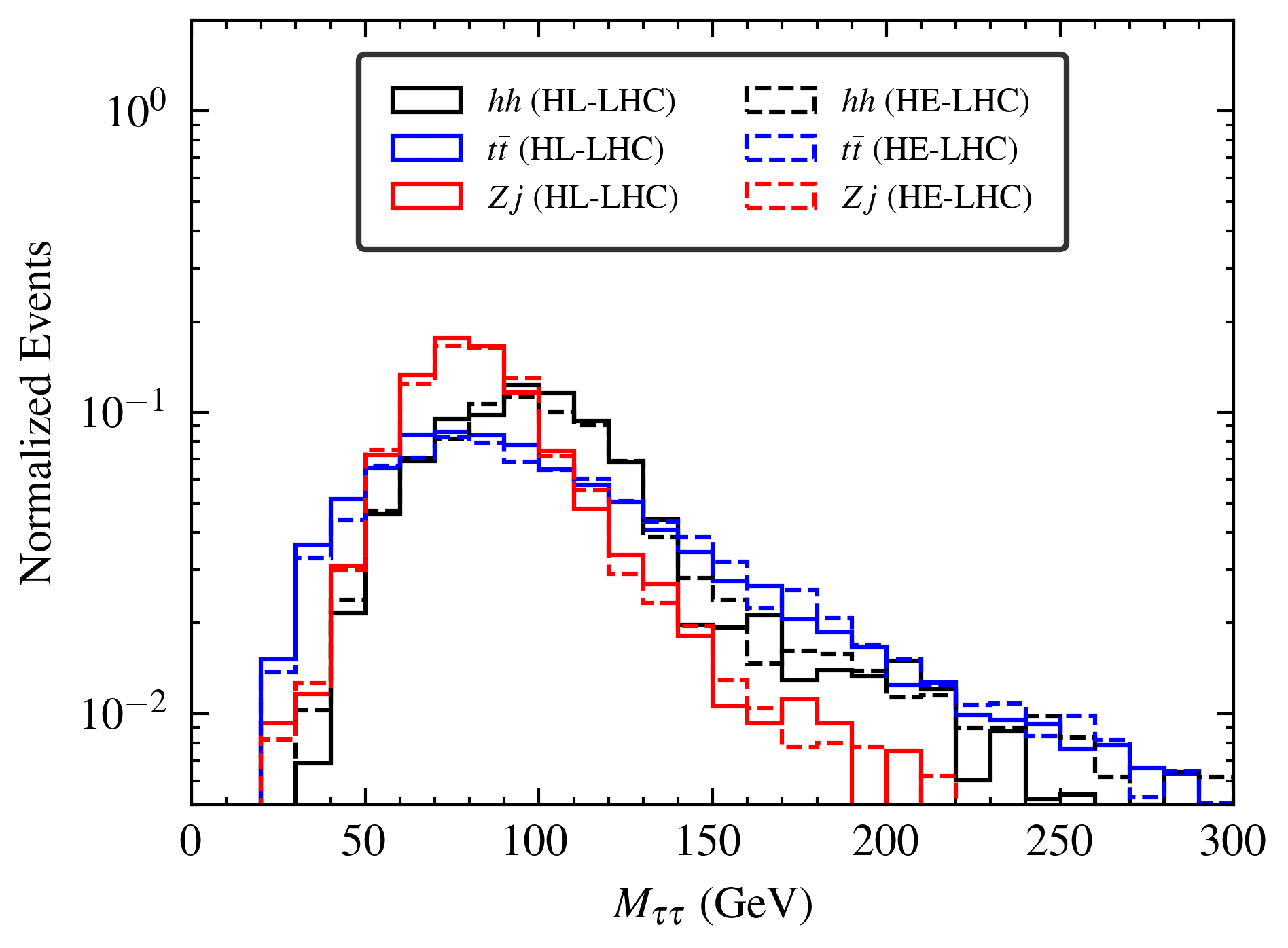}
    \includegraphics[width=0.325\linewidth]{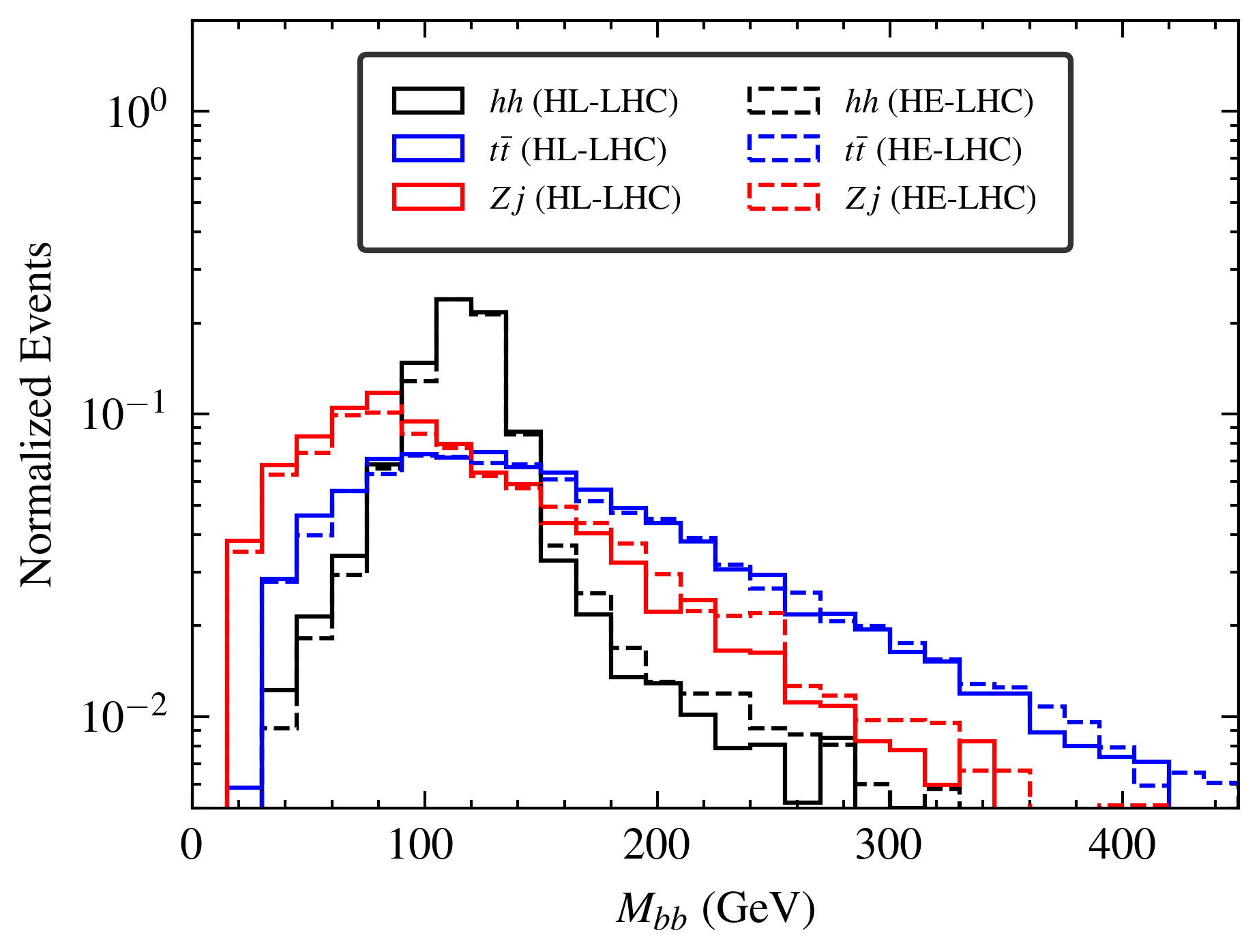} \\
    \includegraphics[width=0.325\linewidth]{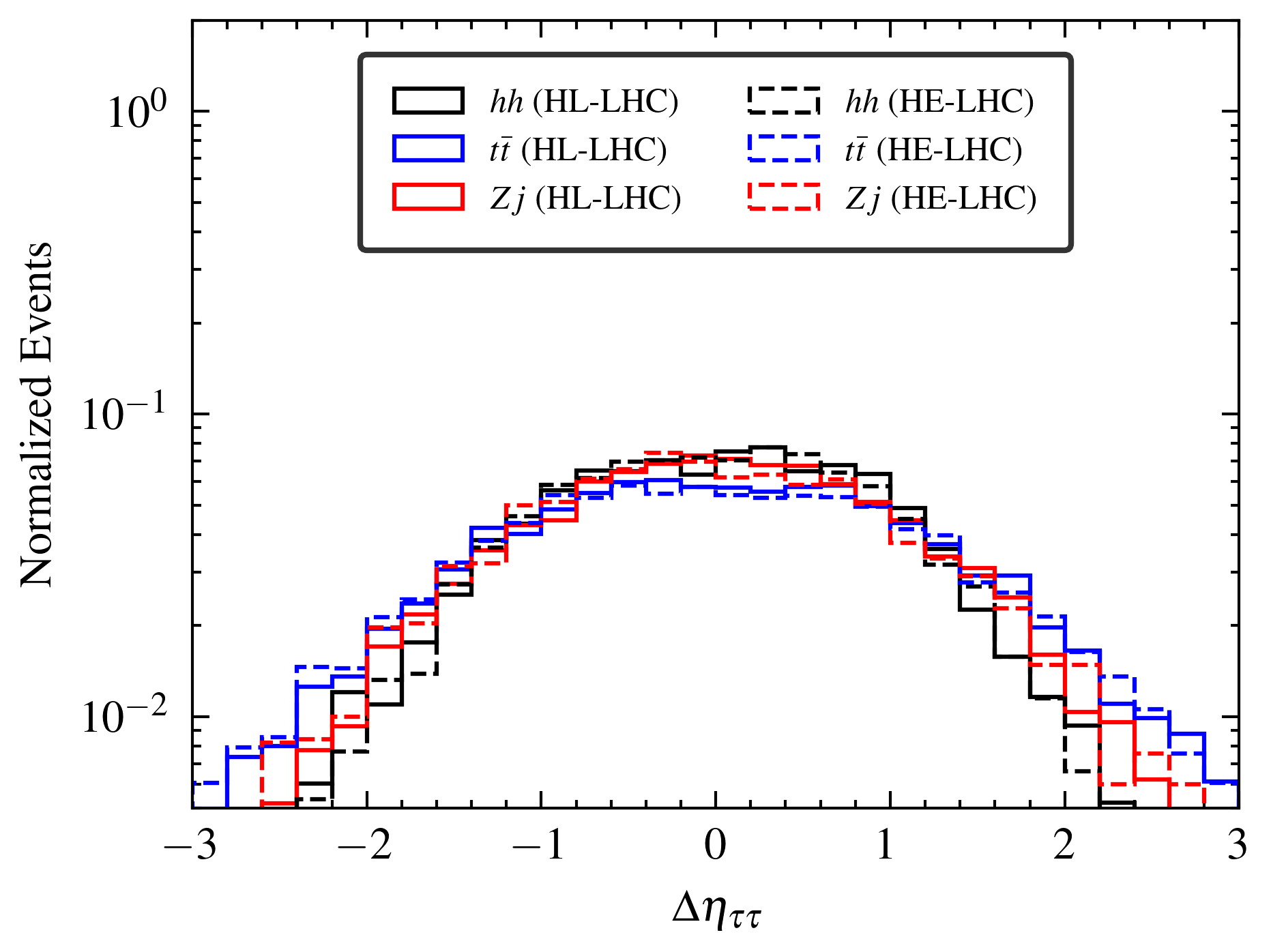}
    \includegraphics[width=0.325\linewidth]{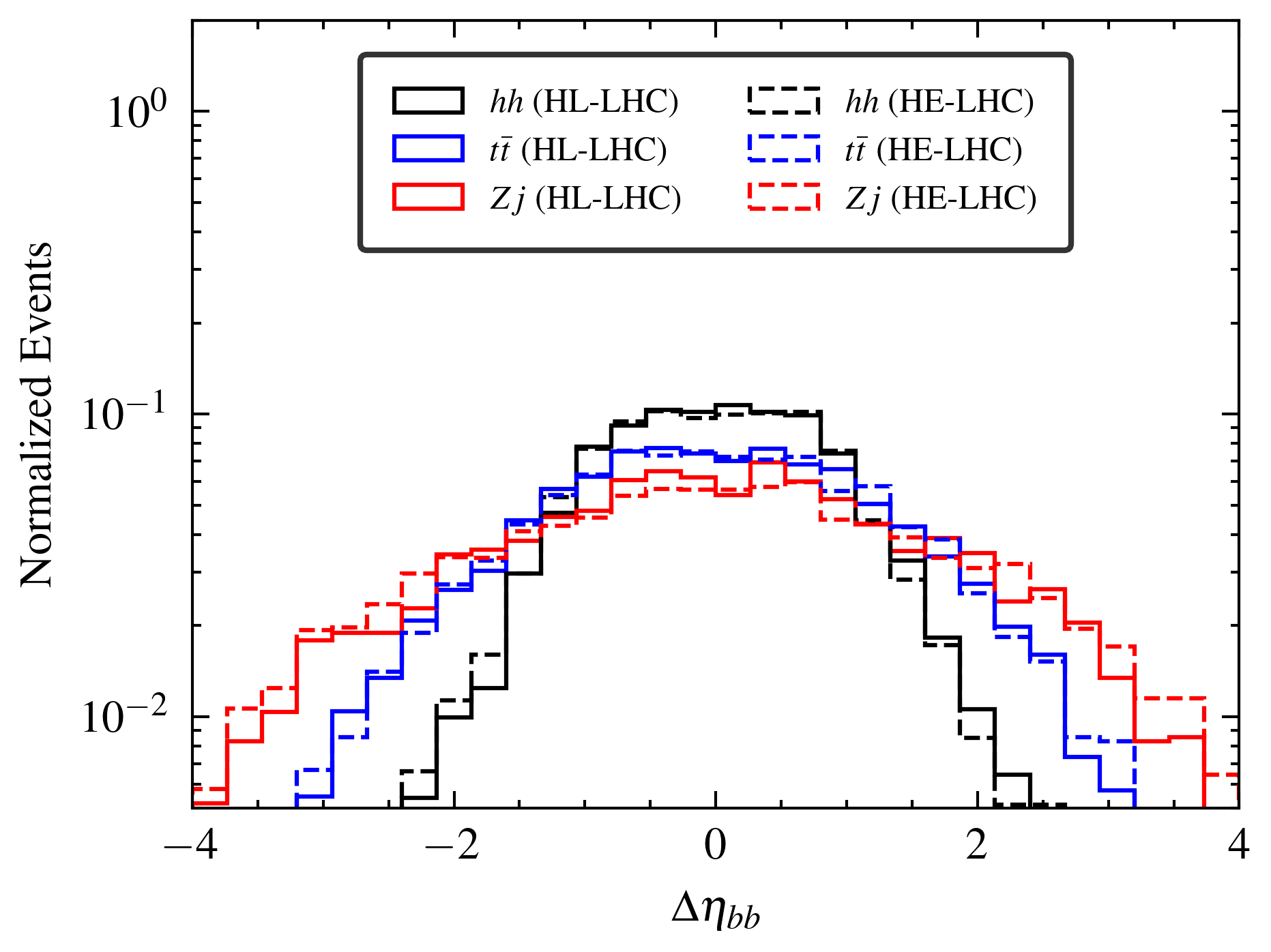}
    \includegraphics[width=0.325\linewidth]{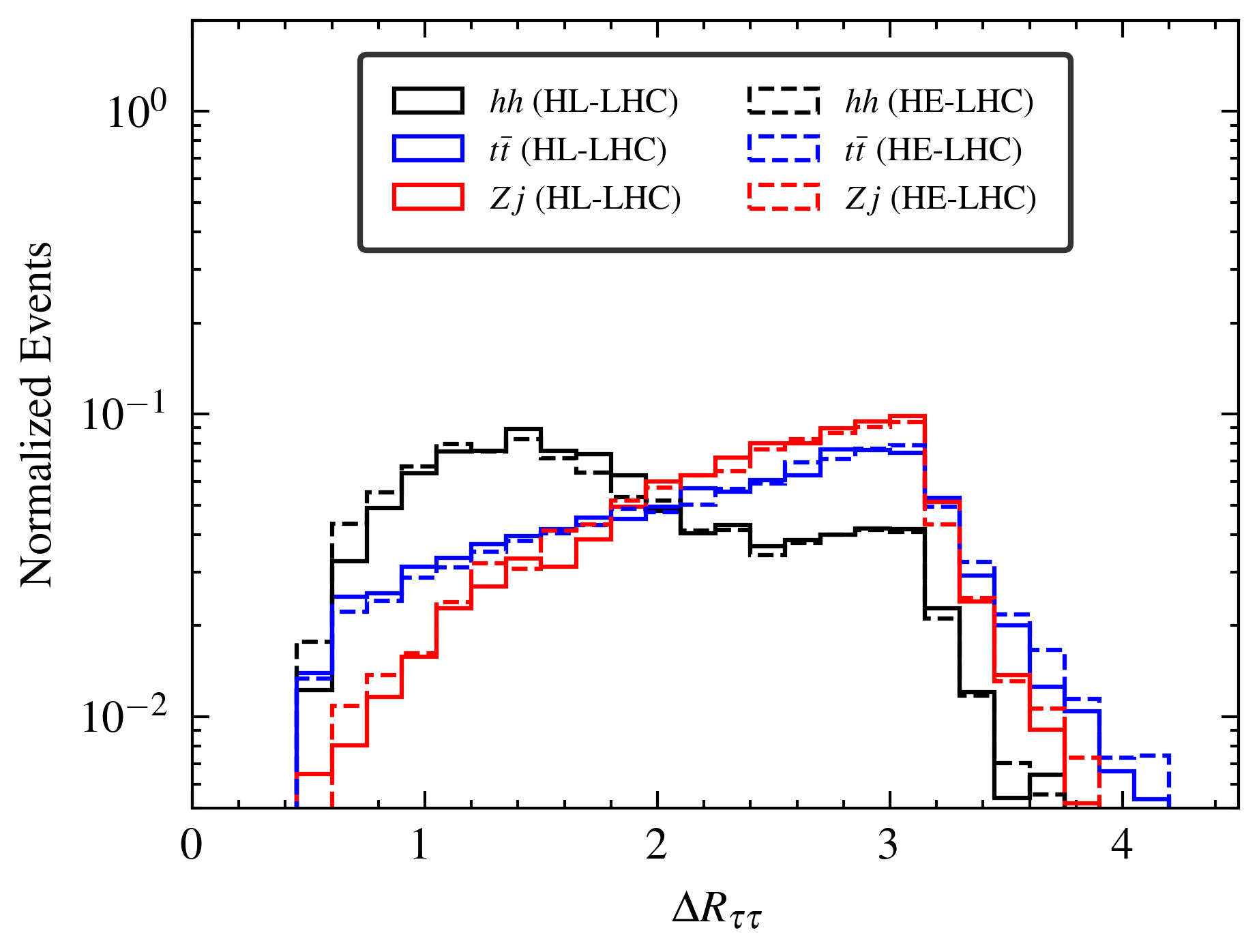} \\
    \includegraphics[width=0.325\linewidth]{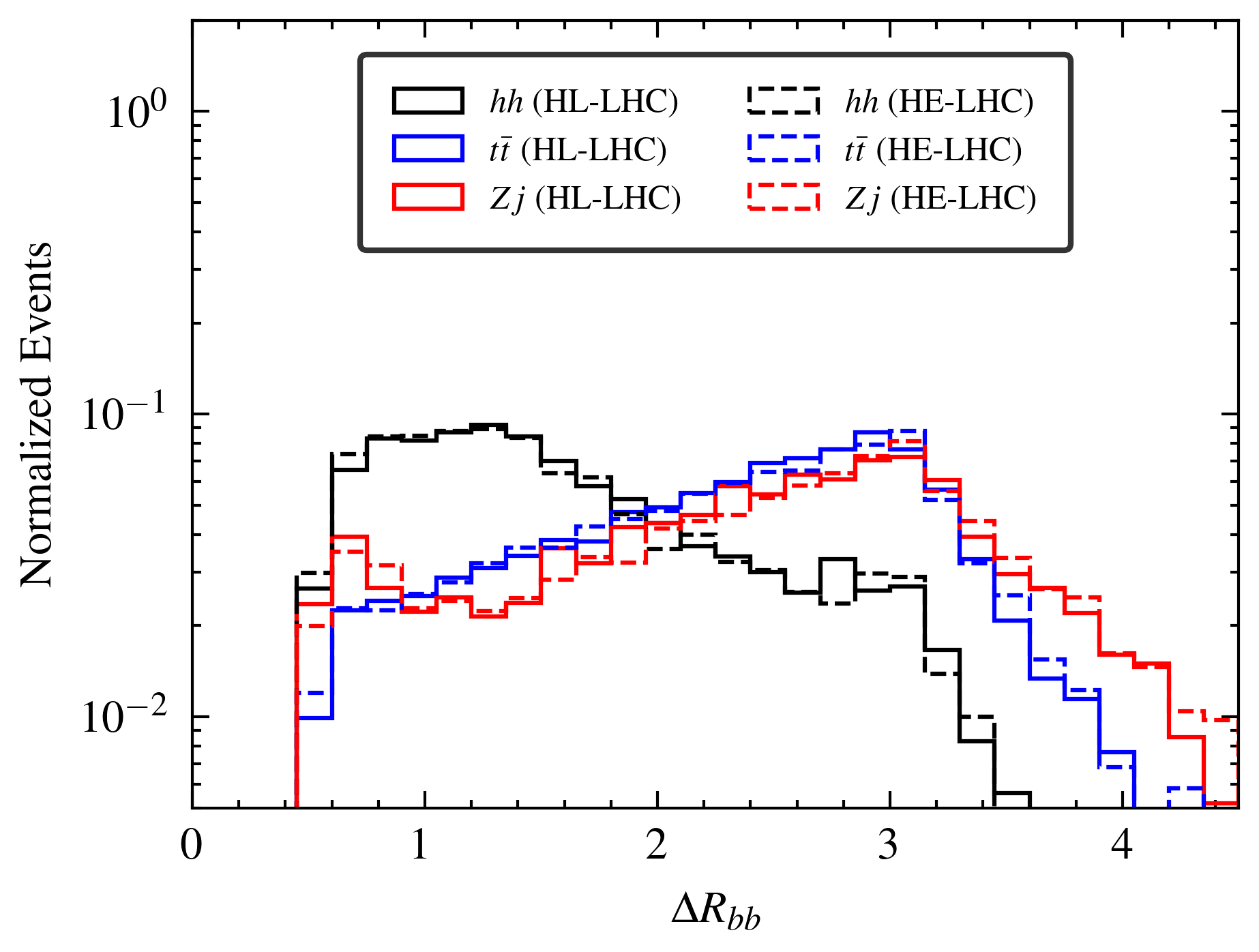}
    \includegraphics[width=0.325\linewidth]{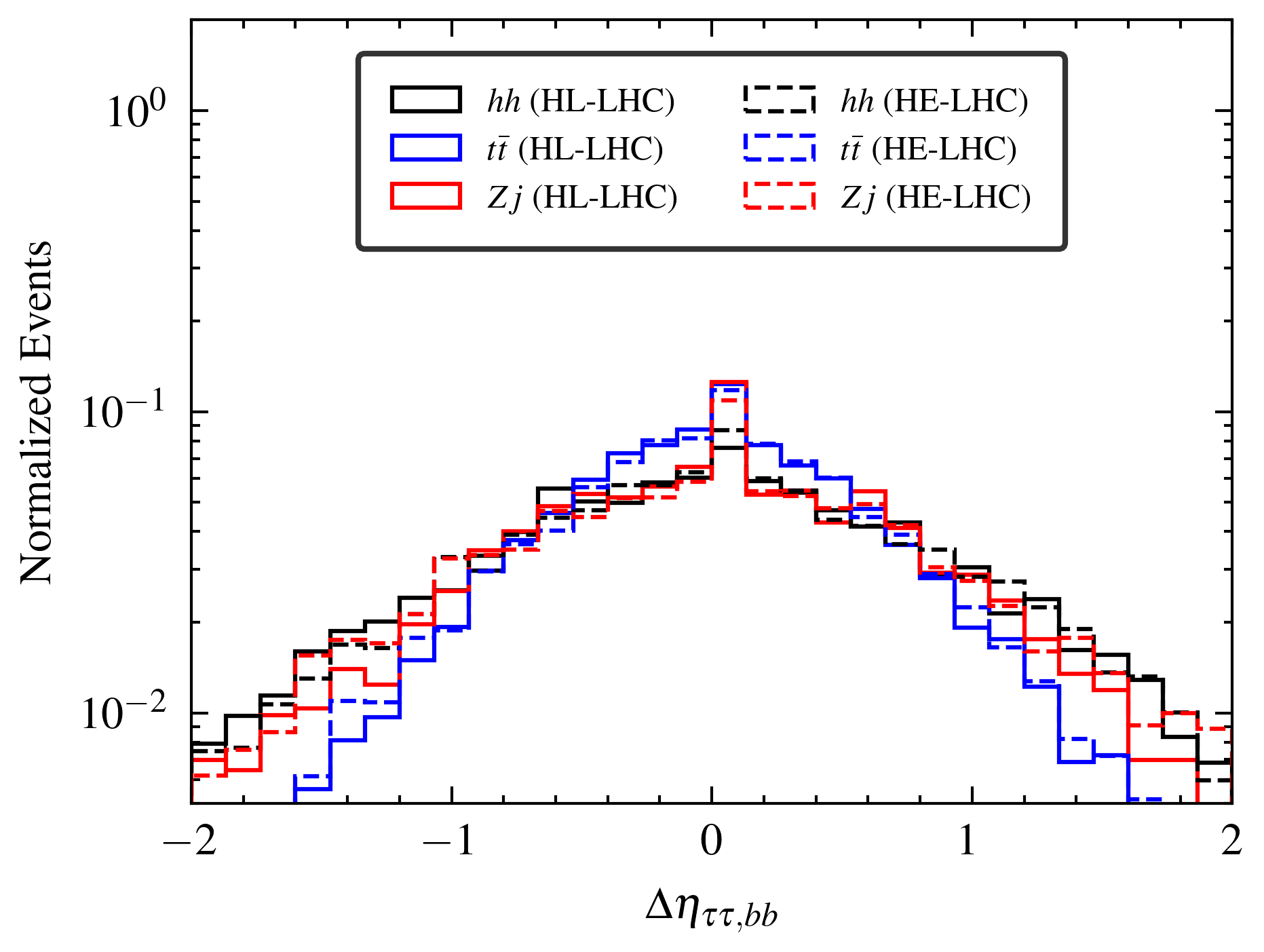}
    \includegraphics[width=0.325\linewidth]{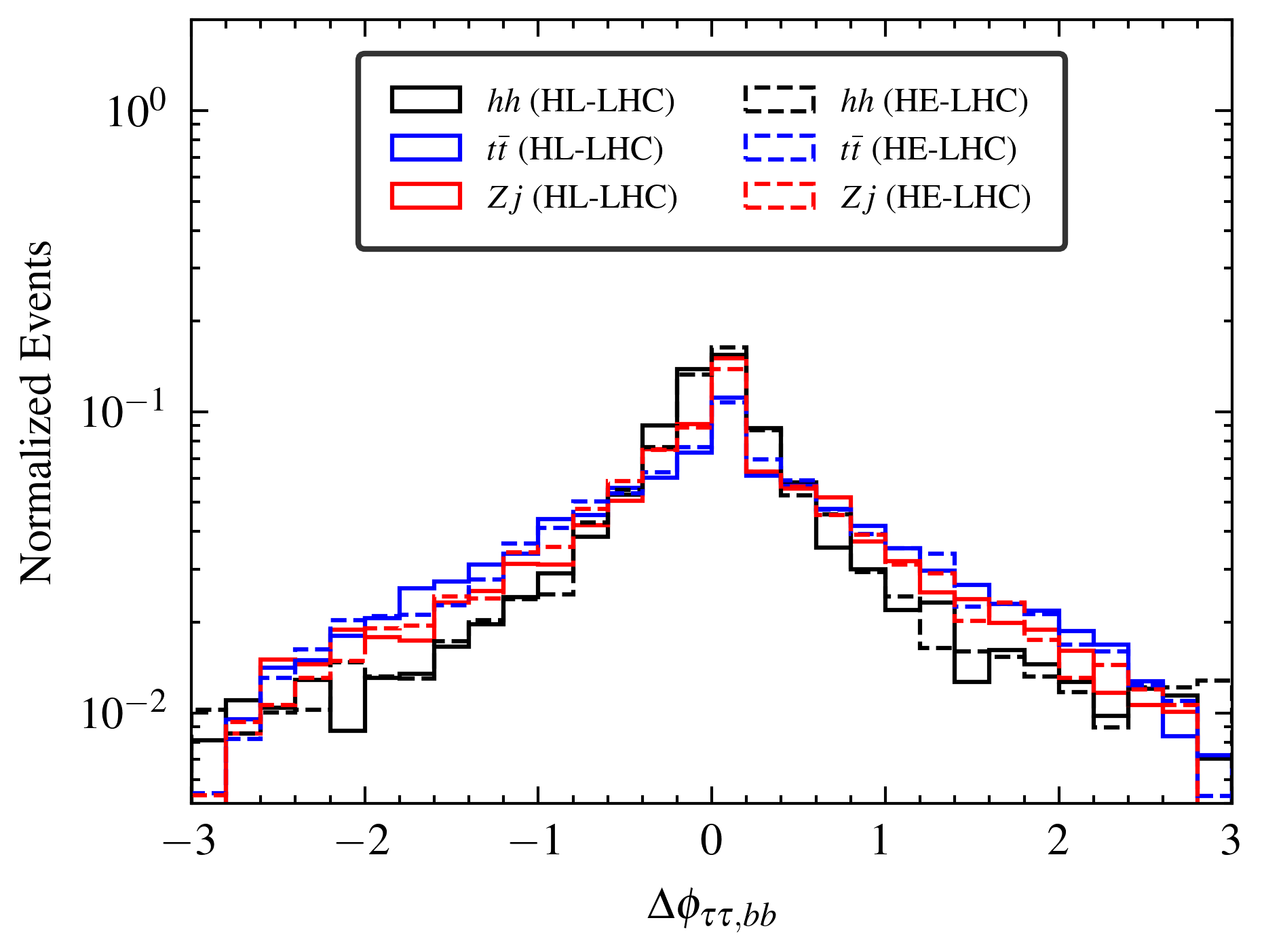}
    \caption{Kinematic distributions of signal and background processes for HL-LHC (\textit{solid}) and HE-LHC (\textit{dashed}). $Zj$ refers to as $Z + \text{jets}$ background.}
    \label{fig:dist}
\end{figure}

In this section, we compare the performance of our ANN discriminator against traditional linear kinematic cuts. The relevant signal and background kinematic distributions for the HL-LHC (14~TeV, 3~ab$^{-1}$) and HE-LHC (27~TeV, 15~ab$^{-1}$) scenarios are shown in figure~\ref{fig:dist}.
We apply the following optimized kinematic cuts, $\mathcal{C}$, to reduce the signal yield $S$ to a level comparable with the ANN output, and evaluate the resulting background yield $B$ and signal-to-background ratio $S{:}B$ as metrics for comparison:
\begin{equation}
\begin{split}
    &\text{MET} > 25 \text{ GeV}\,,\hspace{0.25cm}\Delta R_{bb} < 2.5\,,\hspace{0.25cm}\Delta R_{\tau\tau} < 2.5\,,\\
    &105 \text{ GeV} < M_{bb} < 145 \text{ GeV}\,,\hspace{0.25cm} -2 <\Delta \eta_{bb} < 2\,,\\
    &105 \text{ GeV} < M_{\tau\tau} < 145 \text{ GeV}\,,\hspace{0.25cm} -2 <\Delta \eta_{\tau\tau} < 2\,.\\
\end{split}
\end{equation}
The comparison is provided in Table~\ref{tab:sbx}. We observe that the relative improvement in the $S{:}B$ ratio is approximately an order of magnitude for a comparable background yield, demonstrating the superior signal-background discrimination achieved by the ANN compared to the traditional cut-based approach.
\begin{table}[htb!]
    \centering
    \begin{tabular}{cccc}
    \hline \hline
        Setup & Class & Events (after $\mathcal{C}$ cut) & Events (after ANN cut) \\ \hline
        \multirow{3}*{HL-LHC (14 TeV, 3 ab$^{-1}$)} 
        & $S$ & $26$  & $38$ \\
        & $B$ & $11805$ & $1562$ \\ \cline{2-4}
        & $S:B$ & $2.204\times 10^{-3}$ & $2.433 \times 10^{-2}$ \\ \hline
        \multirow{3}*{HE-LHC (27 TeV, 15 ab$^{-1}$)} 
        & $S$ & $521$  & $396$ \\
        & $B$ & $333948$ & $19390$ \\ \cline{2-4}
        & $S:B$ & $1.560 \times 10^{-3}$ & $2.042 \times 10^{-2}$ \\ \hline \hline
    \end{tabular}
    \caption{Signal ($S$) and background ($B$) event counts after traditional kinematic cuts ($\mathcal{C}$) and the ANN threshold cut for HL-LHC (14 TeV, 3 ab$^{-1}$) and HE-LHC (27 TeV, 15 ab$^{-1}$) scenarios.}
    \label{tab:sbx}
\end{table}

\section{Robustness of ANN models}
\label{app:2}
\begin{figure}[t]
    \centering
    \includegraphics[width=0.475\linewidth]{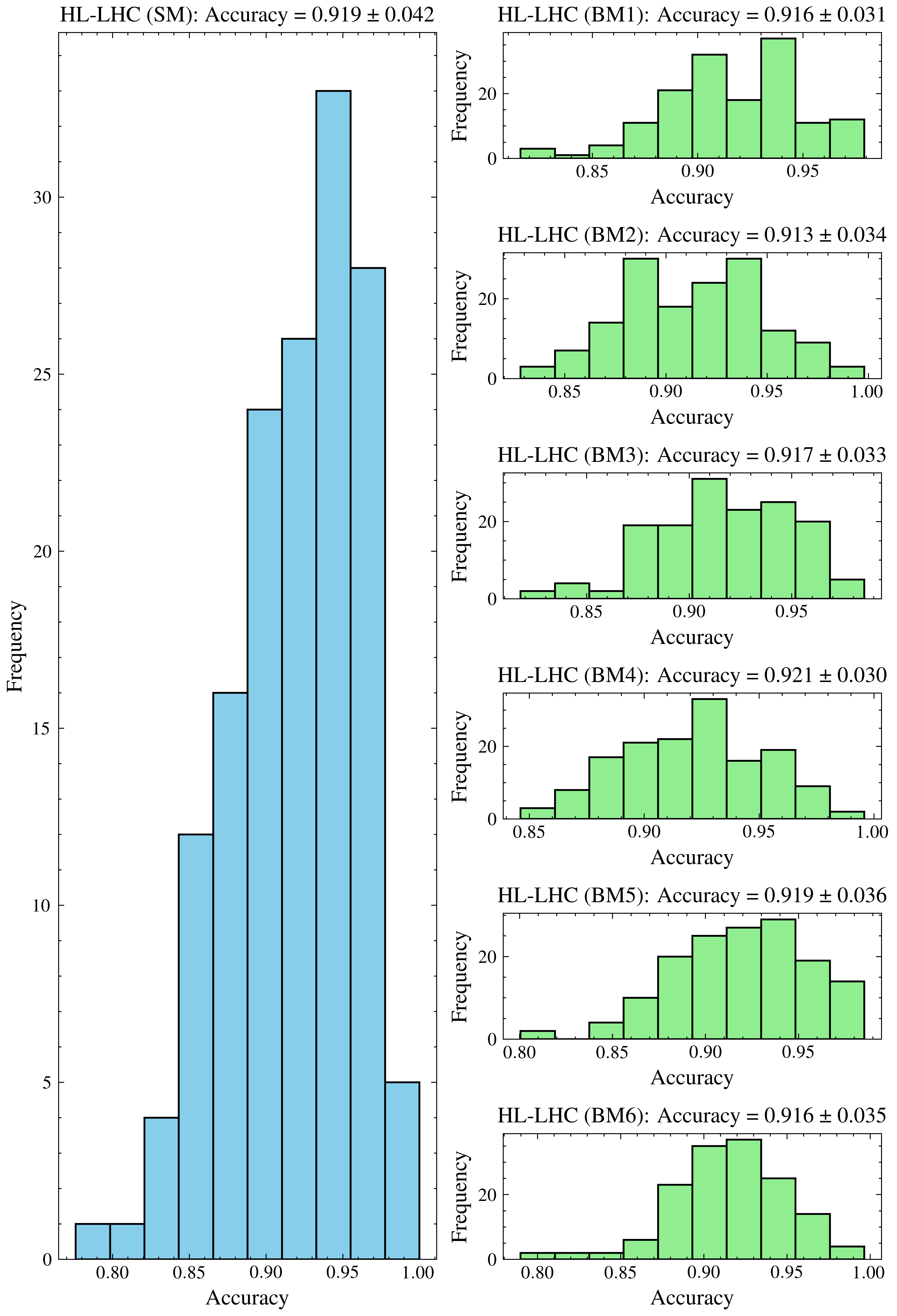}
    \includegraphics[width=0.475\linewidth]{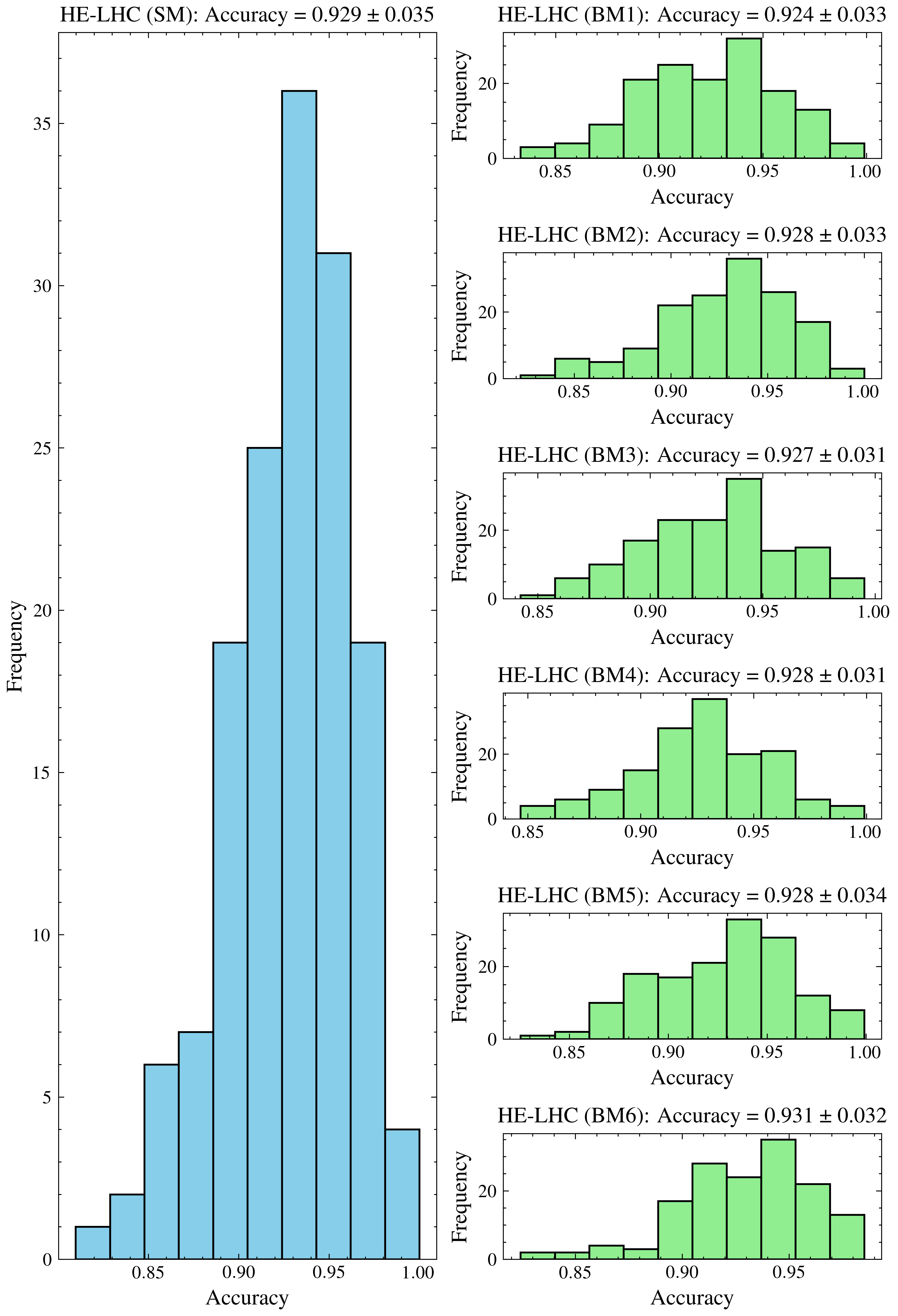}
    \caption{Accuracy score distributions for the \textit{SM-only} and EFT benchmark ensembles (100 events each, repeated over 250 iterations) for the HL-LHC (\textit{left}) and HE-LHC (\textit{right}) ANN models.}
    \label{fig:accscores}
\end{figure}
To assess the robustness of the ANN models, we compute the accuracy scores and their associated uncertainties for several EFT benchmark working points and compare them with the accuracy obtained from the \textit{SM only} dataset used for training. For this purpose, we construct small ensembles of 100 randomly selected events for both the \textit{SM only} sample and each EFT benchmark sample, containing signal and background in equal proportion, and evaluate the ANN accuracy for both HL-LHC and HE-LHC configurations. This procedure is repeated for 150 iterations, from which we extract the mean accuracy and the corresponding $1\sigma$ uncertainty. Robustness is established if the uncertainty bands around the mean accuracy for all benchmarks overlap. We consider the following EFT benchmark points $(C_{H}/\Lambda^{2},\,C_{H\Box}/\Lambda^{2},\,C_{HD}/\Lambda^{2},\,C_{tH}/\Lambda^{2})$ (in TeV$^{-2}$):
\begin{equation}
\begin{split}
    &\text{BM1}: (-3,\,3,\,0,\,0)\,, \hspace{0.8cm} \text{BM2}: (-3,\,-3,\,0,\,0)\,,\\
    &\text{BM3}: (-3,\,0,\,3,\,0)\,, \hspace{0.8cm} \text{BM4}: (-3,\,0,\,-3,\,0)\,,\\
    &\text{BM5}: (-3,\,0,\,0,\,3)\,, \hspace{0.8cm} \text{BM6}: (-3,\,0,\,0,\,-3)\,.
\end{split}
\end{equation}
The corresponding accuracy values for the \textit{SM only} dataset are
\begin{equation}
    \text{Accuracy (SM)}:\quad 0.919 \pm 0.042~(\text{HL-LHC})\,, \qquad 0.929 \pm 0.035~(\text{HE-LHC})\,,
\end{equation}
while the benchmark accuracies are
\begin{equation}
\begin{split}
    \text{Accuracy (BM1)}:&\quad 0.916 \pm 0.031~(\text{HL-LHC})\,, \qquad 0.924 \pm 0.033~(\text{HE-LHC})\,;\\
    \text{Accuracy (BM2)}:&\quad 0.913 \pm 0.034~(\text{HL-LHC})\,, \qquad 0.928 \pm 0.033~(\text{HE-LHC})\,;\\
    \text{Accuracy (BM3)}:&\quad 0.917 \pm 0.033~(\text{HL-LHC})\,, \qquad 0.927 \pm 0.031~(\text{HE-LHC})\,;\\
    \text{Accuracy (BM4)}:&\quad 0.921 \pm 0.030~(\text{HL-LHC})\,, \qquad 0.928 \pm 0.031~(\text{HE-LHC})\,;\\
    \text{Accuracy (BM5)}:&\quad 0.919 \pm 0.036~(\text{HL-LHC})\,, \qquad 0.928 \pm 0.034~(\text{HE-LHC})\,;\\
    \text{Accuracy (BM6)}:&\quad 0.916 \pm 0.035~(\text{HL-LHC})\,, \qquad 0.931 \pm 0.032~(\text{HE-LHC})\,.
\end{split}
\end{equation}
The accuracy frequency distribution histograms are plotted in figure~\ref{fig:accscores}. In all cases, the mean accuracies lie within one another’s $1\sigma$ ranges, confirming the robustness of the ANN models. This behaviour is expected, as the most sensitive discriminating variables (see figure~\ref{fig:dist}) predominantly separate Higgs final states from non-Higgs backgrounds, leaving the EFT-induced modifications effectively unaltered in these features.

\section{Validation of EFT framework}
\label{sec:uv.com}
A simple and well-motivated ultraviolet completion of the dimension-6 SMEFT operator $\mathcal{O}_H=(H^\dagger H)^3$ is provided by the SM extended with a non-$\mathbb{Z}_2$ symmetric heavy real singlet scalar ($S$) \cite{Krasnikov:1997nh,OConnell:2006rsp}. The most general renormalizable scalar potential with $H$ and $S$ is
\begin{equation}
\begin{split}
    V(H,S) =& -\mu_H^2 (H^\dagger H) + \lambda_H (H^\dagger H)^2 + \frac12 m_S^2 S^2 + \frac{\mu_3}{3!} S^3 \\&+ \frac{\lambda_S}{4!} S^4
          + \mu_S \, S (H^\dagger H) + \frac{\kappa}{2} \, S^2 (H^\dagger H).
\end{split}
\end{equation}
Here $\mu_H^2<0$ is the Higgs mass parameter that triggers EWSB, and $\lambda_H$ denotes the SM Higgs self-coupling. The parameter $m_S$ represents the mass of $S$, while $\mu_3$ and $\lambda_S$ correspond to its trilinear and quartic self-interactions, respectively. The interaction between the Higgs doublet and the singlet scalar is mediated by two portal couplings: a dimensionful coupling $\mu_S$ and a dimensionless coupling $\kappa$, which parametrize the linear and quadratic Higgs–singlet interactions in the scalar potential. In the limit $m_S \gg v$, the singlet scalar can be integrated out at tree level. Using the classical equation of motion for $S$ and performing straightforward algebraic manipulations, one obtains the corresponding tree-level matching between the singlet scalar model and the SMEFT with the relation:
\begin{equation}
    \left|\frac{C_H}{\Lambda^2}\right|=\frac{\kappa \mu_S^2}{2m_S^4}.
\end{equation} 
The validity of the EFT can be ensured by the perturbative validity which requires:
    \begin{equation}
        |\kappa| < 4\pi\,, \hspace{1cm} \frac{\mu_{S}}{m_{S}} < 4\pi\,.
    \end{equation}
Hence, proper tuning of the BSM parameters $\kappa$, $\mu_{S}$ and $m_{S}$ can easily accommodate $|C_{H}/\Lambda^{2}| \sim \mathcal{O}(1)$ TeV$^{-2}$.
\bibliographystyle{JHEP} 
\bibliography{eft_gw}

\providecommand{\href}[2]{#2}\begingroup\raggedright\begin{thebibliography}{100}

\bibitem{ATLAS:2012yve}
{\scshape ATLAS} collaboration, \emph{{Observation of a new particle in the
  search for the Standard Model Higgs boson with the ATLAS detector at the
  LHC}}, \href{https://doi.org/10.1016/j.physletb.2012.08.020}{\emph{Phys.
  Lett. B} {\bfseries 716} (2012) 1}
  [\href{https://arxiv.org/abs/1207.7214}{{\ttfamily 1207.7214}}].

\bibitem{CMS:2012qbp}
{\scshape CMS} collaboration, \emph{{Observation of a New Boson at a Mass of
  125 GeV with the CMS Experiment at the LHC}},
  \href{https://doi.org/10.1016/j.physletb.2012.08.021}{\emph{Phys. Lett. B}
  {\bfseries 716} (2012) 30} [\href{https://arxiv.org/abs/1207.7235}{{\ttfamily
  1207.7235}}].

\bibitem{Kuzmin:1985mm}
V.~A. Kuzmin, V.~A. Rubakov and M.~E. Shaposhnikov, \emph{{On the Anomalous
  Electroweak Baryon Number Nonconservation in the Early Universe}},
  \href{https://doi.org/10.1016/0370-2693(85)91028-7}{\emph{Phys. Lett. B}
  {\bfseries 155} (1985) 36}.

\bibitem{Gavela:1993ts}
M.~B. Gavela, P.~Hernandez, J.~Orloff and O.~Pene, \emph{{Standard model CP
  violation and baryon asymmetry}},
  \href{https://doi.org/10.1142/S0217732394000629}{\emph{Mod. Phys. Lett. A}
  {\bfseries 9} (1994) 795}
  [\href{https://arxiv.org/abs/hep-ph/9312215}{{\ttfamily hep-ph/9312215}}].

\bibitem{Gurtler:1997hr}
M.~Gurtler, E.-M. Ilgenfritz and A.~Schiller, \emph{{Where the electroweak
  phase transition ends}},
  \href{https://doi.org/10.1103/PhysRevD.56.3888}{\emph{Phys. Rev. D}
  {\bfseries 56} (1997) 3888}
  [\href{https://arxiv.org/abs/hep-lat/9704013}{{\ttfamily hep-lat/9704013}}].

\bibitem{Laine:1998jb}
M.~Laine and K.~Rummukainen, \emph{{What's new with the electroweak phase
  transition?}},
  \href{https://doi.org/10.1016/S0920-5632(99)85017-8}{\emph{Nucl. Phys. B
  Proc. Suppl.} {\bfseries 73} (1999) 180}
  [\href{https://arxiv.org/abs/hep-lat/9809045}{{\ttfamily hep-lat/9809045}}].

\bibitem{Csikor:1998eu}
F.~Csikor, Z.~Fodor and J.~Heitger, \emph{{Endpoint of the hot electroweak
  phase transition}},
  \href{https://doi.org/10.1103/PhysRevLett.82.21}{\emph{Phys. Rev. Lett.}
  {\bfseries 82} (1999) 21}
  [\href{https://arxiv.org/abs/hep-ph/9809291}{{\ttfamily hep-ph/9809291}}].

\bibitem{Aoki:1999fi}
Y.~Aoki, F.~Csikor, Z.~Fodor and A.~Ukawa, \emph{{The Endpoint of the first
  order phase transition of the SU(2) gauge Higgs model on a four-dimensional
  isotropic lattice}},
  \href{https://doi.org/10.1103/PhysRevD.60.013001}{\emph{Phys. Rev. D}
  {\bfseries 60} (1999) 013001}
  [\href{https://arxiv.org/abs/hep-lat/9901021}{{\ttfamily hep-lat/9901021}}].

\bibitem{Kajantie:1996mn}
K.~Kajantie, M.~Laine, K.~Rummukainen and M.~E. Shaposhnikov, \emph{{Is there
  a~ hot electroweak phase transition at $m_H \gtrsim m_W$?}},
  \href{https://doi.org/10.1103/PhysRevLett.77.2887}{\emph{Phys. Rev. Lett.}
  {\bfseries 77} (1996) 2887}
  [\href{https://arxiv.org/abs/hep-ph/9605288}{{\ttfamily hep-ph/9605288}}].

\bibitem{Caprini:2015zlo}
C.~Caprini et~al., \emph{{Science with the space-based interferometer eLISA.
  II: Gravitational waves from cosmological phase transitions}},
  \href{https://doi.org/10.1088/1475-7516/2016/04/001}{\emph{JCAP} {\bfseries
  04} (2016) 001} [\href{https://arxiv.org/abs/1512.06239}{{\ttfamily
  1512.06239}}].

\bibitem{LISA:2017pwj}
{\scshape LISA} collaboration, \emph{{Laser Interferometer Space Antenna}},
  \href{https://arxiv.org/abs/1702.00786}{{\ttfamily 1702.00786}}.

\bibitem{Seto:2001qf}
N.~Seto, S.~Kawamura and T.~Nakamura, \emph{{Possibility of direct measurement
  of the acceleration of the universe using 0.1-Hz band laser interferometer
  gravitational wave antenna in space}},
  \href{https://doi.org/10.1103/PhysRevLett.87.221103}{\emph{Phys. Rev. Lett.}
  {\bfseries 87} (2001) 221103}
  [\href{https://arxiv.org/abs/astro-ph/0108011}{{\ttfamily
  astro-ph/0108011}}].

\bibitem{Kawamura:2006up}
S.~Kawamura et~al., \emph{{The Japanese space gravitational wave antenna
  DECIGO}}, \href{https://doi.org/10.1088/0264-9381/23/8/S17}{\emph{Class.
  Quant. Grav.} {\bfseries 23} (2006) S125}.

\bibitem{Crowder:2005nr}
J.~Crowder and N.~J. Cornish, \emph{{Beyond LISA: Exploring future
  gravitational wave missions}},
  \href{https://doi.org/10.1103/PhysRevD.72.083005}{\emph{Phys. Rev. D}
  {\bfseries 72} (2005) 083005}
  [\href{https://arxiv.org/abs/gr-qc/0506015}{{\ttfamily gr-qc/0506015}}].

\bibitem{Corbin:2005ny}
V.~Corbin and N.~J. Cornish, \emph{{Detecting the cosmic gravitational wave
  background with the big bang observer}},
  \href{https://doi.org/10.1088/0264-9381/23/7/014}{\emph{Class. Quant. Grav.}
  {\bfseries 23} (2006) 2435}
  [\href{https://arxiv.org/abs/gr-qc/0512039}{{\ttfamily gr-qc/0512039}}].

\bibitem{Buchmuller:1985jz}
W.~Buchmuller and D.~Wyler, \emph{{Effective Lagrangian Analysis of New
  Interactions and Flavor Conservation}},
  \href{https://doi.org/10.1016/0550-3213(86)90262-2}{\emph{Nucl. Phys. B}
  {\bfseries 268} (1986) 621}.

\bibitem{Grzadkowski:2010es}
B.~Grzadkowski, M.~Iskrzynski, M.~Misiak and J.~Rosiek, \emph{{Dimension-Six
  Terms in the Standard Model Lagrangian}},
  \href{https://doi.org/10.1007/JHEP10(2010)085}{\emph{JHEP} {\bfseries 10}
  (2010) 085} [\href{https://arxiv.org/abs/1008.4884}{{\ttfamily 1008.4884}}].

\bibitem{Zhang:1992fs}
X.-m. Zhang, \emph{{Operators analysis for Higgs potential and cosmological
  bound on Higgs mass}},
  \href{https://doi.org/10.1103/PhysRevD.47.3065}{\emph{Phys. Rev. D}
  {\bfseries 47} (1993) 3065}
  [\href{https://arxiv.org/abs/hep-ph/9301277}{{\ttfamily hep-ph/9301277}}].

\bibitem{Grojean:2004xa}
C.~Grojean, G.~Servant and J.~D. Wells, \emph{{First-order electroweak phase
  transition in the standard model with a low cutoff}},
  \href{https://doi.org/10.1103/PhysRevD.71.036001}{\emph{Phys. Rev. D}
  {\bfseries 71} (2005) 036001}
  [\href{https://arxiv.org/abs/hep-ph/0407019}{{\ttfamily hep-ph/0407019}}].

\bibitem{Bodeker:2004ws}
D.~Bodeker, L.~Fromme, S.~J. Huber and M.~Seniuch, \emph{{The Baryon asymmetry
  in the standard model with a low cut-off}},
  \href{https://doi.org/10.1088/1126-6708/2005/02/026}{\emph{JHEP} {\bfseries
  02} (2005) 026} [\href{https://arxiv.org/abs/hep-ph/0412366}{{\ttfamily
  hep-ph/0412366}}].

\bibitem{Damgaard:2015con}
P.~H. Damgaard, A.~Haarr, D.~O'Connell and A.~Tranberg, \emph{{Effective Field
  Theory and Electroweak Baryogenesis in the Singlet-Extended Standard Model}},
  \href{https://doi.org/10.1007/JHEP02(2016)107}{\emph{JHEP} {\bfseries 02}
  (2016) 107} [\href{https://arxiv.org/abs/1512.01963}{{\ttfamily
  1512.01963}}].

\bibitem{Harman:2015gif}
C.~P.~D. Harman and S.~J. Huber, \emph{{Does zero temperature decide on the
  nature of the electroweak phase transition?}},
  \href{https://doi.org/10.1007/JHEP06(2016)005}{\emph{JHEP} {\bfseries 06}
  (2016) 005} [\href{https://arxiv.org/abs/1512.05611}{{\ttfamily
  1512.05611}}].

\bibitem{deVries:2017ncy}
J.~de~Vries, M.~Postma, J.~van~de Vis and G.~White, \emph{{Electroweak
  Baryogenesis and the Standard Model Effective Field Theory}},
  \href{https://doi.org/10.1007/JHEP01(2018)089}{\emph{JHEP} {\bfseries 01}
  (2018) 089} [\href{https://arxiv.org/abs/1710.04061}{{\ttfamily
  1710.04061}}].

\bibitem{DeVries:2018aul}
J.~De~Vries, M.~Postma and J.~van~de Vis, \emph{{The role of leptons in
  electroweak baryogenesis}},
  \href{https://doi.org/10.1007/JHEP04(2019)024}{\emph{JHEP} {\bfseries 04}
  (2019) 024} [\href{https://arxiv.org/abs/1811.11104}{{\ttfamily
  1811.11104}}].

\bibitem{Kanemura:2020yyr}
S.~Kanemura and M.~Tanaka, \emph{{Higgs boson coupling as a probe of the
  sphaleron property}},
  \href{https://doi.org/10.1016/j.physletb.2020.135711}{\emph{Phys. Lett. B}
  {\bfseries 809} (2020) 135711}
  [\href{https://arxiv.org/abs/2005.05250}{{\ttfamily 2005.05250}}].

\bibitem{Zhu:2025pht}
Y.~Zhu, J.~Liu, R.~Qin and L.~Bian, \emph{{Theoretical uncertainties in
  first-order electroweak phase transitions}},
  \href{https://doi.org/10.1103/f4gr-hycg}{\emph{Phys. Rev. D} {\bfseries 112}
  (2025) 015018} [\href{https://arxiv.org/abs/2503.19566}{{\ttfamily
  2503.19566}}].

\bibitem{Delaunay:2007wb}
C.~Delaunay, C.~Grojean and J.~D. Wells, \emph{{Dynamics of Non-renormalizable
  Electroweak Symmetry Breaking}},
  \href{https://doi.org/10.1088/1126-6708/2008/04/029}{\emph{JHEP} {\bfseries
  04} (2008) 029} [\href{https://arxiv.org/abs/0711.2511}{{\ttfamily
  0711.2511}}].

\bibitem{Cai:2017tmh}
R.-G. Cai, M.~Sasaki and S.-J. Wang, \emph{{The gravitational waves from the
  first-order phase transition with a dimension-six operator}},
  \href{https://doi.org/10.1088/1475-7516/2017/08/004}{\emph{JCAP} {\bfseries
  08} (2017) 004} [\href{https://arxiv.org/abs/1707.03001}{{\ttfamily
  1707.03001}}].

\bibitem{Chala:2018ari}
M.~Chala, C.~Krause and G.~Nardini, \emph{{Signals of the electroweak phase
  transition at colliders and gravitational wave observatories}},
  \href{https://doi.org/10.1007/JHEP07(2018)062}{\emph{JHEP} {\bfseries 07}
  (2018) 062} [\href{https://arxiv.org/abs/1802.02168}{{\ttfamily
  1802.02168}}].

\bibitem{Ellis:2019flb}
S.~A.~R. Ellis, S.~Ipek and G.~White, \emph{{Electroweak Baryogenesis from
  Temperature-Varying Couplings}},
  \href{https://doi.org/10.1007/JHEP08(2019)002}{\emph{JHEP} {\bfseries 08}
  (2019) 002} [\href{https://arxiv.org/abs/1905.11994}{{\ttfamily
  1905.11994}}].

\bibitem{Zhou:2019uzq}
R.~Zhou, L.~Bian and H.-K. Guo, \emph{{Connecting the electroweak sphaleron
  with gravitational waves}},
  \href{https://doi.org/10.1103/PhysRevD.101.091903}{\emph{Phys. Rev. D}
  {\bfseries 101} (2020) 091903}
  [\href{https://arxiv.org/abs/1910.00234}{{\ttfamily 1910.00234}}].

\bibitem{Banerjee:2024qiu}
U.~Banerjee, S.~Chakraborty, S.~Prakash and S.~U. Rahaman, \emph{{Feasibility
  of ultrarelativistic bubbles in SMEFT}},
  \href{https://doi.org/10.1103/PhysRevD.110.055002}{\emph{Phys. Rev. D}
  {\bfseries 110} (2024) 055002}
  [\href{https://arxiv.org/abs/2402.02914}{{\ttfamily 2402.02914}}].

\bibitem{Gazi:2024boc}
D.~Gazi, A.~Mukherjee, S.~Niyogi and S.~Poddar, \emph{{Search for Stochastic GW
  Signal as a Complementary Approach to Multi-Higgs Productions at the Hadron
  Colliders to Probe Dimension Six Operator}},
  \href{https://arxiv.org/abs/2408.13326}{{\ttfamily 2408.13326}}.

\bibitem{Athron:2023xlk}
P.~Athron, C.~Bal{\'a}zs, A.~Fowlie, L.~Morris and L.~Wu, \emph{{Cosmological
  phase transitions: From perturbative particle physics to gravitational
  waves}}, \href{https://doi.org/10.1016/j.ppnp.2023.104094}{\emph{Prog. Part.
  Nucl. Phys.} {\bfseries 135} (2024) 104094}
  [\href{https://arxiv.org/abs/2305.02357}{{\ttfamily 2305.02357}}].

\bibitem{Patel:2011th}
H.~H. Patel and M.~J. Ramsey-Musolf, \emph{{Baryon Washout, Electroweak Phase
  Transition, and Perturbation Theory}},
  \href{https://doi.org/10.1007/JHEP07(2011)029}{\emph{JHEP} {\bfseries 07}
  (2011) 029} [\href{https://arxiv.org/abs/1101.4665}{{\ttfamily 1101.4665}}].

\bibitem{Camargo-Molina:2021zgz}
J.~E. Camargo-Molina, R.~Enberg and J.~L{\"o}fgren, \emph{{A new perspective on
  the electroweak phase transition in the Standard Model Effective Field
  Theory}}, \href{https://doi.org/10.1007/JHEP10(2021)127}{\emph{JHEP}
  {\bfseries 10} (2021) 127}
  [\href{https://arxiv.org/abs/2103.14022}{{\ttfamily 2103.14022}}].

\bibitem{Qin:2024idc}
R.~Qin and L.~Bian, \emph{{First-order electroweak phase transition at finite
  density}}, \href{https://doi.org/10.1007/JHEP08(2024)157}{\emph{JHEP}
  {\bfseries 08} (2024) 157}
  [\href{https://arxiv.org/abs/2407.01981}{{\ttfamily 2407.01981}}].

\bibitem{Qin:2024dfp}
R.~Qin and L.~Bian, \emph{{First-order electroweak phase transition with a
  gauge-invariant approach}},
  \href{https://doi.org/10.1103/PhysRevD.111.L051702}{\emph{Phys. Rev. D}
  {\bfseries 111} (2025) L051702}
  [\href{https://arxiv.org/abs/2408.09677}{{\ttfamily 2408.09677}}].

\bibitem{Camargo-Molina:2024sde}
E.~Camargo-Molina, R.~Enberg and J.~L{\"o}fgren, \emph{{A catalog of
  first-order electroweak phase transitions in the Standard Model Effective
  Field Theory}}, \href{https://doi.org/10.1007/JHEP08(2025)113}{\emph{JHEP}
  {\bfseries 08} (2025) 113}
  [\href{https://arxiv.org/abs/2410.23210}{{\ttfamily 2410.23210}}].

\bibitem{Chala:2025aiz}
M.~Chala and G.~Guedes, \emph{{The high-temperature limit of the SM(EFT)}},
  \href{https://doi.org/10.1007/JHEP07(2025)085}{\emph{JHEP} {\bfseries 07}
  (2025) 085} [\href{https://arxiv.org/abs/2503.20016}{{\ttfamily
  2503.20016}}].

\bibitem{Chala:2025cya}
M.~Chala, A.~Dashko and G.~Guedes, \emph{{Running Couplings in High-Temperature
  Effective Field Theory}},  \href{https://arxiv.org/abs/2510.26878}{{\ttfamily
  2510.26878}}.

\bibitem{ATLAS:2022jtk}
{\scshape ATLAS} collaboration, \emph{{Constraints on the Higgs boson
  self-coupling from single- and double-Higgs production with the ATLAS
  detector using pp collisions at $\sqrt{s}=13$ TeV}},
  \href{https://doi.org/10.1016/j.physletb.2023.137745}{\emph{Phys. Lett. B}
  {\bfseries 843} (2023) 137745}
  [\href{https://arxiv.org/abs/2211.01216}{{\ttfamily 2211.01216}}].

\bibitem{CMS:2022dwd}
{\scshape CMS} collaboration, \emph{{A portrait of the Higgs boson by the CMS
  experiment ten years after the discovery.}},
  \href{https://doi.org/10.1038/s41586-022-04892-x}{\emph{Nature} {\bfseries
  607} (2022) 60} [\href{https://arxiv.org/abs/2207.00043}{{\ttfamily
  2207.00043}}].

\bibitem{Feickert:2021ajf}
M.~Feickert and B.~Nachman, \emph{{A Living Review of Machine Learning for
  Particle Physics}},  \href{https://arxiv.org/abs/2102.02770}{{\ttfamily
  2102.02770}}.

\bibitem{Hashino:2022ghd}
K.~Hashino and D.~Ueda, \emph{{SMEFT effects on the gravitational wave spectrum
  from an electroweak phase transition}},
  \href{https://doi.org/10.1103/PhysRevD.107.095022}{\emph{Phys. Rev. D}
  {\bfseries 107} (2023) 095022}
  [\href{https://arxiv.org/abs/2210.11241}{{\ttfamily 2210.11241}}].

\bibitem{Coleman:1973jx}
S.~R. Coleman and E.~J. Weinberg, \emph{{Radiative Corrections as the Origin of
  Spontaneous Symmetry Breaking}},
  \href{https://doi.org/10.1103/PhysRevD.7.1888}{\emph{Phys. Rev. D} {\bfseries
  7} (1973) 1888}.

\bibitem{Elias-Miro:2014pca}
J.~Elias-Miro, J.~R. Espinosa and T.~Konstandin, \emph{{Taming Infrared
  Divergences in the Effective Potential}},
  \href{https://doi.org/10.1007/JHEP08(2014)034}{\emph{JHEP} {\bfseries 08}
  (2014) 034} [\href{https://arxiv.org/abs/1406.2652}{{\ttfamily 1406.2652}}].

\bibitem{Espinosa:2016uaw}
J.~R. Espinosa, M.~Garny and T.~Konstandin, \emph{{Interplay of Infrared
  Divergences and Gauge-Dependence of the Effective Potential}},
  \href{https://doi.org/10.1103/PhysRevD.94.055026}{\emph{Phys. Rev. D}
  {\bfseries 94} (2016) 055026}
  [\href{https://arxiv.org/abs/1607.08432}{{\ttfamily 1607.08432}}].

\bibitem{Dolan:1973qd}
L.~Dolan and R.~Jackiw, \emph{{Symmetry Behavior at Finite Temperature}},
  \href{https://doi.org/10.1103/PhysRevD.9.3320}{\emph{Phys. Rev. D} {\bfseries
  9} (1974) 3320}.

\bibitem{Weinberg:1974hy}
S.~Weinberg, \emph{{Gauge and Global Symmetries at High Temperature}},
  \href{https://doi.org/10.1103/PhysRevD.9.3357}{\emph{Phys. Rev. D} {\bfseries
  9} (1974) 3357}.

\bibitem{Linde:1978px}
A.~D. Linde, \emph{{Phase Transitions in Gauge Theories and Cosmology}},
  \href{https://doi.org/10.1088/0034-4885/42/3/001}{\emph{Rept. Prog. Phys.}
  {\bfseries 42} (1979) 389}.

\bibitem{Linde:1980ts}
A.~D. Linde, \emph{{Infrared Problem in Thermodynamics of the Yang-Mills Gas}},
  \href{https://doi.org/10.1016/0370-2693(80)90769-8}{\emph{Phys. Lett. B}
  {\bfseries 96} (1980) 289}.

\bibitem{Carrington:1991hz}
M.~E. Carrington, \emph{{The Effective potential at finite temperature in the
  Standard Model}}, \href{https://doi.org/10.1103/PhysRevD.45.2933}{\emph{Phys.
  Rev. D} {\bfseries 45} (1992) 2933}.

\bibitem{Parwani:1991gq}
R.~R. Parwani, \emph{{Resummation in a hot scalar field theory}},
  \href{https://doi.org/10.1103/PhysRevD.45.4695}{\emph{Phys. Rev. D}
  {\bfseries 45} (1992) 4695}
  [\href{https://arxiv.org/abs/hep-ph/9204216}{{\ttfamily hep-ph/9204216}}].

\bibitem{Arnold:1992rz}
P.~B. Arnold and O.~Espinosa, \emph{{The Effective potential and first order
  phase transitions: Beyond leading-order}},
  \href{https://doi.org/10.1103/PhysRevD.47.3546}{\emph{Phys. Rev. D}
  {\bfseries 47} (1993) 3546}
  [\href{https://arxiv.org/abs/hep-ph/9212235}{{\ttfamily hep-ph/9212235}}].

\bibitem{Wainwright:2011kj}
C.~L. Wainwright, \emph{{CosmoTransitions: Computing Cosmological Phase
  Transition Temperatures and Bubble Profiles with Multiple Fields}},
  \href{https://doi.org/10.1016/j.cpc.2012.04.004}{\emph{Comput. Phys. Commun.}
  {\bfseries 183} (2012) 2006}
  [\href{https://arxiv.org/abs/1109.4189}{{\ttfamily 1109.4189}}].

\bibitem{Ellis:2018mja}
J.~Ellis, M.~Lewicki and J.~M. No, \emph{{On the Maximal Strength of a
  First-Order Electroweak Phase Transition and its Gravitational Wave Signal}},
  \href{https://doi.org/10.1088/1475-7516/2019/04/003}{\emph{JCAP} {\bfseries
  04} (2019) 003} [\href{https://arxiv.org/abs/1809.08242}{{\ttfamily
  1809.08242}}].

\bibitem{Chala:2025xlk}
M.~Chala, M.~C. Fiore and L.~Gil, \emph{{Hot news on the phase-structure of the
  SMEFT}},  \href{https://arxiv.org/abs/2507.16905}{{\ttfamily 2507.16905}}.

\bibitem{Turner:1990rc}
M.~S. Turner and F.~Wilczek, \emph{{Relic gravitational waves and extended
  inflation}}, \href{https://doi.org/10.1103/PhysRevLett.65.3080}{\emph{Phys.
  Rev. Lett.} {\bfseries 65} (1990) 3080}.

\bibitem{Kosowsky:1991ua}
A.~Kosowsky, M.~S. Turner and R.~Watkins, \emph{{Gravitational radiation from
  colliding vacuum bubbles}},
  \href{https://doi.org/10.1103/PhysRevD.45.4514}{\emph{Phys. Rev. D}
  {\bfseries 45} (1992) 4514}.

\bibitem{Kosowsky:1992rz}
A.~Kosowsky, M.~S. Turner and R.~Watkins, \emph{{Gravitational waves from first
  order cosmological phase transitions}},
  \href{https://doi.org/10.1103/PhysRevLett.69.2026}{\emph{Phys. Rev. Lett.}
  {\bfseries 69} (1992) 2026}.

\bibitem{Kosowsky:1992vn}
A.~Kosowsky and M.~S. Turner, \emph{{Gravitational radiation from colliding
  vacuum bubbles: envelope approximation to many bubble collisions}},
  \href{https://doi.org/10.1103/PhysRevD.47.4372}{\emph{Phys. Rev. D}
  {\bfseries 47} (1993) 4372}
  [\href{https://arxiv.org/abs/astro-ph/9211004}{{\ttfamily
  astro-ph/9211004}}].

\bibitem{Turner:1992tz}
M.~S. Turner, E.~J. Weinberg and L.~M. Widrow, \emph{{Bubble nucleation in
  first order inflation and other cosmological phase transitions}},
  \href{https://doi.org/10.1103/PhysRevD.46.2384}{\emph{Phys. Rev. D}
  {\bfseries 46} (1992) 2384}.

\bibitem{Hindmarsh:2013xza}
M.~Hindmarsh, S.~J. Huber, K.~Rummukainen and D.~J. Weir, \emph{{Gravitational
  waves from the sound of a first order phase transition}},
  \href{https://doi.org/10.1103/PhysRevLett.112.041301}{\emph{Phys. Rev. Lett.}
  {\bfseries 112} (2014) 041301}
  [\href{https://arxiv.org/abs/1304.2433}{{\ttfamily 1304.2433}}].

\bibitem{Giblin:2014qia}
J.~T. Giblin and J.~B. Mertens, \emph{{Gravitional radiation from first-order
  phase transitions in the presence of a fluid}},
  \href{https://doi.org/10.1103/PhysRevD.90.023532}{\emph{Phys. Rev. D}
  {\bfseries 90} (2014) 023532}
  [\href{https://arxiv.org/abs/1405.4005}{{\ttfamily 1405.4005}}].

\bibitem{Hindmarsh:2015qta}
M.~Hindmarsh, S.~J. Huber, K.~Rummukainen and D.~J. Weir, \emph{{Numerical
  simulations of acoustically generated gravitational waves at a first order
  phase transition}},
  \href{https://doi.org/10.1103/PhysRevD.92.123009}{\emph{Phys. Rev. D}
  {\bfseries 92} (2015) 123009}
  [\href{https://arxiv.org/abs/1504.03291}{{\ttfamily 1504.03291}}].

\bibitem{Hindmarsh:2017gnf}
M.~Hindmarsh, S.~J. Huber, K.~Rummukainen and D.~J. Weir, \emph{{Shape of the
  acoustic gravitational wave power spectrum from a first order phase
  transition}}, \href{https://doi.org/10.1103/PhysRevD.96.103520}{\emph{Phys.
  Rev. D} {\bfseries 96} (2017) 103520}
  [\href{https://arxiv.org/abs/1704.05871}{{\ttfamily 1704.05871}}].

\bibitem{Kamionkowski:1993fg}
M.~Kamionkowski, A.~Kosowsky and M.~S. Turner, \emph{{Gravitational radiation
  from first order phase transitions}},
  \href{https://doi.org/10.1103/PhysRevD.49.2837}{\emph{Phys. Rev. D}
  {\bfseries 49} (1994) 2837}
  [\href{https://arxiv.org/abs/astro-ph/9310044}{{\ttfamily
  astro-ph/9310044}}].

\bibitem{Kosowsky:2001xp}
A.~Kosowsky, A.~Mack and T.~Kahniashvili, \emph{{Gravitational radiation from
  cosmological turbulence}},
  \href{https://doi.org/10.1103/PhysRevD.66.024030}{\emph{Phys. Rev. D}
  {\bfseries 66} (2002) 024030}
  [\href{https://arxiv.org/abs/astro-ph/0111483}{{\ttfamily
  astro-ph/0111483}}].

\bibitem{Caprini:2006jb}
C.~Caprini and R.~Durrer, \emph{{Gravitational waves from stochastic
  relativistic sources: Primordial turbulence and magnetic fields}},
  \href{https://doi.org/10.1103/PhysRevD.74.063521}{\emph{Phys. Rev. D}
  {\bfseries 74} (2006) 063521}
  [\href{https://arxiv.org/abs/astro-ph/0603476}{{\ttfamily
  astro-ph/0603476}}].

\bibitem{Gogoberidze:2007an}
G.~Gogoberidze, T.~Kahniashvili and A.~Kosowsky, \emph{{The Spectrum of
  Gravitational Radiation from Primordial Turbulence}},
  \href{https://doi.org/10.1103/PhysRevD.76.083002}{\emph{Phys. Rev. D}
  {\bfseries 76} (2007) 083002}
  [\href{https://arxiv.org/abs/0705.1733}{{\ttfamily 0705.1733}}].

\bibitem{Caprini:2009yp}
C.~Caprini, R.~Durrer and G.~Servant, \emph{{The stochastic gravitational wave
  background from turbulence and magnetic fields generated by a first-order
  phase transition}},
  \href{https://doi.org/10.1088/1475-7516/2009/12/024}{\emph{JCAP} {\bfseries
  12} (2009) 024} [\href{https://arxiv.org/abs/0909.0622}{{\ttfamily
  0909.0622}}].

\bibitem{Niksa:2018ofa}
P.~Niksa, M.~Schlederer and G.~Sigl, \emph{{Gravitational Waves produced by
  Compressible MHD Turbulence from Cosmological Phase Transitions}},
  \href{https://doi.org/10.1088/1361-6382/aac89c}{\emph{Class. Quant. Grav.}
  {\bfseries 35} (2018) 144001}
  [\href{https://arxiv.org/abs/1803.02271}{{\ttfamily 1803.02271}}].

\bibitem{Linde:1980tt}
A.~D. Linde, \emph{{Fate of the False Vacuum at Finite Temperature: Theory and
  Applications}},
  \href{https://doi.org/10.1016/0370-2693(81)90281-1}{\emph{Phys. Lett. B}
  {\bfseries 100} (1981) 37}.

\bibitem{Guada:2020xnz}
V.~Guada, M.~Nemev{\v{s}}ek and M.~Pintar, \emph{{FindBounce: Package for
  multi-field bounce actions}},
  \href{https://doi.org/10.1016/j.cpc.2020.107480}{\emph{Comput. Phys. Commun.}
  {\bfseries 256} (2020) 107480}
  [\href{https://arxiv.org/abs/2002.00881}{{\ttfamily 2002.00881}}].

\bibitem{Ellis:2020nnr}
J.~Ellis, M.~Lewicki and V.~Vaskonen, \emph{{Updated predictions for
  gravitational waves produced in a strongly supercooled phase transition}},
  \href{https://doi.org/10.1088/1475-7516/2020/11/020}{\emph{JCAP} {\bfseries
  11} (2020) 020} [\href{https://arxiv.org/abs/2007.15586}{{\ttfamily
  2007.15586}}].

\bibitem{Caprini:2024hue}
{\scshape LISA Cosmology Working Group} collaboration, \emph{{Gravitational
  waves from first-order phase transitions in LISA: reconstruction pipeline and
  physics interpretation}},
  \href{https://doi.org/10.1088/1475-7516/2024/10/020}{\emph{JCAP} {\bfseries
  10} (2024) 020} [\href{https://arxiv.org/abs/2403.03723}{{\ttfamily
  2403.03723}}].

\bibitem{Bodeker:2009qy}
D.~Bodeker and G.~D. Moore, \emph{{Can electroweak bubble walls run away?}},
  \href{https://doi.org/10.1088/1475-7516/2009/05/009}{\emph{JCAP} {\bfseries
  05} (2009) 009} [\href{https://arxiv.org/abs/0903.4099}{{\ttfamily
  0903.4099}}].

\bibitem{Caprini:2019egz}
C.~Caprini et~al., \emph{{Detecting gravitational waves from cosmological phase
  transitions with LISA: an update}},
  \href{https://doi.org/10.1088/1475-7516/2020/03/024}{\emph{JCAP} {\bfseries
  03} (2020) 024} [\href{https://arxiv.org/abs/1910.13125}{{\ttfamily
  1910.13125}}].

\bibitem{Guo:2020grp}
H.-K. Guo, K.~Sinha, D.~Vagie and G.~White, \emph{{Phase Transitions in an
  Expanding Universe: Stochastic Gravitational Waves in Standard and
  Non-Standard Histories}},
  \href{https://doi.org/10.1088/1475-7516/2021/01/001}{\emph{JCAP} {\bfseries
  01} (2021) 001} [\href{https://arxiv.org/abs/2007.08537}{{\ttfamily
  2007.08537}}].

\bibitem{Espinosa:2010hh}
J.~R. Espinosa, T.~Konstandin, J.~M. No and G.~Servant, \emph{{Energy Budget of
  Cosmological First-order Phase Transitions}},
  \href{https://doi.org/10.1088/1475-7516/2010/06/028}{\emph{JCAP} {\bfseries
  06} (2010) 028} [\href{https://arxiv.org/abs/1004.4187}{{\ttfamily
  1004.4187}}].

\bibitem{Seto:2005qy}
N.~Seto, \emph{{Correlation analysis of stochastic gravitational wave
  background around 0.1-1 Hz}},
  \href{https://doi.org/10.1103/PhysRevD.73.063001}{\emph{Phys. Rev. D}
  {\bfseries 73} (2006) 063001}
  [\href{https://arxiv.org/abs/gr-qc/0510067}{{\ttfamily gr-qc/0510067}}].

\bibitem{Hashino:2018wee}
K.~Hashino, R.~Jinno, M.~Kakizaki, S.~Kanemura, T.~Takahashi and M.~Takimoto,
  \emph{{Selecting models of first-order phase transitions using the synergy
  between collider and gravitational-wave experiments}},
  \href{https://doi.org/10.1103/PhysRevD.99.075011}{\emph{Phys. Rev. D}
  {\bfseries 99} (2019) 075011}
  [\href{https://arxiv.org/abs/1809.04994}{{\ttfamily 1809.04994}}].

\bibitem{Ai:2023see}
W.-Y. Ai, B.~Laurent and J.~van~de Vis, \emph{{Model-independent bubble wall
  velocities in local thermal equilibrium}},
  \href{https://doi.org/10.1088/1475-7516/2023/07/002}{\emph{JCAP} {\bfseries
  07} (2023) 002} [\href{https://arxiv.org/abs/2303.10171}{{\ttfamily
  2303.10171}}].

\bibitem{Ekstedt:2024fyq}
A.~Ekstedt, O.~Gould, J.~Hirvonen, B.~Laurent, L.~Niemi, P.~Schicho et~al.,
  \emph{{How fast does the WallGo? A package for computing wall velocities in
  first-order phase transitions}},
  \href{https://doi.org/10.1007/JHEP04(2025)101}{\emph{JHEP} {\bfseries 04}
  (2025) 101} [\href{https://arxiv.org/abs/2411.04970}{{\ttfamily
  2411.04970}}].

\bibitem{vandeVis:2025plm}
J.~van~de Vis, P.~Schicho, L.~Niemi, B.~Laurent, J.~Hirvonen and O.~Gould,
  \emph{{WallGo investigates: Theoretical uncertainties in the bubble wall
  velocity}},  \href{https://arxiv.org/abs/2510.27691}{{\ttfamily 2510.27691}}.

\bibitem{Croon:2020cgk}
D.~Croon, O.~Gould, P.~Schicho, T.~V.~I. Tenkanen and G.~White,
  \emph{{Theoretical uncertainties for cosmological first-order phase
  transitions}}, \href{https://doi.org/10.1007/JHEP04(2021)055}{\emph{JHEP}
  {\bfseries 04} (2021) 055}
  [\href{https://arxiv.org/abs/2009.10080}{{\ttfamily 2009.10080}}].

\bibitem{Andreassen:2014gha}
A.~Andreassen, W.~Frost and M.~D. Schwartz, \emph{{Consistent Use of the
  Standard Model Effective Potential}},
  \href{https://doi.org/10.1103/PhysRevLett.113.241801}{\emph{Phys. Rev. Lett.}
  {\bfseries 113} (2014) 241801}
  [\href{https://arxiv.org/abs/1408.0292}{{\ttfamily 1408.0292}}].

\bibitem{Andreassen:2014eha}
A.~Andreassen, W.~Frost and M.~D. Schwartz, \emph{{Consistent Use of Effective
  Potentials}}, \href{https://doi.org/10.1103/PhysRevD.91.016009}{\emph{Phys.
  Rev. D} {\bfseries 91} (2015) 016009}
  [\href{https://arxiv.org/abs/1408.0287}{{\ttfamily 1408.0287}}].

\bibitem{ATLAS:2022vkf}
{\scshape ATLAS} collaboration, \emph{{A detailed map of Higgs boson
  interactions by the ATLAS experiment ten years after the discovery}},
  \href{https://doi.org/10.1038/s41586-022-04893-w}{\emph{Nature} {\bfseries
  607} (2022) 52} [\href{https://arxiv.org/abs/2207.00092}{{\ttfamily
  2207.00092}}].

\bibitem{ATL-PHYS-PUB-2022-037}
{\scshape ATLAS} collaboration, \emph{{Combined effective field theory
  interpretation of Higgs boson and weak boson production and decay with ATLAS
  data and electroweak precision observables}},  tech. rep., CERN, Geneva,
  2022.

\bibitem{CMS:2021nnc}
{\scshape CMS} collaboration, \emph{{Constraints on anomalous Higgs boson
  couplings to vector bosons and fermions in its production and decay using the
  four-lepton final state}},
  \href{https://doi.org/10.1103/PhysRevD.104.052004}{\emph{Phys. Rev. D}
  {\bfseries 104} (2021) 052004}
  [\href{https://arxiv.org/abs/2104.12152}{{\ttfamily 2104.12152}}].

\bibitem{CMS:2024bua}
{\scshape CMS} collaboration, \emph{{Constraints on anomalous Higgs boson
  couplings from its production and decay using the WW channel in
  proton{\textendash}proton collisions at $\sqrt{s} = 13~\text {TeV}$}},
  \href{https://doi.org/10.1140/epjc/s10052-024-12925-0}{\emph{Eur. Phys. J. C}
  {\bfseries 84} (2024) 779}
  [\href{https://arxiv.org/abs/2403.00657}{{\ttfamily 2403.00657}}].

\bibitem{Alloul:2013bka}
A.~Alloul, N.~D. Christensen, C.~Degrande, C.~Duhr and B.~Fuks,
  \emph{{FeynRules 2.0 - A complete toolbox for tree-level phenomenology}},
  \href{https://doi.org/10.1016/j.cpc.2014.04.012}{\emph{Comput. Phys. Commun.}
  {\bfseries 185} (2014) 2250}
  [\href{https://arxiv.org/abs/1310.1921}{{\ttfamily 1310.1921}}].

\bibitem{Degrande:2014vpa}
C.~Degrande, \emph{{Automatic evaluation of UV and R2 terms for beyond the
  Standard Model Lagrangians: a proof-of-principle}},
  \href{https://doi.org/10.1016/j.cpc.2015.08.015}{\emph{Comput. Phys. Commun.}
  {\bfseries 197} (2015) 239}
  [\href{https://arxiv.org/abs/1406.3030}{{\ttfamily 1406.3030}}].

\bibitem{Alwall:2011uj}
J.~Alwall, M.~Herquet, F.~Maltoni, O.~Mattelaer and T.~Stelzer, \emph{{MadGraph
  5 : Going Beyond}},
  \href{https://doi.org/10.1007/JHEP06(2011)128}{\emph{JHEP} {\bfseries 06}
  (2011) 128} [\href{https://arxiv.org/abs/1106.0522}{{\ttfamily 1106.0522}}].

\bibitem{Hirschi:2011pa}
V.~Hirschi, R.~Frederix, S.~Frixione, M.~V. Garzelli, F.~Maltoni and R.~Pittau,
  \emph{{Automation of one-loop QCD corrections}},
  \href{https://doi.org/10.1007/JHEP05(2011)044}{\emph{JHEP} {\bfseries 05}
  (2011) 044} [\href{https://arxiv.org/abs/1103.0621}{{\ttfamily 1103.0621}}].

\bibitem{ATLAS:2024pov}
{\scshape ATLAS} collaboration, \emph{{Search for the nonresonant production of
  Higgs boson pairs via gluon fusion and vector-boson fusion in the
  $b\bar{b}\tau^+\tau^-$ final state in proton-proton collisions at
  $\sqrt{s}=13$ TeV with the ATLAS detector}},
  \href{https://doi.org/10.1103/PhysRevD.110.032012}{\emph{Phys. Rev. D}
  {\bfseries 110} (2024) 032012}
  [\href{https://arxiv.org/abs/2404.12660}{{\ttfamily 2404.12660}}].

\bibitem{AH:2022elh}
A.~A~H and H.-S. Shao, \emph{{N$^{3}$LO+N$^{3}$LL QCD improved Higgs pair cross
  sections}}, \href{https://doi.org/10.1007/JHEP02(2023)067}{\emph{JHEP}
  {\bfseries 02} (2023) 067}
  [\href{https://arxiv.org/abs/2209.03914}{{\ttfamily 2209.03914}}].

\bibitem{Bierlich:2022pfr}
C.~Bierlich et~al., \emph{{A comprehensive guide to the physics and usage of
  PYTHIA 8.3}},
  \href{https://doi.org/10.21468/SciPostPhysCodeb.8}{\emph{SciPost Phys.
  Codeb.} {\bfseries 2022} (2022) 8}
  [\href{https://arxiv.org/abs/2203.11601}{{\ttfamily 2203.11601}}].

\bibitem{deFavereau:2013fsa}
{\scshape DELPHES 3} collaboration, \emph{{DELPHES 3, A modular framework for
  fast simulation of a generic collider experiment}},
  \href{https://doi.org/10.1007/JHEP02(2014)057}{\emph{JHEP} {\bfseries 02}
  (2014) 057} [\href{https://arxiv.org/abs/1307.6346}{{\ttfamily 1307.6346}}].

\bibitem{Cacciari:2011ma}
M.~Cacciari, G.~P. Salam and G.~Soyez, \emph{{FastJet User Manual}},
  \href{https://doi.org/10.1140/epjc/s10052-012-1896-2}{\emph{Eur. Phys. J. C}
  {\bfseries 72} (2012) 1896}
  [\href{https://arxiv.org/abs/1111.6097}{{\ttfamily 1111.6097}}].

\bibitem{Ellis:2016jkw}
J.~Ellis, \emph{{TikZ-Feynman: Feynman diagrams with TikZ}},
  \href{https://doi.org/10.1016/j.cpc.2016.08.019}{\emph{Comput. Phys. Commun.}
  {\bfseries 210} (2017) 103}
  [\href{https://arxiv.org/abs/1601.05437}{{\ttfamily 1601.05437}}].

\bibitem{Krasnikov:1997nh}
N.~V. Krasnikov, \emph{{Influence of SU(2) x U(1) singlet scalars on Higgs
  boson signal at LHC}},
  \href{https://doi.org/10.1142/S0217732398000978}{\emph{Mod. Phys. Lett. A}
  {\bfseries 13} (1998) 893}
  [\href{https://arxiv.org/abs/hep-ph/9709467}{{\ttfamily hep-ph/9709467}}].

\bibitem{OConnell:2006rsp}
D.~O'Connell, M.~J. Ramsey-Musolf and M.~B. Wise, \emph{{Minimal Extension of
  the Standard Model Scalar Sector}},
  \href{https://doi.org/10.1103/PhysRevD.75.037701}{\emph{Phys. Rev. D}
  {\bfseries 75} (2007) 037701}
  [\href{https://arxiv.org/abs/hep-ph/0611014}{{\ttfamily hep-ph/0611014}}].

\end{thebibliography}\endgroup
\end{document}